\documentclass[amsmath,amssymb,aps,prb,reprint,twocolumn,superscriptaddress,longbibliography,floatfix]{revtex4-2}

\usepackage{amsmath}
\usepackage{amsfonts}
\usepackage{graphicx}
\usepackage{xcolor}
\usepackage{ulem}
\usepackage{units}  
\usepackage{float}
\usepackage{xspace}
\usepackage{subfigure}
\usepackage{hyperref}
\usepackage{braket}
\usepackage{url}
\usepackage{booktabs}
\usepackage{multirow}

\begin{document}

\title{Interacting ground states of moir{\'e} ladders}

\author{Paban Kumar Patra$^\flat$}
\email{paban.patra@iitb.ac.in}
\affiliation{Department of Physics, Indian Institute of Technology Bombay, Powai, Mumbai 400076, India}
\author{Ranjith R. Kumar$^\flat$}
\email{ranjith.btd6@gmail.com}
\affiliation{Department of Physics, Indian Institute of Technology Bombay, Powai, Mumbai 400076, India}
\author{Yixuan Huang}
\affiliation{Theoretical Division, Los Alamos National Laboratory, Los Alamos, New Mexico 87545, USA}
\affiliation{Center for Integrated Nanotechnologies, Los Alamos National Laboratory, Los Alamos, New Mexico 87545, USA}
\author{Hridis K. Pal}
\email{hridis.pal@iitb.ac.in}
\affiliation{Department of Physics, Indian Institute of Technology Bombay, Powai, Mumbai 400076, India}

\date{\today}

\begin{abstract}
Moir\'e materials have emerged as a rich platform for exploring strong correlation effects in low dimensions, with twisted bilayer graphene (TBG) as a paradigmatic example. To distill the essential ingredients driving moir\'e-induced phases, a simplified one-dimensional analog—a two-leg ladder with spatially modulated interleg hopping and a uniform magnetic flux—was recently introduced. This model, which we refer to as the moir\'e ladder, features a nearly flat lowest-energy band in a suitable parameter regime, capturing the band-flattening mechanism of TBG. We investigate the ground-state phase diagram of the moir\'e ladder using a combination of bosonization and density matrix renormalization group (DMRG) techniques, and systematically disentangle the respective roles of the flux and the hopping modulation. At half filling, previous numerical work identified a metal-insulator transition at finite interaction strength and an unexpected ferromagnetic ground state. Revisiting this, we show that the metal-insulator transition can be understood perturbatively within bosonization, governed by the number of Fermi points. In contrast, the ferromagnetic correlations are nonperturbative and require both flux and spatial modulation—neither alone is sufficient. We extend our analysis to other fillings: one-quarter, three-quarters, slightly above half-filling (half-filling plus two electrons), and slightly below half-filling (half-filling minus two electrons). At moderate interactions, we observe ferromagnetism below half-filling and antiferromagnetism above; at stronger interactions, ferromagnetism dominates across all studied fillings. Reducing the interaction strength relative to the bandwidth induces a transition back to antiferromagnetic behavior. These results reveal a robust and unconventional mechanism for one-dimensional ferromagnetism rooted in moir\'e band structure effects and establish the moir\'e ladder as a minimal model for probing correlation-driven phases in moir\'e systems. Crucially, the analysis demonstrates that periodic interleg hopping alone does not engender new correlated phases; the magnetic flux is essential for the observed unconventional behavior.

\end{abstract}

\maketitle

%\tableofcontents

%%%%%%%

\section{\label{intro}Introduction}

In recent years, moir\'{e} materials have emerged as a versatile platform for studying strongly correlated electronic phenomena in two dimensions. A paradigmatic example is twisted bilayer graphene (TBG), where a small relative twist between two graphene layers generates a long-wavelength moir\'{e} pattern. The resulting moir\'{e} potential significantly reconstructs the electronic band structure, leading to the appearance of nearly flat bands at certain twist angles called ``magic angles''~\cite{ PhysRevB.84.035440,Cao2018}. The associated quenching of the kinetic energy amplifies electron-electron interactions, giving rise to a rich array of correlated phases~\cite{Cao2018corr,Yankowitz2019,sharpe2019,Lu2019,Andrei_2020,2020NatPh..16..725B,bhowmik2023emergentphasesgrapheneflat}. The phase diagram is unusually rich, highly tunable, and experimentally accessible. Similar physics has been observed in trilayer and multilayer moir{\'e} graphene systems, as well as in twisted transition metal dichalcogenides and other moir\'e van der Waals heterostructures~\cite{Wang2020,Regan2020,tang2020,Mak2022}, all of which exhibit tunable moir\'{e} band structures and various interaction-driven ground states.

A range of theoretical models and techniques have been employed to study interaction effects in these systems. The prevailing approach has been to construct models that faithfully reproduce the single-particle band structure of specific moir\'e systems, and then incorporate interactions to explore correlated phases. While this strategy is fruitful in generating detailed phase diagrams and enabling comparisons with experiments, it often obscures the underlying physical mechanisms. The models tend to be highly specific and complex, limiting their ability to reveal general organizing principles. Certain fundamental questions remain unresolved: How much of the observed richness
stems from the specific electronic structure of the host material---e.g., graphene in TBG---and how much of the behavior is generic to the moir\'e potential itself? More specifically, which features of the moir\'e potential are responsible for driving particular correlation-driven phases? The complexity of full moir\'{e} Hamiltonians makes it difficult to disentangle these factors. To gain deeper insight, it is essential to study simplified models that isolate key features of moir\'{e} physics while remaining analytically and numerically tractable.

To this end, a one-dimensional (1D) model was recently proposed~\cite{PhysRevB.102.155429} that captures key aspects of moir\'{e} band formation. The model, which we dub a moir\'e ladder, consists of a two-leg ladder pierced by a uniform magnetic flux, with a spatially modulated interleg hopping amplitude. In a certain parameter regime, the lowest-energy band of this system becomes nearly flat, effectively mimicking the essential band-flattening mechanism of TBG. Interaction effects at half-filling of this quasiflat band were numerically studied, revealing two striking departures from conventional 1D behavior: (i) a metal-insulator transition was found to occur at a finite critical interaction strength, in contrast to the generic result in 1D where infinitesimal interaction suffices; and (ii) the insulating ground state exhibited ferromagnetic order at large interaction strength, rather than the antiferromagnetic order typical of 1D Mott insulators. Importantly, these results arise purely from moir\'e-like ingredients and are independent of graphene-specific physics. They highlight how many-body phenomena can be qualitatively different in moir\'e systems at a fundamental level. Following this, several subsequent studies have explored 1D analogs of moir\'e physics in a variety of models~\cite{PhysRevLett.126.036803,PhysRevLett.130.143801,Gonçalves2024,zhang2025,vorobev2025,wang2024flat}, uncovering a rich variety of emergent phenomena.  These  developments have significantly broadened the scope of moir\'e physics, opening a new frontier for the exploration of moir\'e-driven phenomena in one dimension.

The objective of this work is twofold. First, we further dissect the moir\'e ladder model in order to identify which specific features of the moir\'e potential are responsible for the emergence of the unconventional many-body phases. While Ref.~\cite{PhysRevB.102.155429} successfully separated moir\'e-driven phenomena from the specific electronic structure of graphene, it did not disentangle the roles of the two key ingredients: the spatially modulated interleg hopping and the uniform magnetic flux. Here, we isolate and compare the effects of these two ingredients vis-\`a-vis interactions.  Second, we extend our analysis to a range of electron fillings beyond half-filling, with the aim of probing the stability and evolution of the unusual ferromagnetic ground state. In particular, we ask whether ferromagnetism is unique to half-filling, or whether it persists—and under what conditions—at other dopings.

To pursue these goals, we adopt a two-step approach. First, we revisit the model at half filling and examine the correlated phases in greater detail, uncovering new insights into previously reported results.  We demonstrate that the metal-insulator transition at finite interaction strength can be understood within a weak-coupling bosonization framework, governed by the number of Fermi points, but the ferromagnetic correlations that arise at strong coupling are genuinely nonperturbative, lacking a straightforward explanation within conventional 1D paradigms. Importantly, we show that these unconventional behaviors require both the spatially modulated interleg hopping and the uniform flux; they vanish when the flux is removed, indicating that the periodic modulation alone does not give rise to qualitatively new many-body physics. Rather, it is the presence of flux that plays a crucial role in enabling the novel correlated phases.  Second, we extend our analysis to other electron fillings to investigate the robustness and evolution of the magnetic ground state. Specifically, we study quarter filling, three-quarter filling, and slightly above and below half filling in the full model with flux and periodic interleg hopping. We find that, for moderate interaction strengths where interband mixing is negligible, the system exhibits ferromagnetic correlations below half filling and antiferromagnetic correlations above half filling. At larger interaction strengths, where interband mixing becomes important, ferromagnetic correlations persist across all studied fillings. In addition, whenever the system is ferromagnetic, reducing the interaction strength relative to the bandwidth induces a transition to antiferromagnetic correlations. 

Beyond its implications for moir\'{e} physics, our work contributes to a broader understanding of correlated states in one dimension. A well-known theorem by Lieb and Mattis~\cite{PhysRev.125.164} asserts that the ground state of a generic 1D system has the smallest possible total spin, thereby ruling out ferromagnetism under standard conditions. However, certain models that include beyond-nearest-neighbor hopping terms can evade this constraint, leading to ferromagnetic ground states~\cite{MIELKE1993443,10.1143/PTP.99.489,Mielke_1991,PhysRevB.40.2354}. Our findings are directly relevant to this line of inquiry. The emergence of ferromagnetic spin correlations in the moir\'{e} ladder across a range of fillings suggests a new mechanism for realizing ferromagnetism in 1D systems, grounded in band structure effects induced by the moir\'{e} potential.

The remainder of this paper is organized as follows. In Section~\ref{model}, we introduce the moir\'e ladder model and outline how it captures the main ingredients of quasiflat moir\'e band formation in TBG. Section~\ref{bandstructure} presents an analysis of the single-particle band structure and its key properties. In Section~\ref{halffilling}, we study interaction-driven ground states at half filling, both at weak and strong coupling limits, using a combination of analytical and numerical approaches. This includes a careful reexamination of the results of Ref.~\cite{PhysRevB.102.155429}, providing new insight into their origin. Section~\ref{otherfillings} extends our analysis to a range of other fillings, exploring how magnetic correlations evolve with doping and whether ferromagnetic order persists beyond half filling. Section~\ref{discussion} offers a broad discussion of our results in the context of both moir\'{e} and 1D correlated physics. Finally, Section~\ref{conclusion} summarizes our main findings.

%%%%%%%%%%%%%%%%%%%%%                

\section{\label{model}Model and motivation}

\begin{figure*}
\centering
\includegraphics[width=1.6\columnwidth]{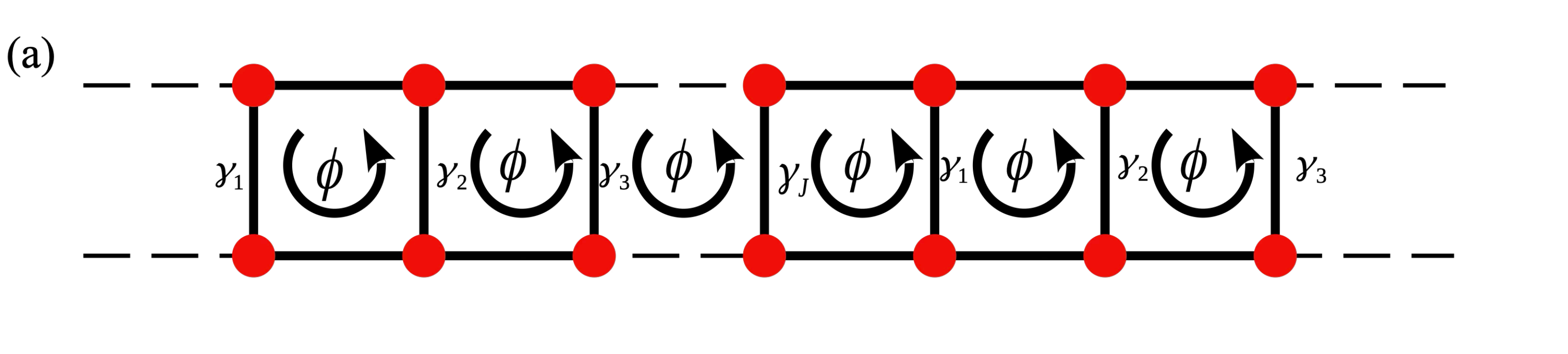}\vspace{2mm}
\includegraphics[width=1.6\columnwidth]{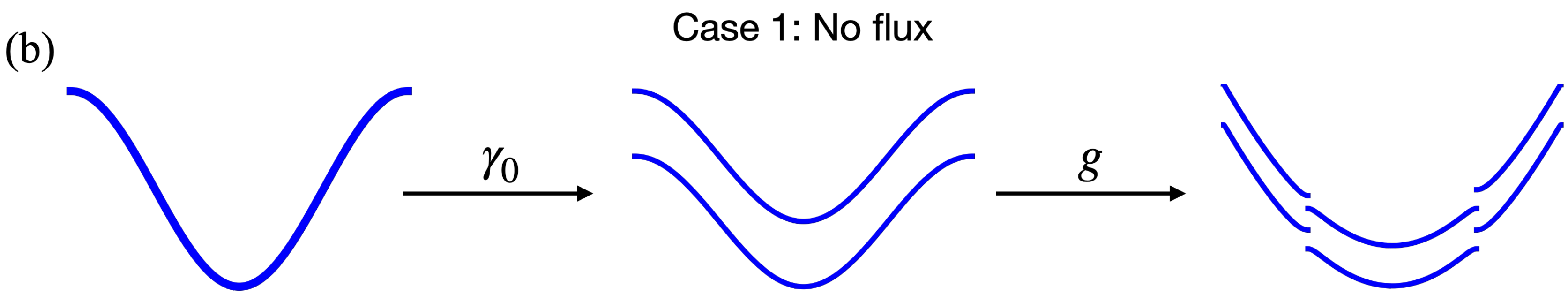}\vspace{4mm}
\includegraphics[width=1.6\columnwidth]{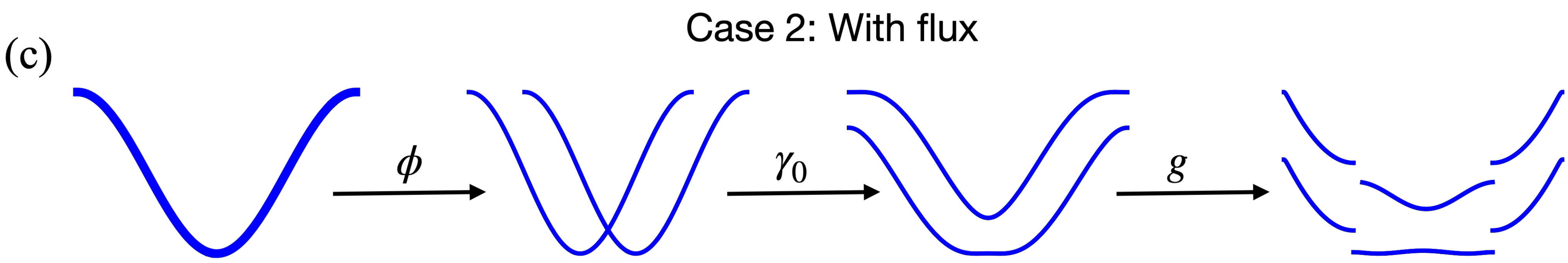}\vspace{4mm}
 \includegraphics[width=1.6\columnwidth]{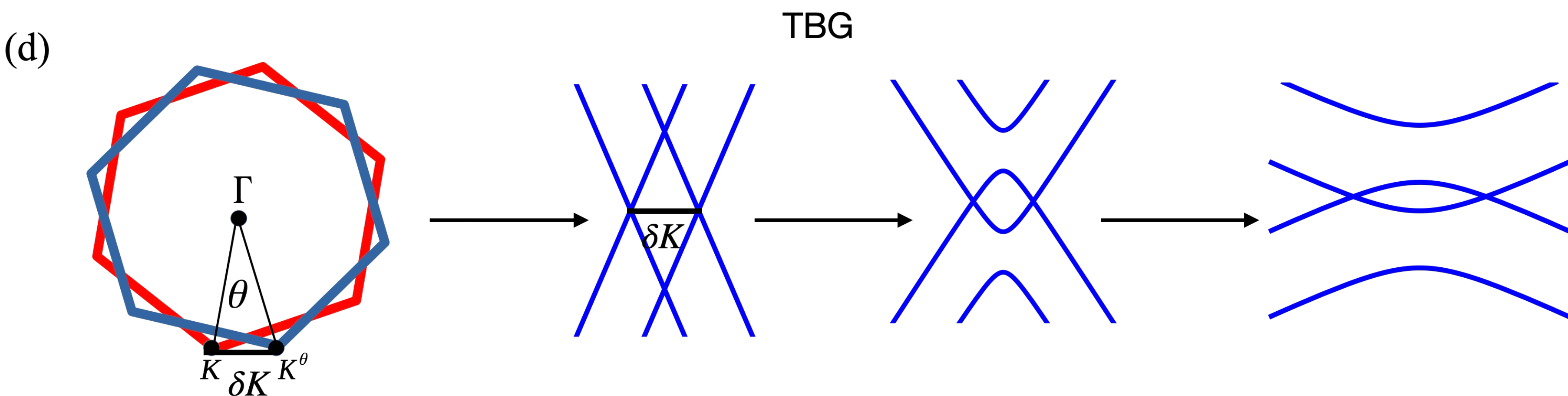}\vspace{4mm}
\caption{(a) Lattice representation of the single-particle Hamiltonian (Eq.~\ref{nonintham}). (b) Schematic representation of the origin of minibands with only periodic interleg hopping and no flux. (c) Schematic representation of the origin of minibands with both flux and periodic interleg hopping. (d) Schematic representation of origin of low-energy minibands in TBG, where $K$ and $K^{\theta}$ represent the Dirac points of the unrotated and rotated layers of graphene, respectively, and $\delta K=K^{\theta}-K$.}\label{fig1}
\end{figure*} 

We consider a two-leg ladder model, as shown in Fig. \ref{fig1}(a),  with the following single-particle Hamiltonian \cite{PhysRevB.102.155429}:
\begin{equation}
\label{nonintham}
H_0 =-t\sum_{j,\sigma ,\lambda } c_{j+1\sigma \lambda }^{\dagger } c_{j\sigma \lambda } + \sum_{j,\sigma }\gamma_{j} c_{j\sigma 2 }^{\dagger } c_{j\sigma 1 } + \mathrm{H.c.}.
\end{equation} 
Here, $c_{j\sigma \lambda }^\dagger$ ($c_{j\sigma \lambda }$) represents the creation (annihilation) operator for an electron at site $j$ with spin $\sigma =\downarrow, \uparrow$ and leg index $\lambda = 1, 2$. The parameters $t$ and $\gamma_j$ describe nearest-neighbor hopping within a leg and between two legs, respectively. Importantly, while $t$ is the same at all sites, $\gamma_j$ depends on the site index as
\begin{equation}
  \gamma_j=\gamma_0 \left(1+g\ \mathrm{cos}\frac{2\pi j}{J}\right) e^{i \Gamma\frac{2\pi j}{J}},    \label{intleghop}
\end{equation}
where $\gamma_0$ and $g$ are real constants, $J$ is some integer, and $\Gamma=0,1$. There are two ingredients in this model that set it apart from the usual two-leg ladder model: a periodic interleg hopping, defined by strength $g$ and periodicity $J$, and a constant flux $\phi=2\pi/J$ piercing through each plaquette, which can be turned on or off by $\Gamma$. We dub this model a moir{\'e} ladder.

Before analyzing the band structure in detail, it is useful to first understand the qualitative nature of the expected bands and how these two ingredients influence them. In the absence of interleg hopping, the two legs of the ladder give rise to doubly degenerate bands. When interleg hopping is introduced, this degeneracy is lifted in different ways depending on the specific form of the hopping term.  
\begin{enumerate}
     \item $g\ne0, \Gamma=0$: The interleg hopping changes  periodically, but there is no flux. In this limit, $\gamma_{0}$ splits the doubly degenerate band along the energy axis and $g$ opens up a gap due to Bragg scattering, creating minibands. One expects $2J$ minibands in a  reduced Brillouin zone (BZ) of length $2\pi/J$ (the intersite distance $a$ is taken to be unity), as shown schematically in Fig. \ref{fig1}(b).
    \item $g\ne0, \Gamma=1$: There is a constant flux per plaquette, in addition to the periodically modulated interleg hopping. In this case, the flux splits the doubly degenerate band  along the momentum axis and introduces a momentum shift between the minima of the two bands, $\gamma_0$ opens a gap at the band touching point, followed by Bragg scattering due to the periodic part of the interleg hopping. This also creates $2J$ minibands in a reduced BZ of length $2\pi/J$. This is illustrated in Fig. \ref{fig1}(c).
    \end{enumerate}    
Thus, moir\'e minibands emerge in both Case 1 and 2, but they exhibit qualitatively distinct characteristics due to fundamentally different forms of the interleg moir{\'e} hopping. These differences, in turn, give rise to markedly different behavior at both single-particle and many-body levels. We aim to systematically analyze and clarify these differences in this work, focusing on the lowest-energy miniband.

Our motivation for studying the moir{\'e} ladder model stems from its qualitative resemblance to TBG, the paradigmatic example of moir{\'e} materials. To demonstrate the analogy, we refer to Fig.~\ref{fig1}(d), where we schematically depict the origin of the moir{\'e} minibands in TBG. When two graphene layers are rotated by a small angle relative to each other, a long-wavelength moir{\'e} pattern emerges in real space. This real-space rotation results in a corresponding rotation of the Brillouin zones (BZs) of each layer. The low-energy physics is governed by states near the $K$ point, where Dirac cones are located (here, we restrict to only one valley). The relative rotation introduces a momentum shift $\delta K$, displacing the Dirac cones of the two layers with respect to each other, as shown in the figure. These shifted Dirac cones are then hybridized by the spatially varying interlayer hopping with a periodicity defined by the same scale, $\sim 1/\delta K$. As a result, a gap opens up at the band-touching points along with  the formation of minibands in a reduced BZ, as depicted in the figure. Two key ingredients underlie the process: (i) a relative momentum shift between initially degenerate bands and (ii) interlayer hybridization accompanied by BZ folding—-together giving rise to the characteristic moir{\'e} minibands. In comparing Figs.~\ref{fig1}(c) and~\ref{fig1}(d), it is evident that these are exactly the same two features that define our ladder model, with the flux introducing ingredient (i) and the modulated interleg hopping capturing ingredient (ii). Thus, our model captures the core mechanisms of moir\'e-induced band formation. However, it is important to notice that our model deliberately omits several other features of TBG, such as Dirac dispersion, two-dimensionality, sublattice degrees of freedom, and possible nontrivial band topology. These ingredients arise from the specific structure of graphene as the contributing layer degree of freedom, which we have  simplified to a pair of chains. Further details can be found in Ref.~\cite{PhysRevB.102.155429}.

What does our model offer? First, it provides a framework in which moir{\'e} physics can be disentangled from the specific features of graphene, enabling a focused investigation of moir{\'e}--induced phenomena in isolation. The insights obtained from this simplified setting may offer guidance for exploring moir{\'e} effects in a wider class of physical systems beyond graphene.  Second, it enables us to independently and controllably tune the two essential components of the moir{\'e} potential--namely, the momentum shift (flux) and the periodic interleg hopping--across both noninteracting and interacting regimes. This, in turn, allows us to identify which specific features of the moir{\'e} hopping are responsible for particular characteristics of the ground state, especially in the presence of interactions. Such a separation is not accessible in the actual TBG setting, where the two features are always present together. Third, by formulating the problem in one dimension, we gain access to powerful theoretical and numerical tools for studying interactions, such as bosonization and the density matrix renormalization group (DMRG). These methods offer a more controlled and accurate treatment of correlations than the mean-field approaches typically employed in two-dimensional systems like TBG. This makes it possible to achieve a deeper and more reliable understanding of the interplay between interactions and moir{\'e} hopping. Finally, as a one-dimensional analogue of two-dimensional moir{\'e} materials, the model is of intrinsic theoretical interest. It also holds experimental relevance, given the increasing feasibility of realizing such systems in engineered platforms, including ultracold atoms in optical lattices.

In passing, we note that our model can be easily generalized by rewriting Eq.~(\ref{intleghop}) as $\gamma_j=\gamma_0 \left(1+g\ \mathrm{cos}\frac{2\pi j}{J'}\right) e^{i \Gamma\frac{2\pi j}{J}}$, with $J\ne J'$, allowing for an incommensurate relationship between the flux and the periodic interleg hopping. Since our main goal is to distill the essential physics of moir{\'e} systems, we focus on the simpler case. However, exploring the more general scenario would be an interesting direction for future work, both when $J/J'$ is rational as well as irrational.

%%%%%%%%%%%%%%%%%%%%%%

\section{\label{bandstructure}Band structure and fermiology}

\begin{figure*}
\includegraphics[width=1.7\columnwidth]{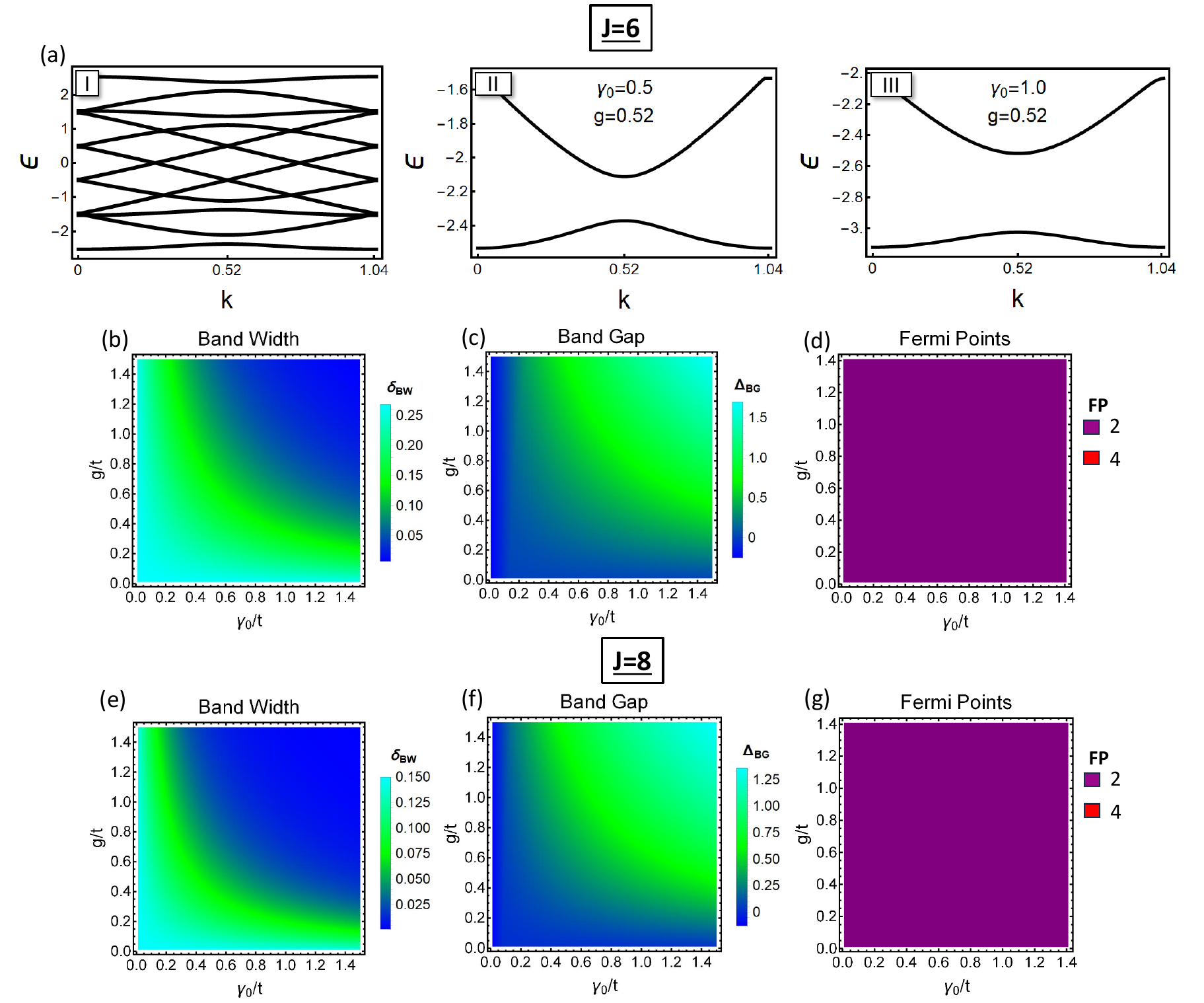}

\caption{Single-particle characteristics of the Hamiltonian in Eq.~\ref{nonintham} for Case 1: $g \neq 0$, $\Gamma = 0$. The first panel shows the band structure for $J = 6$: full band diagram (I), followed by band diagrams of the lowest two bands for different parameter values (II and III). Here, $\epsilon$ represents the energy of an electron while $k$ represents the crystal momentum. The second panel presents detailed properties of the lowest band, namely the bandwidth, the band gap, and the number of Fermi points at half-filling, as shown in plots (b)–(d). Similar plots for $J = 8$ are shown in the third panel from (e)-(g). \label{fig2}
}
\end{figure*}

\begin{figure*}
\includegraphics[width=1.8\columnwidth]{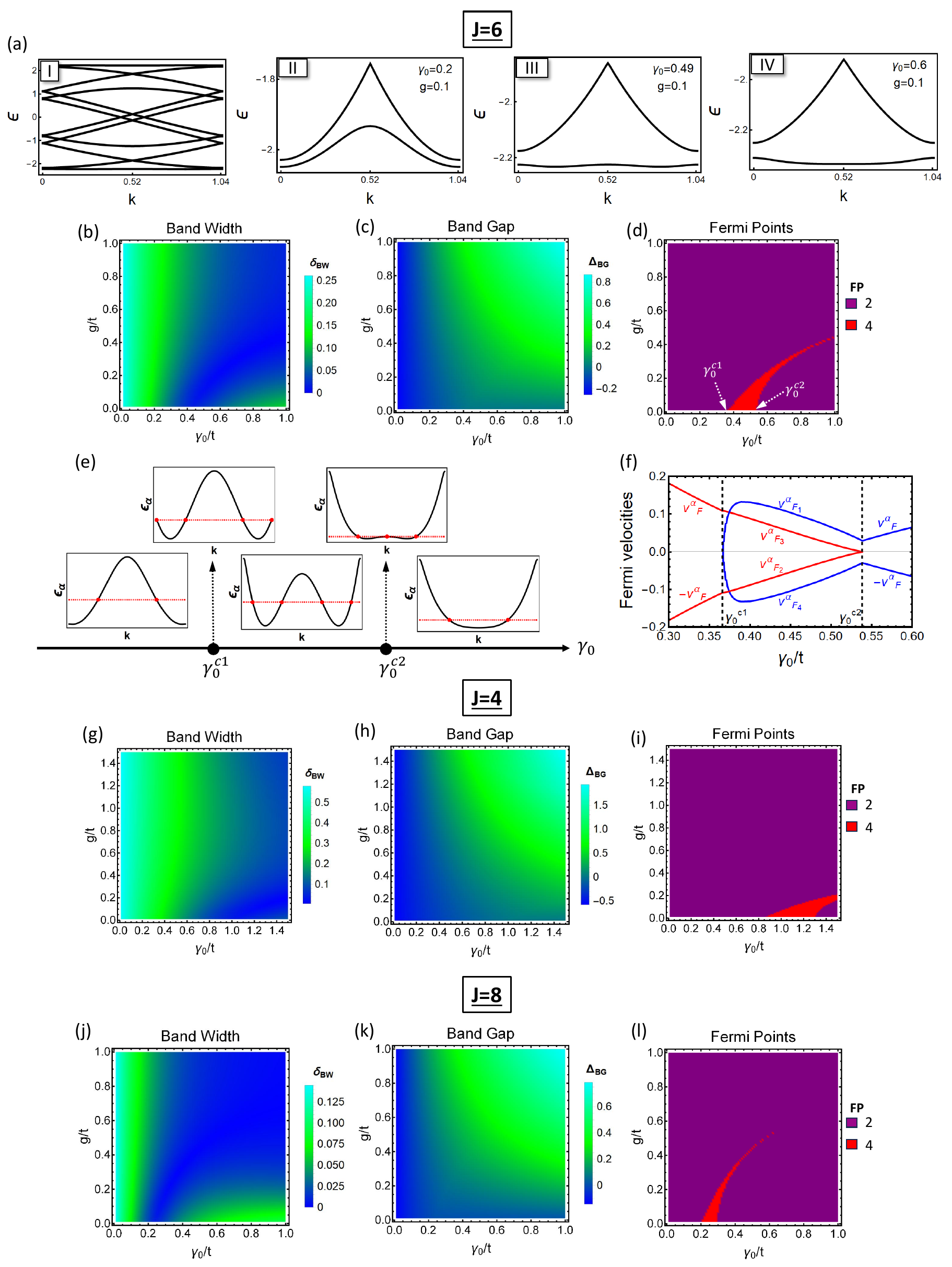}
\caption{Single-particle characteristics of the Hamiltonian in Eq.~\ref{nonintham} for Case 2: $g \neq 0$, $\Gamma = 1$. The first panel shows the band structure for $J = 6$: full band diagram (I), followed by band diagrams of the lowest two bands for various parameter values (II-IV). Here, $\epsilon$ represents the energy of an electron while $k$ represents the crystal momentum. The second panel shows properties of the lowest band, namely the bandwidth, the band gap, and the number of Fermi points at half-filling, as shown in plots (b)–(d). Fig.(e) shows how the number of Fermi points at half-filling of the lowest band, $\epsilon_{\alpha}$, changes by varying $\gamma_{0}$ with fixed $g=0.01$. The transition between two and four Fermi points occurs at $\gamma_0^{c1}$ and $\gamma_0^{c2}$. Fig.(f) shows Fermi velocities at two- and four-Fermi points regions. The transitions are governed by different scales: near $\gamma_0^{c1}$, the Fermi velocities ($v_{F_1}^{\alpha}$ and $v_{F_4}^{\alpha}$) go to zero due to the effect of $g$; near $\gamma_0^{2c}$, the Fermi velocities ($v_{F_2}^{\alpha}$ and $v_{F_3}^{\alpha}$) go to zero due to $\gamma_0$. Note that, in the four Fermi point regime $v_{F_1}=-v_{F_4}$ and $v_{F_2}=-v_{F_3}$. Similar to the second panel, the characteristics of the lowest band for $J=4$ and $J=8$ are shown in the fourth (Figs.(g)-(i)) and fifth (Figs.(j)-(l)) panels, respectively.}\label{fig3}
\end{figure*}

We now study in detail the low-energy band structure of our model in different parameter regimes. We consider the two cases discussed above and systematically analyze the following features: the bandwidth of the lowest band ($\delta_{BW}$), the bandgap between the lowest and the next-to-lowest bands ($\Delta_{BG}$), and the number of Fermi points crossed by the Fermi level at half-filling of the lowest band. We also present in each case, typical band diagrams for the low-energy sector. 

\subsection{Case 1: $g \neq 0$,$\Gamma=0$ (No flux)}

This scenario is illustrated in Fig.~\ref{fig2}. Since there are $2J$ minibands due to zone folding, the typical bandwidth of each band $\sim t/J$. With increase in the parameters $\gamma_0$ and $g$ (scaled with respect to $t$), the bandwidth of the lowest-energy band decreases along with a concomitant increase in the bandgap between the two low-energy bands. In the most of the parameter space, the two bands are well separated with a band gap; however, at low values of $\gamma_0$, the bands may overlap. Importantly, the Fermi level always crosses two points at half filling of the lowest-energy band, irrespective of the values of the interleg hopping parameters $\gamma_0$ and $g$. These qualitative features remain unchanged with varying $J$, as confirmed by comparing the cases $J=6$ and $J=8$.

\subsection{Case 2: $g \neq 0$,$\Gamma=1$ (With flux)}

This scenario is illustrated in Fig.~\ref{fig3}. Similar to Case 1, the spectrum consists of $2J$ minibands. However, unlike in Case 1, these minibands are no longer uniform in bandwidth. As seen in the representative band diagrams, for certain parameter values, the lowest-energy band becomes nearly flat while the next-lowest band remains dispersive. Indeed, the bandwidth exhibits a more intricate dependence than in Case 1--it becomes small even for arbitrarily small $g$, as long as $\gamma_0$ is nonzero. The band gap, however, shows qualitatively similar behavior to that in Case 1: it increases with larger values of $\gamma_0$ and $g$, and closes for smaller $\gamma_0$. A particularly notable feature is the evolution of the number of Fermi points at half filling. In an intermediate range of $\gamma_0$ (i.e., $\gamma_0^{c1}<\gamma_0<\gamma_0^{c2}$) and small $g$, we observe four Fermi points, as shown in Fig.~\ref{fig3}(e). As $\gamma_0$ is increased or decreased from this regime, the number of Fermi points undergoes a transition from 4 to 2 at $\gamma_0^{c1}$ and $\gamma_0^{c2}$. This behavior differs from Case 1, where four Fermi points never appear. At $\gamma_0^{1c}$ and $\gamma_0^{2c}$, the Fermi velocities of a pair of Fermi points remain zero, as shown in Fig.~\ref{fig3}(f). This property is induced by both $g$ and $\gamma_0$. The parameter $g$ causes a sharp curve in the dispersion at the edges of the reduced BZ, while $\gamma_0$ facilitates the coalescence of two minima of the lower band. These features can be visualized from the dispersions shown in Fig.~\ref{fig3}(e). At $\gamma_0^{c1}$, a pair of Fermi points lie at the edges of the reduced BZ, where dispersion features extrema due to $g$. Therefore, the Fermi velocity ($v_{F_1}^{\alpha}$ and $v_{F_4}^{\alpha}$ where $\alpha$ represents the lowest band) is zero at this point, as shown in Fig.~\ref{fig3}(f). However, it rises quickly since the $g$ is kept small and then slowly decreases with increasing $\gamma_0$. On the other side, at $\gamma_0^{c2}$, the Fermi velocities ($v_{F_2}^{\alpha}$ and $v_{F_3}^{\alpha}$) of another pair of Fermi points hit zero while the other remains low as a consequence coalescence of two minima of the band due to $\gamma_0$. 
Interestingly, this transition in the number of Fermi points is achievable for arbitrarily small $g$, and is largely dictated by $\gamma_0$. Nevertheless, $g$  remains essential for the formation of minibands in the first place: when $g$ is identically zero, the model reduces to a two-leg ladder model with flux, which hosts two bands, instead of $2J$ minibands \cite{Carr_2006,PhysRevB.76.195105}. We also note that the parameter regime in which four Fermi points appear coincides with the region of minimal bandwidth, where the band becomes quasiflat. Upon varying $J$, the qualitative features remain unchanged, with only quantitative modifications. Specifically, as $J$ increases, the region in which four Fermi points appear shifts to lower values of $\gamma_0$ and shrinks in size, yet it remains present for all values of $J$.

%%%%%%%%%%%%%%%%%%%%%

\section{\label{halffilling}Interacting ground states at half filling}

We now study the effects of interactions. To this end, we introduce an onsite Hubbard term to the noninteracting Hamiltonian in Eq.~(\ref{nonintham}):
\begin{equation}
\label{intham}
    H_\mathrm{int} = U \sum_{j,\lambda} n_{j\uparrow\lambda} n_{j\downarrow\lambda}.
\end{equation}
Here, \( U>0 \) denotes the strength of the onsite repulsive interaction, and \( n_{j\sigma\lambda} \) counts the number of electrons with spin \( \sigma \) at site \( j \) on chain $\lambda$. Since our focus is on the physics of the lowest energy band, we restrict our analysis to the parameter regime where this band is well separated from the next one, ensuring a nonzero bandgap \( \Delta_{BG} > 0 \). We explore both weak and strong coupling limits, which are characterized by the ratio of the effective interaction strength \( \langle U \rangle \) to the bandwidth of the lowest band \( \delta_{BW} \), where $\langle U \rangle=U\sum_{j,\lambda}\langle n_{j\uparrow\lambda} n_{j\downarrow\lambda}\rangle$, calculated in the non-interacting limit (because we are dealing with a miniband, $\langle U\rangle$ in lieu of $U$ is the relevant interaction scale to compare with the mini-bandwidth). We treat the two cases of moir\'e bands separately, aiming to identify how different components of the moir\'e potential influence various aspects of correlated physics.

We begin with the case where the lowest band is half filled, corresponding to an electron density of $n=\frac{1}{2J}$. Employing a combination of bosonization and DMRG techniques~\cite{PhysRevLett.69.2863,PhysRevB.48.10345,SCHOLLWOCK201196} with open boundary conditions, we perform a detailed study of the interacting ground state and map out the resulting phase diagram. This specific filling was previously examined numerically in Ref.~\cite{PhysRevB.102.155429}. The analysis presented here offers a deeper theoretical understanding of the earlier results and, additionally, yields new findings.

\subsection{\label{half} Weak coupling limit}
In the weak-coupling limit, $\langle U \rangle / \delta_{BW} \ll 1$, we employ bosonization to examine the ground state in the presence of interactions. For analytical tractability, we adopt the following approach: we assume that the interleg hopping modulation amplitude $g$ is small enough that its effect is appreciable only near the reduced BZ edges--where it opens a gap and forms the miniband--but its effect on the single-particle eigenenergies and eigenstates near the Fermi level at half-filling is negligible, [cf.  Fig.~\ref{fig1}]. This allows us to perform the bosonization analysis on the unmodulated (i.e., $g = 0$) parent Hamiltonian, which is analytically solvable, while accounting for $g$ only through its role in enabling relevant (Umklapp) scattering processes at half-filling (half-filling of the miniband corresponds to a very low filling of the parent band, where such processes would be absent otherwise). As shown in Fig.~\ref{fig3}, this approximation preserves the system's Fermiology and provides a controlled framework that captures the essential physics.

\begin{figure*}
\includegraphics[width=2.1\columnwidth]{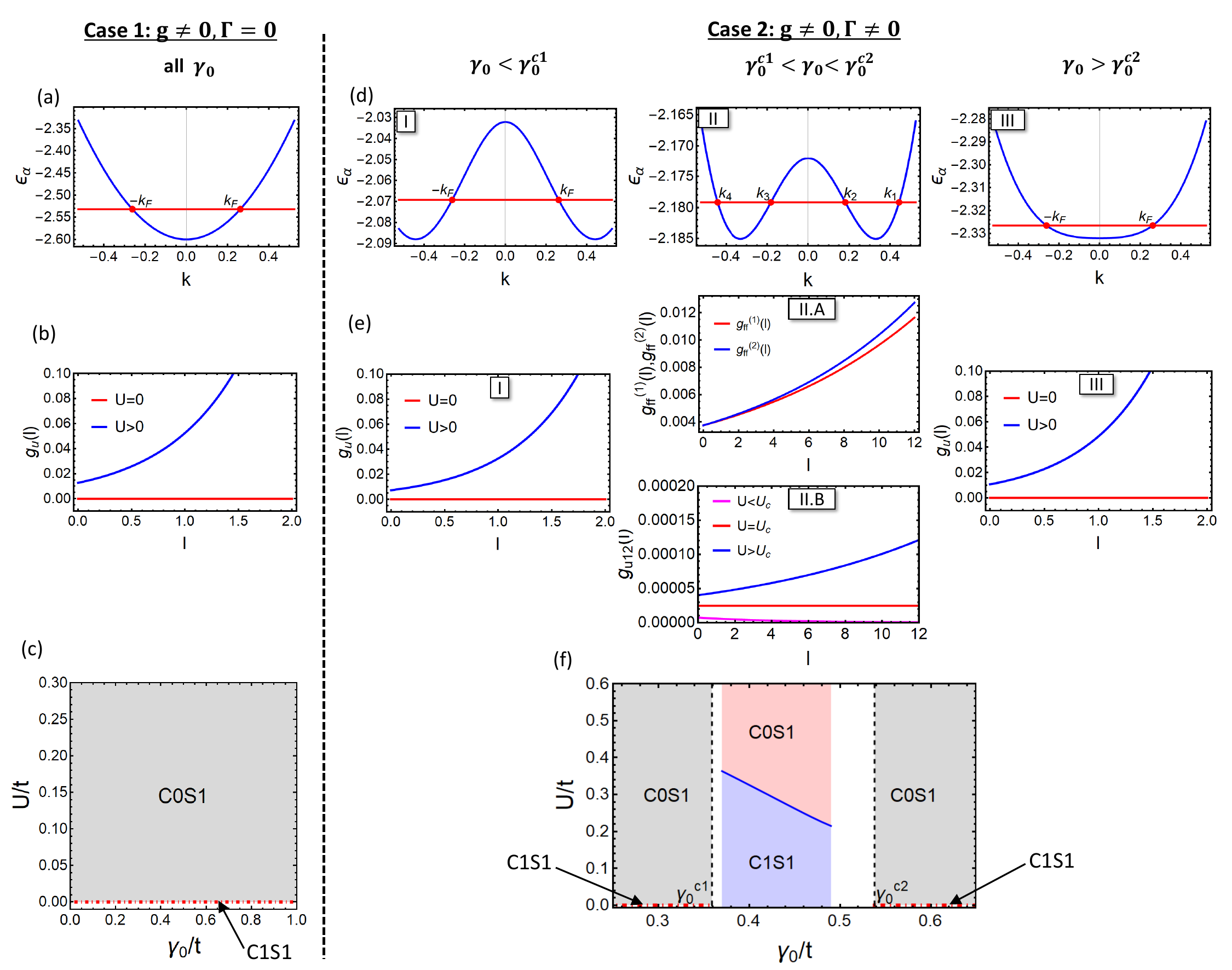}
\caption{Details of the dispersion, RG flow and phase diagrams at half-filling of the lowest band $\epsilon_{\alpha}$ in both Case 1 (no flux) and Case 2 (with flux) with $J=6$ and $t=1$. (a) Dispersion of the lower band in the reduced BZ for Case 1 with $\gamma_0=1.0$. (b) RG flow of the Umklapp coupling with initial value $g_{u}(0)=1.2\times10^{-2}$ for Case 1 with $\gamma_0=1.0$. For $U>0$, the flow increases with $l$ to the strong coupling limit. (c) Phase diagram for Case 1. The Luttinger liquid phase C1S1 at $U=0$, opens up a charge gap due to Umklapp term $g_u$ for $U>0$, leading to an insulating phase C0S1 with gapless spin sector. Figures (d)-(I) and (e)-(I) are the dispersion and RG flow for Case 2, respectively, and $\gamma_0<\gamma_0^{c1}$ with $\gamma_0=0.3$, $\phi=\pi/3$. For $U>0$, the RG flows, with $g_u(0)=0.7\times10^{-2}$, to the strong coupling limit leading to C0S1 phase. Figures (d)-(II) and (e)-(II.A, II.B) are the dispersion and RG flow, for Case 2, respectively, and $\gamma_0^{c1}<\gamma_0<\gamma_0^{c2}$ with $\gamma_0=0.44$, $\phi=\pi/3$ in the four Fermi point region. II.A shows the RG flow of the couplings $g_{ff}^{(1)}$ and $g_{ff}^{(1)}$ with $g_{ff}^{(1)}(0)=g_{ff}^{(2)}(0)=0.37\times10^{-2}$, that are increasing with length scale for all $U>0$. II.B shows the RG flow of the Umklapp coupling $g_{u12}(l)$ with the length scale. At the critical value $U_c=0.26$, the RG flow, with $g_{u12}(0)=0.24\times10^{-4}$, remains constant. Above $U_c$ the flow, with $g_{u12}(0)=0.4\times10^{-4}$, increases with $l$, while below $U_c$, with $g_{u12}(0)=0.073\times10^{-4}$, it decreases. Figures (d)-(III) and (e)-(III) are the dispersion and RG flow, for Case 2, respectively, and $\gamma_0>\gamma_0^{c2}$ with $\gamma_0=0.6$, $\phi=\pi/3$. For $U>0$, the RG flows, with $g_u(0)=1\times10^{-2}$, to the strong coupling limit leading to C0S1 phase. (f) Phase diagram for Case 2 with $\phi=\pi/3$. We obtain $U_c$ in the four Fermi point region from the scaling dimension of the Umklapp term $g_{u12(34)}$, which dictates the transition line (solid blue line) between metal C1S1 and insulator C0S1. At the white regions in the parameter space, the bosonization fails to capture the correct $U_c$ values. The two Fermi point region is a C0S1 insulator for $U>0$ and is a C1S1 metal on the line $U=0$ (dot-dashed red line). The transition lines (dashed black lines) at $\gamma_0^{c1}$ and $\gamma_0^{c2}$ mark the change in the number of Fermi points.}
\label{fig4}
\end{figure*}

\subsubsection{Case 1: $g \neq 0$,$\Gamma=0$ (No flux)}
The single-particle parent Hamiltonian (with $g=0$) is that of a standard two-leg ladder model. It is easily diagonalized by Fourier transforming  Eq.~(\ref{nonintham}) and using the linear transformation $c_{1,\sigma}(k)= (\alpha_{k,\sigma}+\beta_{k,\sigma})/\sqrt{2}$ and $c_{2,\sigma}(k)=(\alpha_{k,\sigma}-\beta_{k,\sigma})/\sqrt{2}$. This results in two bands: 
\begin{equation}
    \epsilon_{\alpha(\beta)}(k)=-2t\cos k\mp\gamma_0,
    \label{EDC1}
\end{equation}
with $\alpha (\beta)$ corresponding to $-(+)$. Now, introducing periodic modulation ($g\neq0$) will induce a gap at $\pm\pi/aJ$, which results in the lowest miniband derived from $\epsilon_{\alpha}$. The half-filling of this band in the reduced BZ always has two Fermi points at $k_F=\pm\frac{\pi}{2aJ}$, as shown in Fig.~\ref{fig4}(a). 

We perform the bosonization by linearizing the spectrum near the Fermi points. The right ($R$) and left ($L$) moving branches, with the fermionic fields $\alpha_{R\sigma}(x)$ and $\alpha_{L\sigma}(x)$ (with $x=aj$), respectively, are bosonized using the usual bosonization rules. The non-interacting Hamiltonian yields charge and spin Luttinger liquids, while the interacting Hamiltonian gives diagonal ($H_{d}$), off-diagonal ($H_{od}$), backscattering ($H_b$), and Umklapp ($H_u$) terms. These are summarized in Table.~\ref{tabboso2}, with the details of the calculations relegated to Appendix.~\ref{boso2fp}. The $H_d$ and $H_{od}$ terms with couplings $g_d$ and $g_{od}$ respectively, renormalize the Luttinger parameters in both charge and spin sectors. The $H_b$ term, with coupling $g_b$, contains the cosine of the spin bosonic field $\phi_s$, which is marginally irrelevant for repulsive interaction leading to a gapless spin sector.
The $H_u$ scattering term, with coupling $g_u$, appears only at half-filling and is obtained as,
\begin{align}
    H_u&=g_u\int dx \cos[2\sqrt{2\pi}\phi_c(x)]
\end{align}
where $\phi_c(x)$ is the bosonic field in the charge sector. The scaling dimension of the Umklapp term is $\Delta_{g_u}=(2/K_c)$, where $K_{c}=\sqrt{(1+(U/v_F^{\alpha}\pi))}$ is the Luttinger liquid parameter for the charge sector. For $U>0$, $K_c>1$, therefore, the scaling dimension is relevant, i.e., $\Delta_{g_u}<2$ and induces a charge gap. Therefore, this term dictates the metal-insulator transition at infinitesimal $U$, similar to the standard Hubbard model. 
This can be further identified from the renormalization group (RG) flow, up to first order in the coupling constants with the length scale ($l$).
The RG equation for the Umklapp term can be obtained as
\begin{equation}
\frac{dg_u}{dl}=(2-2K_c)g_u
\end{equation}
As shown in Fig.~\ref{fig4}(b), the RG flow of the Umklapp coupling grows with $l$ for any $U>0$, leading to the insulating phase. 

In the usual notation for a phase C\textit{p}S\textit{q} with $p$ and $q$ gapless charge and spin modes, respectively, the system initially has gapless charge and spin Luttinger liquid phase in the absence of interaction, i.e., C1S1. As the interaction is turned on, the Umklapp term opens up a charge gap while the spin sector remains gapless, leading to the C0S1 insulating phase, as shown in Fig.~\ref{fig4}(c). Moreover, 
the spin correlation (with logarithmic corrections) shows dominant $2k_F$ terms, with the slowest decay leading to the antiferromagnetic correlations with the exponent $\frac{1}{2}\left(K_c+K_s\right)$. Therefore, this case is not different from a standard Hubbard chain, as far as interactions are concerned.

\begin{table*}
	\begin{tabular}{|c|c|c|c|c|}
		\hline
		\hline
		\multicolumn{2}{|c|}{Hamiltonian: Coupling} &  \multirow{3}{*}{Scattering terms} & \multirow{3}{*}{Bosonized forms}  & \multirow{3}{*}{Scaling dimension}\\
		\multicolumn{2}{|c|}{(Hubbard value)}     &                  &                 & \multirow{3}{*}{(relevance)}      \\
		Case 1: $g \neq 0$, $\Gamma=0$  & Case 2: $g \neq 0$, $\Gamma=1$ & & &\\
		\hline
		\hline
		\multirow{3}{*}{$H_d$: $g_{d}=\frac{U}{4\pi}$} & \multirow{3}{*}{$g_{d}=\frac{U(u_{k_F}^2+v_{k_F}^2)}{4\pi}$} & \multirow{3}{*}{$\alpha_{R\uparrow}^{\dagger}\alpha_{R\uparrow}\alpha_{R\downarrow}^{\dagger}\alpha_{R\downarrow}$,} & \multirow{3}{*}{$[(\partial_x\phi_c)^2+(\partial_x\theta_c)^2],$}  & \multirow{3}{*}{--}\\
		& & \multirow{3}{*}{$\alpha_{L\uparrow}^{\dagger}\alpha_{L\uparrow}\alpha_{L\downarrow}^{\dagger}\alpha_{L\downarrow}$.}& \multirow{3}{*}{$[(\partial_x\phi_s)^2+(\partial_x\theta_s)^2]$} & \\
		& & & & \\
		\hline
		\multirow{3}{*}{$H_{od}$: $g_{od}=\frac{U}{4\pi}$} & \multirow{3}{*}{$g_{od}=\frac{U(u_{k_F}v_{k_F})}{2\pi}$} & \multirow{3}{*}{$\alpha_{R\uparrow}^{\dagger}\alpha_{R\uparrow}\alpha_{L\downarrow}^{\dagger}\alpha_{L\downarrow}$,}  & \multirow{3}{*}{$[(\partial_x\phi_c)^2-(\partial_x\theta_c)^2],$} & \multirow{3}{*}{--}\\
		& & \multirow{3}{*}{$\alpha_{L\uparrow}^{\dagger}\alpha_{L\uparrow}\alpha_{R\downarrow}^{\dagger}\alpha_{R\downarrow}$.}& \multirow{3}{*}{$[(\partial_x\phi_s)^2-(\partial_x\theta_s)^2]$} & \\
		& & & & \\
		\hline
		\multirow{3}{*}{$H_{b}$: $g_{b}=\frac{U}{(2\pi\eta)^2}$} & \multirow{3}{*}{$g_{b}=\frac{U(u_{k_F}v_{k_F})^2}{\pi^2\eta^2}$} &  \multirow{3}{*}{$\alpha_{R\uparrow}^{\dagger}\alpha_{L\uparrow}\alpha_{L\downarrow}^{\dagger}\alpha_{R\downarrow}$,}  & \multirow{3}{*}{$\cos[2\sqrt{2\pi}\phi_s]$} & \multirow{3}{*}{$\Delta_{g_b}=\left( \frac{2}{K_s}\right) $}\\
		& & 
		\multirow{3}{*}{$H.c.$} & & \multirow{3}{*}{(irrelevant)}\\
		& & & & \\
		\hline
		\multirow{3}{*}{$H_{u}$: $g_{u}=\frac{U}{(2\pi\eta)^2}$} & \multirow{3}{*}{$g_{u}=\frac{U(u_{k_F}v_{k_F})^2}{\pi^2\eta^2}$} &  \multirow{3}{*}{$\alpha_{R\uparrow}^{\dagger}\alpha_{L\uparrow}\alpha_{R\downarrow}^{\dagger}\alpha_{L\downarrow}$,}  & \multirow{3}{*}{$\cos[2\sqrt{2\pi}\phi_c]$} & \multirow{3}{*}{$\Delta_{g_u}=\left( \frac{2}{K_c}\right)$}\\
		& & \multirow{3}{*}{$H.c.$} & & \multirow{3}{*}{(relevant)}\\
		& & & & \\
		\hline
		\hline
	\end{tabular}
\caption{g-ology and bosonization details in the two Fermi point regime of both cases, i.e., Case 1 (no flux) and Case 2 (with flux). Here $c,s$ represents charge and spin sectors respectively with corresponding bosonic fields $\phi_{c,s}$, $\theta_{c,s}$ and Luttinger parameter $K_{c,s}=\sqrt{\frac{v_F^{\alpha}\pm2g_d\pm2g_{od}}{v_F^{\alpha}\pm2g_d\mp2g_{od}}}$. In Case 2, $u_{k_F}$ and $v_{k_F}$ are obtained from Eq.~\ref{ukvk}, which for Case 1, $u_{k_F}=v_{k_F}=1/\sqrt{2}$.}
\label{tabboso2}
\end{table*}

\subsubsection{Case 2: $g \neq 0$,$\Gamma=1$ (With flux)}
The single-particle parent Hamiltonian (with $g=0$) is that of a two-leg ladder model with a constant flux per plaquette. It is diagonalized by Fourier transforming  Eq.~(\ref{nonintham}) and using the linear transformation $c_{1,\sigma}(k)=u_k \alpha_{k,\sigma}+v_k\beta_{k,\sigma}$ and $
c_{2,\sigma}(k)=v_k \alpha_{k,\sigma}-u_k\beta_{k,\sigma}$,
where the coherence factors are given by
\begin{align}
u_k(v_k)=\sqrt{\frac{1}{2}\left( 1\mp\frac{\sin k \sin \left(\frac{\phi}{2}\right)}{\sqrt{\sin^2 k \sin^2 \left(\frac{\phi}{2}\right)+\left( \frac{\gamma_0}{2t}\right)^2}}\right) },
\label{ukvk}
\end{align}
with $u_k(v_k)$ corresponding to $-(+)$. This results in two bands \cite{PhysRevB.76.195105,PhysRevB.71.161101}:
{\small
\begin{align}
\epsilon_{\alpha(\beta)}(k) 
&= -2t \left[ \cos k \cos\left(\frac{\phi}{2}\right) \pm 
\sqrt{ \sin^2 k \sin^2\left(\frac{\phi}{2}\right) + \left(\frac{\gamma_0}{2t}\right)^2 } \right],
\label{EDC2}
\end{align}}
with $\alpha (\beta)$ corresponding to $+(-)$. Now, introducing periodic modulation ($g\neq0$) will induce a gap at $\pm\pi/aJ$, which results in the lowest miniband derived from $\epsilon_{\alpha}$. The half-filling of this band, $\epsilon_{\alpha}$ in the reduced BZ, has both two and four Fermi points depending on the parameter $\gamma_0$, as shown in Fig.~\ref{fig4}(d)-(I),(II),(III). As stated before, two Fermi points appear when $\gamma_0<\gamma_0^{c1}$ and $\gamma_0>\gamma_0^{c2}$, whereas four Fermi points occur when $\gamma_0^{c1}<\gamma_0<\gamma_0^{c2}$ (here, $\gamma_0^{c1}<\gamma_0^{c2}$).

The bosonization procedure, for both interacting and non-interacting Hamiltonians, in the two Fermi point regimes, with the dispersions shown in Fig.~\ref{fig4}(d)-(I) and Fig.~\ref{fig4}(d)-(III), can be found similar to Case 1 (no flux), as summarized in Table.~\ref{tabboso2} and detailed in Appendix.~\ref{boso2fp}. The scaling dimension of the Umklapp term $(2/K_c)$ becomes relevant for $U>0$ since $K_c>1$, similar to the standard Hubbard model and that considered in Case 1, leading to a metal-insulator transition at infinitesimal $U$. The spin sector remain gapless as the scaling dimension of $H_b$ is irrelevant for $U>0$. Therefore, we obtain the C1S1 phase for $U=0$ and the C0S1 phase for $U>0$, similar to Case 1. The only difference is that the presence of flux renormalizes the couplings and Luttinger parameter $K_c$. At lower flux values (including the one considered in this work), however, this effect is negligibly small, which can be seen by comparing the RG flow in Fig.~\ref{fig4}(e)-(I) and Fig.~\ref{fig4}(e)-(III) with Fig.~\ref{fig4}(b). As far as the two regions with two Fermi points are concerned ($\gamma_0<\gamma_0^{c1}$ and $\gamma_0>\gamma_0^{c2}$), the only difference is in the chirality of the Fermi points: right (left) movers appearing at $-k_F(+k_F)$ in one region simply switch their places in the other region. This difference only flips the signs, but keeps the behavior of bosonic Hamiltonians unchanged. Moreover, in both regions, the charge correlations are short-ranged while spin correlations are associated with slow decay of $2k_F$ terms with antiferromagnetic correlations.
 
In the regime of four Fermi points, as shown in Fig.~\ref{fig4}(d)-(II), linearization at the Fermi points $k_{1,2,3,4}$, gives two right ($k_{1,3}$) and two left ($k_{2,4}$) branches with the corresponding slowly varying fermionic fields $\alpha_{1\sigma,3\sigma}(x)$ and $\alpha_{2\sigma,4\sigma}(x)$. The Fermi velocities at these branches satisfy $v_{F_1}^{\alpha}=-v_{F_4}^{\alpha}$ and $v_{F_2}^{\alpha}=-v_{F_3}^{\alpha}$, as shown in Fig.~\ref{fig3}(f). We perform the bosonization procedure for non-interacting Hamiltonian, which gives the charge and spin Luttinger liquids. In the case of the interacting Hamiltonian, the additional Fermi points can induce more scattering processes. We will categorize them into momentum-conserving terms and Umklapp terms. We identify four kinds of momentum-conserving scatterings that include the usual diagonal ($H_d$), off-diagonal ($H_{od}$), and backward ($H_b$) scatterings, with an additional four Fermi ($H_{ff}$) scattering terms.
The $H_d$ and $H_{od}$ terms renormalize the Luttinger parameters in both charge and spin sectors. 
The $H_b$ yields the cosine operators of only spin bosonic fields and $H_{ff}$ yields the cosine operators of both spin and charge bosonic fields. Therefore, these terms can open up gaps in the charge and spin spectrum. 
Apart from these terms, at half-filling, Umklapp scatterings ($H_u$), where the change in momentum up to a reciprocal lattice vector can be obtained. 
These terms involve the scattering between two Fermi points, with coupling $g_{u12}$ and $g_{u34}$, and the scattering between four Fermi points with coupling $g_{uff}$. 
A summary of various scattering processes is provided in Table.~\ref{tabboso}, while the details of the calculation are given in Appendix.~\ref{boso4fp}. 

The couplings $g_{b12}$, $g_{b13}$, $g_{b14}$, and $g_{b23}$ of the scattering terms in Hamiltonian $H_b$ are found marginally irrelevant as their RG flow decreases with the length scale. Therefore, none of the spin bosonic fields are pinned due to these couplings. In contrast, the four Fermi scattering terms, with couplings $g_{ff}^{(1)}$ and $g_{ff}^{(2)}$, are relevant as their scaling dimensions $\Delta_{g_{ff}^{(1)}}=\Delta_{g_{ff}^{(2)}}<2$ for $U>0$. The bosonized $H_{ff}$ can be written as
\begin{align}
    H_{ff}&=\int dx(2g_{ff}^{(1)} \cos[2\sqrt{\pi}\theta_{c-}(x)]\cos[2\sqrt{\pi}\theta_{s+}(x)]\nonumber\\
&-2g_{ff}^{(2)}\cos[2\sqrt{\pi}\theta_{c-}(x)]\cos[2\sqrt{\pi}\phi_{s+}(x)])\nonumber\\
\end{align}
where $\theta_{c-}(x)$, $\theta_{s+}(x)$ and $\phi_{s+}(x)$ are the bosonic fields in the charge and spin sectors. The corresponding RG equations can be obtained as
{\small\begin{align}
    \frac{dg_{ff}^{(1)}}{dl}&=\left[2-\left( \frac{1}{K_{c}^{-}}+\frac{1}{K_{s}^{+}}\right)\right] g_{ff}^{(1)} - \left(\frac{K_{s}^{+}}{2}\right)g_{ff}^{(2)}(g_{b14}+g_{b23})
    \end{align}}
   { \small \begin{align}
\frac{dg_{ff}^{(2)}}{dl}=\left[2-\left( \frac{1}{K_{c}^{-}}+K_{s}^{+}\right) \right]g_{ff}^{(2)}+\left(\frac{1}{2K_{s}^{+}}\right)g_{ff}^{(1)}(g_{b14}+g_{b23})
\end{align}}
where $K_c^{-}$ and $K_s^{+}$ (given in Eq.~\ref{LLP}) are the Luttinger liquid parameters of the charge and spin sector, respectively. Since the two couplings involve dual fields ($\phi_{s+}(x)\leftrightarrow\theta_{s+}(x)$) and the scaling dimensions of these couplings are equal due to $SU(2)$ symmetry ($K_{s}^{+}=1$), the second order term in the RG reveal the dominating flow among the two. Therefore, upon integrating the RG equations numerically, we identify that the flow for $g_{ff}^{(2)}$ grows to the strong coupling limit faster than $g_{ff}^{(1)}$, as shown in Fig.\ref{fig4}(e)-(II.A). Therefore, these scattering terms pin one charge and one spin bosonic fields, i.e., $\theta_{c-}$ and $\phi_{s+}$, at the minima of their cosine potentials and open up a gap in one of the charge and spin sectors.

Now, we focus on the Umklapp scattering processes to understand the metal-insulator transition. The half-filling of the reduced BZ with four Fermi points implies $|k_1-k_2|=\frac{\pi}{2aJ}$. Therefore, at half-filling, only the higher-order terms of these scatterings are allowed since the momentum transfer $4|k_1-k_2|=\frac{2\pi}{aJ}$ (similarly $4|k_3-k_4|=\frac{2\pi}{aJ}$). The bosonized version of such  Umklapp Hamiltonian reads 
\begin{align}
H_{u}&=\int dx ((g_{u12}+g_{u34}) \cos[2\sqrt{2\pi}\phi_{c+}(x)]\nonumber\\
&\cos[2\sqrt{2\pi}\theta_{c-}(x)] 
+ 2g_{uff} \cos[2\sqrt{2\pi}\phi_{c+}(x)]\nonumber\\
&(\cos[2\sqrt{2\pi}\theta_{s-}(x)] - \cos[2\sqrt{2\pi}\theta_{s+}(x)]))
\end{align}
where, $\phi_{c+}(x)$, $\theta_{c-}(x)$ and $\theta_{s\pm}(x)$ are the bosonic fields in charge and spin sectors. In our model, we have $g_{u12}=g_{u34}$. The scaling dimensions of these Umklapp terms $\Delta_{g_{uff}}=\Delta_{g_{u12}}=\Delta_{g_{u34}}=4$ at $U=0$ and therefore remain strongly irrelevant. However, $\Delta_{g_{u12}}$ and $\Delta_{g_{u34}}$ decreases with increasing $U$ such that $\Delta_{g_{u12(34)}}=2$ for $U=U_{c}$ and $\Delta_{g_{u12(34)}}<2$ for $U>U_{c}$. Therefore, in contrast to the standard Umklapp interaction, these terms become relevant only at a finite interaction strength. The contribution to the charge sector comes from the terms $g_{u12(34)}$ that can open up a charge gap at $U_{c}$, defining a metal-insulator transition. The corresponding RG equation can be obtained as 
\begin{equation}
\frac{dg_{u12(34)}}{dl}=\left[2-2\left( K_{c}^{+}+\frac{1}{K_{c}^{-}}\right)\right] g_{u12(34)}
\end{equation}
Upon integrating the RG equation numerically, we observe the three distinct behaviors of the RG flow, as shown in Fig.~\ref{fig4}(e)-(II.B). It decreases with the length scale for $U<U_{c}$, remains marginal at $U=U_{c}$, and grows large into the strong coupling limit only for $U>U_{c}$. Therefore, the corresponding bosonic fields $\phi_{c+}$ and $\theta_{c-}$ are pinned at the minima of the potential and open up a gap in the charge sector for $U>U_{c}$.

\begin{table*}
	\centering
	\begin{tabular}{|c|c|c|c|}
		\hline
		\hline
		Hamiltonian: Coupling &  Scattering term &  Bosonized form & Scaling dimension\\
		(Hubbard value)       &                  &                 & (relevance)      \\
		\hline
		\hline
		$H_d$: $g_{ii}$ &  & \multirow{3}{*}{[$(\partial_x\phi_{c\pm})^2+(\partial_x\theta_{c\pm})^2$],} &  \\
		($g_{11},g_{22},g_{33},g_{44}$) & \multirow{3}{*}{$\alpha_{i\uparrow}^{\dagger}\alpha_{i\uparrow}\alpha_{i\downarrow}^{\dagger}\alpha_{i\downarrow}$}  & \multirow{3}{*}{[$(\partial_x\phi_{s\pm})^2+(\partial_x\theta_{s\pm})^2$]} & \multirow{3}{*}{--}\\
		$\left(\frac{U(u_{k_{i}}^4 +v_{k_{i}}^4)}{\pi}\right)$& & & \\
		\hline
		$H_{od}$: $g_{ii^{\prime}}$ &  & \multirow{3}{*}{[$(\partial_x\phi_{c\pm})^2-(\partial_x\theta_{c\pm})^2$],} &  \\
		($g_{12},g_{13},g_{14},g_{23},g_{24},g_{34}$)  & \multirow{3}{*}{$\alpha_{i\uparrow}^{\dagger}\alpha_{i\uparrow}\alpha_{i^{\prime}\downarrow}^{\dagger}\alpha_{i^{\prime}\downarrow}$} & \multirow{3}{*}{[$(\partial_x\phi_{s\pm})^2-(\partial_x\theta_{s\pm})^2$]} & \multirow{3}{*}{--} \\
		$\left( \frac{U(u_{k_{i}}^2 u_{k_{i^{\prime}}}^2+v_{k_{i}}^2 v_{k_{i^{\prime}}}^2)}{\pi}\right)$& & & \\
		\hline                               
		$H_{b}$: $g_{bii^{\prime}}$ &  & $\cos[2\sqrt{\pi}\phi_{s+}]\cos[2\sqrt{\pi}\theta_{s-}]$, & $\Delta_{g_{b12},g_{b34}}=[K_s^+ +(1/K_s^-)]$\\
		($g_{b12},g_{b13},g_{b14},g_{b23},g_{b24},g_{b34}$) &  \multirow{4}{*}{$\alpha_{i\uparrow}^{\dagger}\alpha_{i^{\prime}\uparrow}\alpha_{i^{\prime}\downarrow}^{\dagger}\alpha_{i\downarrow}$,} & $\cos[2\sqrt{\pi}\theta_{s+}]\cos[2\sqrt{\pi}\theta_{s-}]$, & $\Delta_{g_{b13},g_{b24}}=[(1/K_s^+) +(1/K_s^-)]$\\
		& \multirow{4}{*}{$H.c.$} & $\cos[2\sqrt{\pi}(\theta_{s+}+\phi_{s+})]$, & $\Delta_{g_{b14}}=[K_s^+ +(1/K_s^+)]$\\
		$\left( \frac{-U(u_{k_{i}}^2 u_{k_{i^{\prime}}}^2 +v_{k_{i}}^2 v_{k_{i^{\prime}}}^2)}{2\pi^2\eta^2}\right) $& & $\cos[2\sqrt{\pi}(\theta_{s+}-\phi_{s+})]$. & $\Delta_{g_{b23}}=[K_s^+ +(1/K_s^+)]$\\
		& & & (marginally irrelevant)\\
		\hline                            
		$H_{ff}$: $g_{ff}$ & $\alpha_{1(3)\uparrow}^{\dagger}\alpha_{2(4)\uparrow}\alpha_{4(2)\downarrow}^{\dagger}\alpha_{3(1)\downarrow}$,& $\cos[2\sqrt{\pi}\theta_{c-}]\cos[2\sqrt{\pi}\theta_{s+}]$, & $\Delta_{g_{ff}^{(1)}}=[(1/K_c^-) +(1/K_s^+)]$\\
		$(g_{ff}^{(1)},g_{ff}^{(2)})$ & $\alpha_{1(2)\uparrow}^{\dagger}\alpha_{3(4)\uparrow}\alpha_{4(3)\downarrow}^{\dagger}\alpha_{2(1)\downarrow}$, & $\cos[2\sqrt{\pi}\theta_{c-}]\cos[2\sqrt{\pi}\phi_{s+}]$. & $\Delta_{g_{ff}^{(2)}}=[(1/K_c^-) +K_s^+]$\\
		$\left( \frac{U(u_{k_1} u_{k_2} u_{k_3} u_{k_4}+v_{k_1} v_{k_2} v_{k_3} v_{k_4})}{2\pi^2\eta^2}\right)$ & $H.c.$ &  & (relevant for all $U$)\\
		\hline                            
		$H_{u}$: $g_{u12(34)}$ & $(\alpha_{1(3)\uparrow}^{\dagger}\alpha_{2(4)\uparrow}\alpha_{1(3)\downarrow}^{\dagger}\alpha_{2(4)\downarrow})^2$, & $\cos[2\sqrt{2\pi}\phi_{c+}]\cos[2\sqrt{2\pi}\theta_{c-}]$ & $\Delta_{g_{u12},g_{u34}}=2[K_c^+ +(1/K_c^-)]$ \\
		$\left( \frac{2U^2(u_{k_1(3)}^2 u_{k_2(4)}^2 +v_{k_1(3)}^2 v_{k_2(4)}^2)^2}{(2\pi\eta)^4}\right) $& $H.c.$ &  & (relevant above $U_c$)\\
		\hline                            
		$H_{u}$: $g_{uff}$  & $(\alpha_{1(3)\uparrow}^{\dagger}\alpha_{2(4)\uparrow}\alpha_{3(1)\downarrow}^{\dagger}\alpha_{4(2)\downarrow})^2$, &$\cos[2\sqrt{2\pi}\phi_{c+}]\cos[2\sqrt{2\pi}\theta_{s-}]$, & $\Delta_{g_{uff}^{(1)}}=2[K_c^+ +(1/K_s^-)]$ \\
		($g_{uff}^{(1)},g_{uff}^{(2)}$) & $(\alpha_{1(3)\uparrow}^{\dagger}\alpha_{4(2)\uparrow}\alpha_{3(1)\downarrow}^{\dagger}\alpha_{2(4)\downarrow})^2$, & $\cos[2\sqrt{2\pi}\phi_{c+}]\cos[2\sqrt{2\pi}\theta_{s+}]$ & $\Delta_{g_{uff}^{(2)}}=2[K_c^+ +(1/K_s^+)]$\\
		$\left( \frac{2U^2(u_{k_1}^2 u_{k_2}^2 u_{k_3}^2 u_{k_4}^2+v_{k_1}^2 v_{k_2}^2 v_{k_3}^2 v_{k_4}^2)^2}{(2\pi\eta)^4}\right) $& $H.c.$ & & (irrelevant) \\
		\hline
		\hline
	\end{tabular}
	\caption{g-ology and bosonization details in the four Fermi point regime. Here $i,i^{\prime}=1,2,3,4$ and $c,s$ refer to charge and spin sectors, respectively, with the corresponding bosonic fields $\phi_{c\pm,s\pm}$ and $\theta_{c\pm,s\pm}$. The detailed form of Luttinger parameters $K_c^{\pm}$ and $K_s^{\pm}$ are given in the Appendix.~\ref{boso4fp} (Eq.~\ref{LLP}).}
	\label{tabboso}
\end{table*}

Based on these analyses, we obtain a phase diagram in Fig.~\ref{fig4}(f) that shows various phases for small $g$ (up to $g=0.01$)--we expect the same qualitative behavior at larger values of $g$ as well. In the four Fermi point regime, initially for $U<U_c$ one of the charge and spin sectors are gapped, as the bosonic fields $\theta_{c-}$ and $\phi_{s+}$ are pinned due to the four Fermi points scatterings, leading to C1S1 metal. For $U>U_c$, additionally, the higher order Umklapp scatterings pins bosonic fields $\phi_{c+}$ and $\theta_{c-}$ that open up the gap in the charge sector, leading to a gapless spin C0S1 insulator. Therefore, the metal-insulator transition is observed between C1S1 and C0S1 at $U=U_c$,  as shown in Fig.~\ref{fig4}(f). In the entire four-Fermi regime, the difference in the Fermi velocities at $k_{1}$ ($v_{F_1}^{\alpha}$) and $k_2$ ($v_{F_2}^{\alpha}$) is sufficiently large to partially suppress the RG flow of various couplings to be irrelevant. Similarly for $k_{3}$ ($v_{F_3}^{\alpha}$) and $k_4$ ($v_{F_4}^{\alpha}$).
This leads to a partially gapless C1S1 metallic phase. In the insulating phase the gap is induced only in the charge sector due to pinning of both $\phi_{c+}$ and $\theta_{c-}$ from the Umklapp terms, while the spin degrees of freedom remain intact, since the different Fermi velocities still suppress spin fields $\phi_{s-}, \theta_{s-}$ from opening a gap. Therefore, we observe a metal-insulator transition at $U=U_c$ between C1S1 and C0S1. 
Apart from the finite $U$ Mott transition, we observe the transitions between two and four Fermi point regimes 
at $\gamma_0^{c1}$ and $\gamma_0^{c2}$ ($\gamma_0^{c1}<\gamma_0^{c2}$).  
At these transitions, the Fermi velocities of a pair of Fermi points remain zero, as shown in Fig.~\ref{fig3}(f). Therefore, bosonization fails to capture the correct $U_c$ near these transitions, as the linear approximation fails in the vicinity of these transitions.  
Typically, $U_c \rightarrow 0$ near the transition between two and four Fermi points as shown in Ref.~\cite{PhysRevB.76.115118} in the context of a different model--we expect the same to happen here. Note that the difference in the width of the undefined $U_c$ region (white regions in Fig.~\ref{fig4}(f)) near both $\gamma_0^{1c}$ and $\gamma_0^{2c}$ is due to the different scales of the parameters $g$ and $\gamma_0$, respectively, deciding the single-particle physics, as discussed earlier [cf. Fig.~\ref{fig3}(f)]. Note that the C1S1 (and C0S1) phases in two and four Fermi point regimes differ by the number of charge and spin gapped modes they possess. Moreover, the correlations at these phases are governed by the different exponents. The transition between these phases occurs at $\gamma_0^{c1}$ and $\gamma_0^{c2}$.

We next investigate the charge correlation, $C=\left\langle \rho(x)\rho(0)\right\rangle$ and the spin correlation, $S=\left\langle S(x)S(0)\right\rangle$. These correlators can be obtained from the density operator \cite{giamarchi2004quantum}, $\rho_{\lambda,\sigma}(x)=\zeta_{\lambda,k}\sum_{\sigma}\alpha^{\dagger}_{\sigma}(x)\alpha_{\sigma}(x)$, where $\zeta_{1,k_{i}}=u_{k_{i}}$ and $\zeta_{2,k_{i}}=v_{k_{i}}$ ($u_{k_{i}}$ and $v_{k_{i}}$ are given in Eq.~\ref{ukvk}). Detailed expression of $\rho_{\lambda,\sigma}(x)$ is given in Appendix.~\ref{denop}. At the half-filling of four Fermi points, below $U_c$, the bosonic fields $\phi_{c-}$ and $\theta_{s+}$, that are the duals of gapped fields $\theta_{c-}$ and $\phi_{s+}$, yields exponentially decaying correlations, while the gapless fields $\phi_{s-}$, $\phi_{c+}$, $\theta_{s-}$ and $\theta_{c+}$ induces power-law decay. 
Upon bosonization, the smooth part in the density operator contains all the bosonic fields and therefore leads to exponential correlations. Among the $2k_F$ terms, $\alpha_{1\sigma}^{\dagger}\alpha_{2\sigma} \sim \alpha_{3\sigma}^{\dagger}\alpha_{4\sigma} \sim e^{i2\sqrt{\pi/2}(\phi_{c+}+\theta_{c-}+\sigma(\phi_{s+}+\theta_{s-}))}$, with wave vector $(k_1-k_2)$, involves only ordered fields in both charge and spin sectors. Hence, it contributes to the power-law decay of the associated correlations. The charge correlations $\left\langle \rho(x)\rho(0)\right\rangle$ thus show $2k_F$ power-law decay with the exponent $\frac{1}{8}(K_{c}^{+}+\frac{1}{K_{s}^-})$.
Similarly, in the Mott insulating phase (above $U_c$), Umklapp terms further pin the charge field $\phi_{c+}$. Therefore, now the disordered fields $\phi_{c-}$, $\theta_{s+}$ and $\theta_{c+}$ yield the exponential decay of correlations. The same $2k_F$ terms show the dominant decay with the exponent $\frac{1}{8K_{s}^{-}}$.
In general, the $z$ component of the spin correlation is $S^z(x)=\rho(x)-(1/2)$ and therefore the charge correlations are mapped to the $z$ component of the spin sector \cite{giamarchi2004quantum}. Therefore, $\left\langle S^z(x)S^z(0)\right\rangle$ shows the power-law decay dominated by the $2k_F$ terms. The corresponding wave vector $(k_1-k_2)$ gives rise to oscillatory decay leading to antiferromagnetic correlations. 

In light of the preceding analysis, we now revisit the results of Ref.~\cite{PhysRevB.102.155429}, where the same interacting model was studied numerically using DMRG at half-filling. One of the key observations reported there was the occurrence of a metal-insulator transition at a finite critical interaction strength $U_c$. The simulations were carried out for $J = 6$ with parameters $\gamma_0 = 0.49$ and $g = 0.1$, chosen to access the regime where the bandwidth is minimized--corresponding to the quasiflat band limit of the model. As seen in Fig.~\ref{fig3} and discussed in Section III, this parameter regime also falls within the region characterized by four Fermi points. Our bosonization analysis predicts a nonzero $U_c$ in this regime, thus providing a natural explanation for the numerical observation in Ref.~\cite{PhysRevB.102.155429}.

To summarize, the interacting phase diagram at weak coupling is highly sensitive to the microscopic origin of the moir\'e miniband—specifically, to whether a uniform flux is present. In the absence of flux, the correlated phases closely resemble those of the conventional 1D Hubbard model, and no qualitatively new features emerge. By contrast, the introduction of flux fundamentally alters the phase diagram by modifying the band structure to include additional Fermi points, thereby enabling new scattering channels that are otherwise inaccessible.

\begin{figure*}
\centering
\includegraphics[width=0.6\columnwidth]{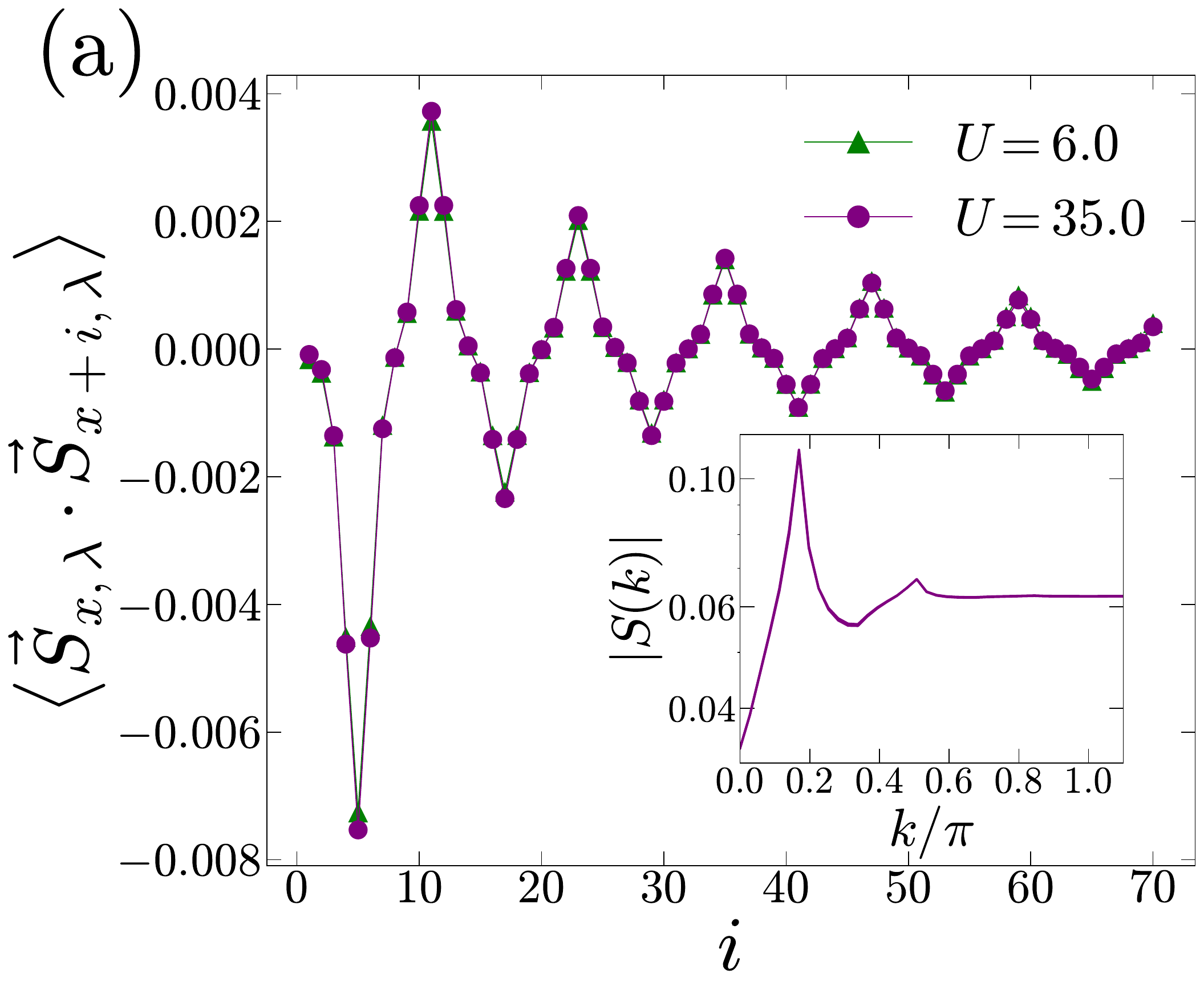}
\includegraphics[width=0.6\columnwidth]{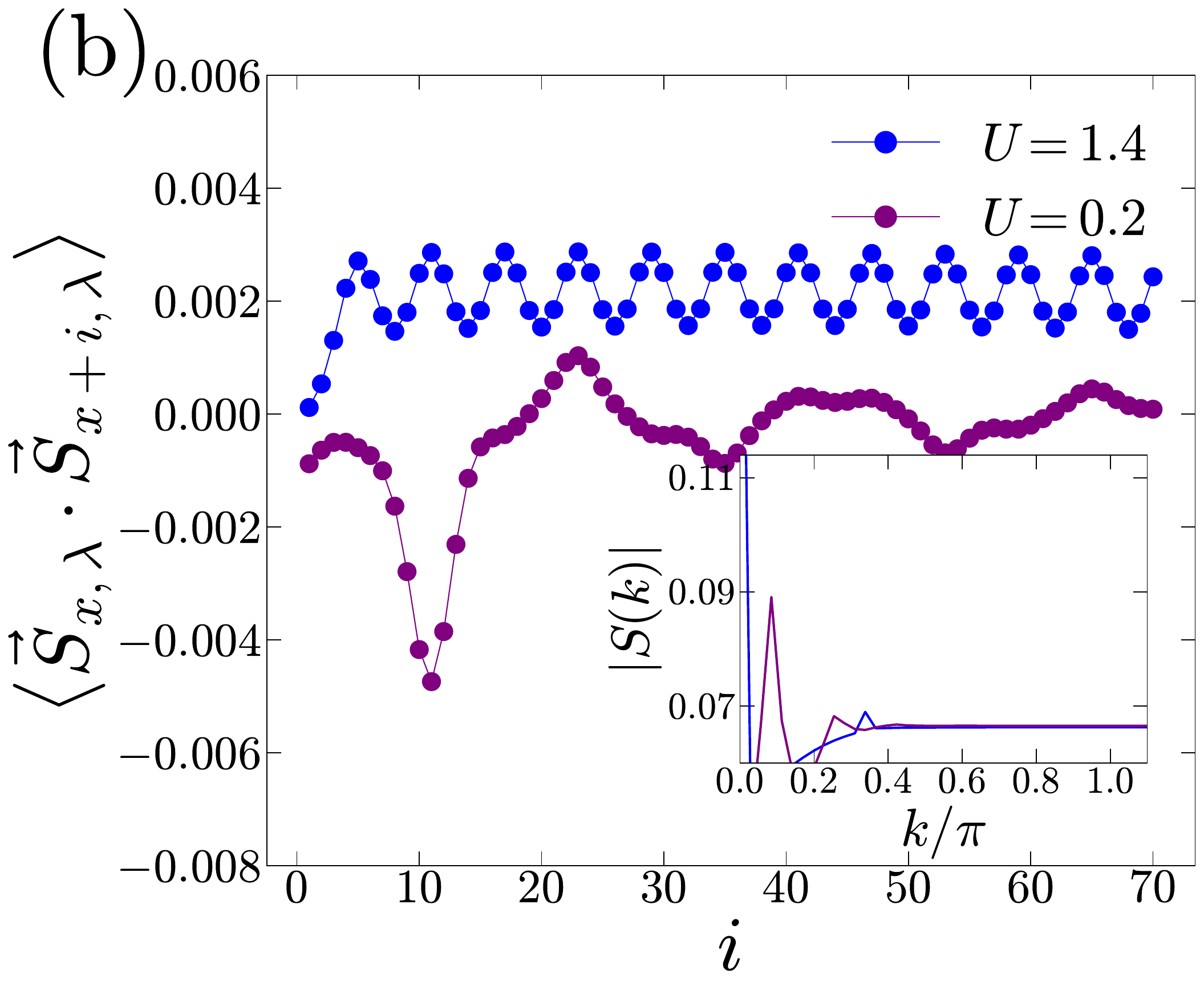}
\includegraphics[width=0.6\columnwidth]{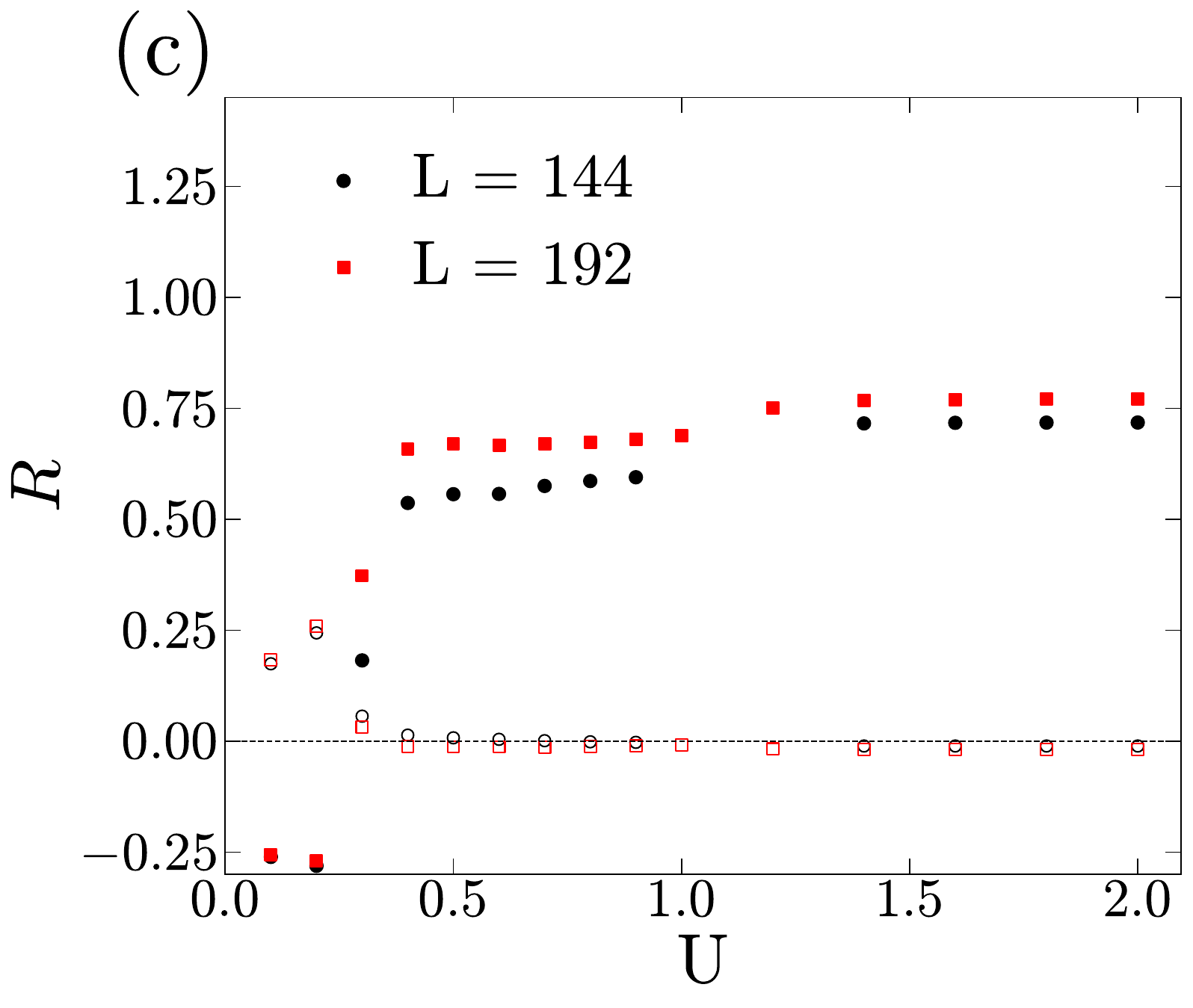}
\includegraphics[width=0.6\columnwidth]{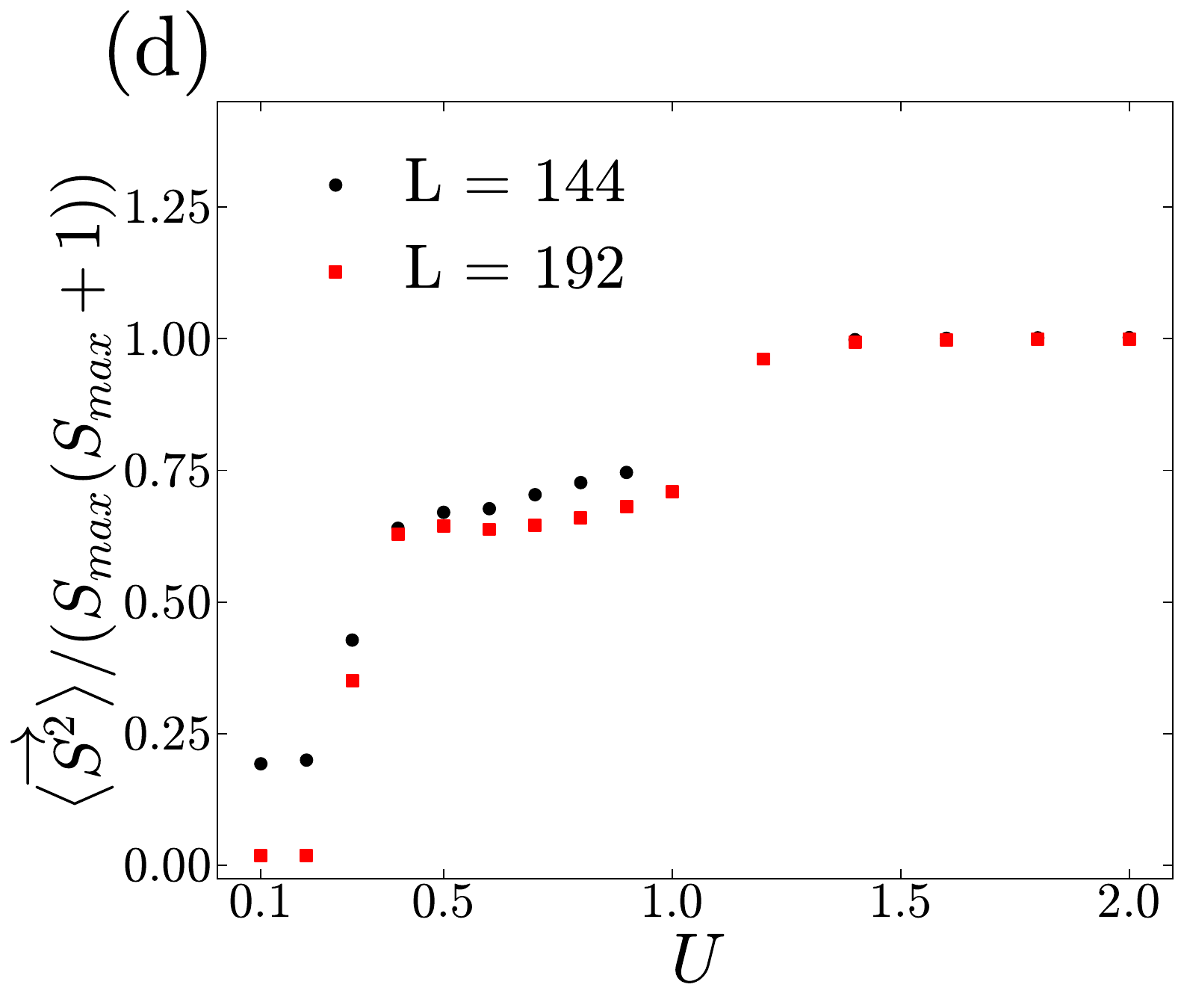}
\includegraphics[width=0.6\columnwidth]{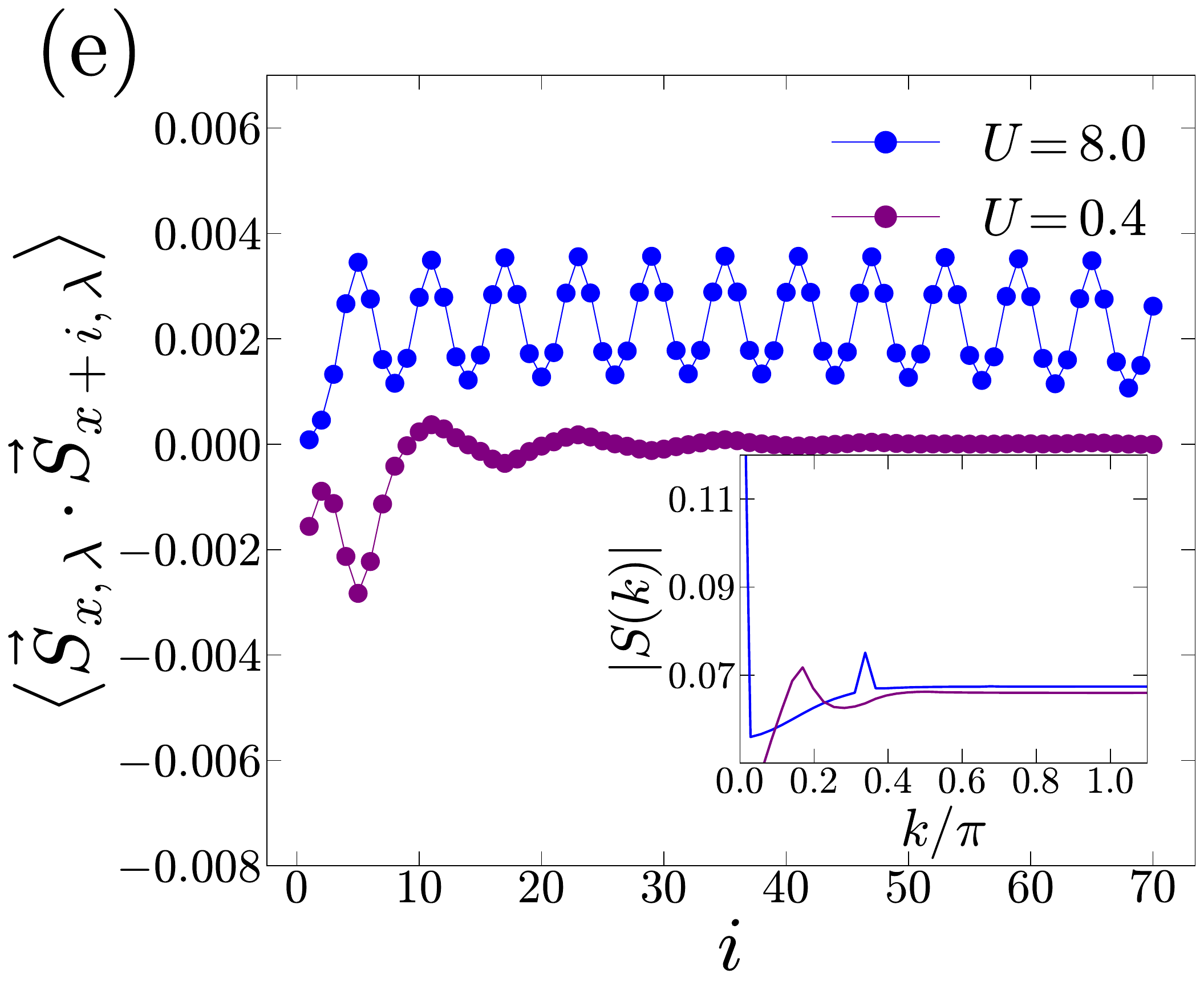}
\includegraphics[width=0.6\columnwidth]{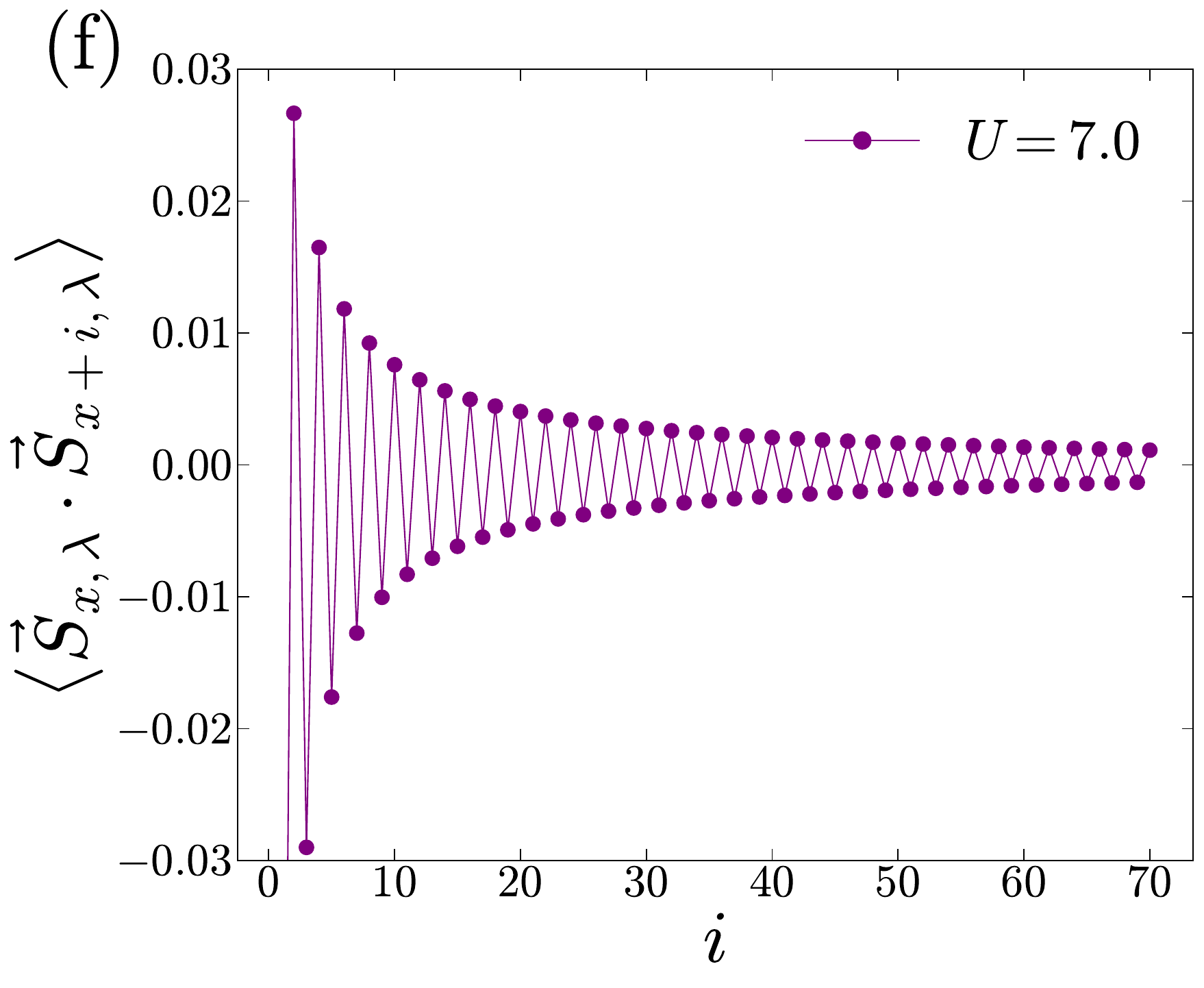}

\caption{Intraleg spin-spin correlations at half-filling of the lowest band are shown for different choices of \( \gamma_j \) in Eq.~\ref{intleghop}: (a) \( g \neq 0, \Gamma = 0 \), representing a system without flux but with periodically varying interleg hopping, for parameters \( \gamma_0 = 1 \) and \( g = 0.52 \); (b) \( g \neq 0, \Gamma = 1 \), where both flux and periodic interleg hopping are present, leading to four Fermi points at half-filling, for \( \gamma_0 = 0.49 \) and \( g = 0.1 \);
(c) Correlation ratio \( R \) as a function of \( U \), using the same parameters as in (b), where filled markers correspond to the  peak at \( k = 0 \), and hollow colored markers indicate the  peak at {\( k = (k_1 - k_2) \)}.
 (d) Spin normalised ratio with  $U$ also corresponds to the same parameters as in (b). (e) $g \neq 0, \Gamma = 1$, again with both flux and periodic interleg hopping, but now yielding two Fermi points at half-filling, paramters used for the plot $\gamma_{0}=0.37$ and $g=0.3$. (f) Intraleg spin-spin correlation for  $\gamma_j=\gamma_0 \mathrm{e}^{i\frac{2\pi j}{J}}$, implying the presence of flux but no periodic interleg hopping, parameters used are $t=0.2$ and$\gamma_{0}=0.49$. For case-(a),(b),(e) and (f), the parameters  and $U$ are chosen to maintain  an interaction-to-bandwidth ratio of $\langle U \rangle / \delta_{\mathrm{BW}} \geq 1$ (see main text for details). It is evident that in cases (a) and (f), the spin-spin correlations are antiferromagnetic, while in cases (b) and (e), they are ferromagnetic. The spin-spin correlation is plotted for one leg only, as the results are identical on the other leg in each case with $x=L/4$. For the ferromagnetic phase in (b), we have also varied $x$ (see Appendix~\ref {sfof} for details).} \label{fig5}
\end{figure*} 

\subsection{\label{halfstrong}  Strong coupling  limit}

In the strong coupling limit $\langle U \rangle / \delta_{\mathrm{BW}} \gtrsim 1$, we use DMRG technique~\cite{PhysRevLett.69.2863,PhysRevB.48.10345,SCHOLLWOCK201196} with open boundary conditions to find the ground state properties of the model at half-filling. The calculations are conducted using the ITensor library~\cite{itensor, itensor-r0.3}. To mitigate finite-size effects, we have investigated systems of different sizes, reaching a maximum of $2L = 2 \times 192$ sites and a cutoff error below $10^{-9}$ with bond dimensions up to 3200 (see Appendix~\ref{converge} for further details). The results presented below,  unless otherwise stated, are for  $2L=2\times 144$ sites. The energy difference between the two iterations remains lower than $10^{-7}$ throughout the final sweeps. Total charge and $S^{z}$  conservation methods are employed to enhance convergence. Throughout this section, we fix $t = 1$ and $J = 6$, unless otherwise stated. 

\subsubsection{Case 1: $g \neq 0$,$\Gamma=0$ (No flux)}

In this case, the ground state retains the same qualitative structure as in the weak-coupling limit. Specifically, the charge sector remains gapped, while the spin sector stays gapless. Throughout the entire range of interaction strengths—both weak and strong coupling—the spin-spin correlations exhibit antiferromagnetic behavior. A representative example is shown in Fig.~\ref{fig5}(a), where the real-space spin correlations display oscillations with wavevector $k = 2\pi/12 = 2k_{F}$, in agreement with expectations from bosonization. This is further confirmed by the spin structure factor, plotted in the inset of Fig.~\ref{fig5}(a), which shows a clear peak at $2k_{F}$. The spin structure factor is defined as $
S(k) = \frac{1}{L} \sum_{i,j} \vec{S}_{i,\lambda} \cdot \vec{S}_{j,\lambda} \, e^{ik(x_i - x_j)}$. Overall, no new qualitative features emerge in either the charge or spin sectors as $\langle U\rangle/\delta_{BW}$ increases.

\subsubsection{Case 2: $g \neq 0$,$\Gamma=1$ (With flux )}

In this scenario the charge sector does not present any new qualitative features on increasing $U$--it remains gapped beyond a certain critical interaction strength, which is zero or nonzero depending on the number of Fermi points as discussed earlier. In the spin sector, the spin gap remains zero independent of the interaction strength, but interestingly, the spin-spin correlation undergoes a transition from antiferromagnetic to ferromagnetic as the ratio $\langle U \rangle / \delta_{BW}$ increases. This interesting behavior was first reported in Ref.~\cite{PhysRevB.102.155429}. In the present work, we revisit this transition with a more systematic and detailed analysis, addressing several aspects not considered previously. Note, while Ref.~\cite{PhysRevB.102.155429} explored the transition by varying $\delta_{BW}$ at fixed $U$, we adopt the complementary approach of holding $\delta_{BW}$ fixed while tuning $U$. This methodology not only serves to corroborate the earlier findings but also ensures that the underlying Fermiology remains unchanged throughout the transition, allowing for a more controlled examination of interaction effects.

We first consider the case when there are four Fermi points at half-filling. This arises, for example,  when $\gamma_0 = 0.49$ and $ g = 0.1$. The corresponding spin-spin correlation is given in Fig.~\ref{fig5}(b),   At smaller values of $U$, the spin-spin correlation exhibits a spin density wave with antiferromagnetic correlations, as expected. On increasing the interaction strength, however, it becomes ferromagnetic. The transition is further corroborated by studying the spin structure factor, shown in the inset to Fig.~\ref{fig5}(b). In the antiferromagnetic regime, two peaks are observed in the structure factor \( S(k) \) one at \( k = 2(k_1 + k_2) \approx 0.26 \) and \( k = k_1 - k_2 = \frac{2\pi}{24} \). The latter is the expected dominant oscillation mode with a power-law decay, whereas the former oscillation mode is expected to be sub-dominant with an exponential decay, according to bosonization calculation. The simultaneous presence of both is attributed to finite-size effects.
Nevertheless, both peaks vanish upon entering the ferromagnetic regime. In ferromagnetic regime, \( S(k) \) exhibits peaks at \( k = 0 \) and \( k = \frac{2\pi}{6} \). The former peak signals the presence of ferromagnetism, whereas the latter peak reflects the moir\'e periodicity. Note that the antiferromagnetic order here is quasi–long-range due to the Mermin–Wagner theorem, which prohibits true long-range order in one dimension even at $T=0$. In contrast, the ferromagnetic order can exhibit true long-range behavior \cite{patrickfazekas,Pitaevskii1991}. 

To unambiguously characterize the onset of ferromagnetism, we compute the correlation ratio introduced in Ref.~\cite{PhysRevLett.117.086404}:
\begin{equation}
    R = 1 - \frac{S(k + \delta k)}{S(k)},
\end{equation}
Here, \( S(k) \) denotes the spin structure factor, and \( \delta k = 2\pi/L \) is the spacing between adjacent momentum values, with \( L \) being the system length.  The correlation ratio \( R \) serves as an indicator of long-range order: it scales to one in an ordered phase and scales to zero in a disordered phase. The relevant phase is identified by the momentum \( k \) at which the peak in the spin structure factor occurs. In  Fig.~\ref{fig5}(c), we plot  \( R \) as a function of the interaction strength \( U \). For the ferromagnetic phase, this quantity is evaluated at \( k = 0 \) (shown as filled marker); it becomes positive around \( U = 0.3 \) and increases toward unity as \( U \) increases. In contrast, for the antiferromagnetic case, it is evaluated at the characteristic antiferromagnetic wavevector {\( k = k_1 - k_2 \); here, $R$ (shown as hollow markers) initially increases for $U \leq 0.2$, but approaches zero around $U \approx 0.3$ and then goes to negative as U increases, indicating the disappearance of the peak in the spin structure factor. 

These observations collectively signal a transition in the ground state from antiferromagnetic to ferromagnetic behavior around \( U \approx 0.3 \), followed by the strengthening of ferromagnetic order as \( U \) increases further. This transition is consistent across large system sizes (\( L = 144 \), \( L = 192 \)). Additionally, for peak at $k=0$, with increasing $U$, the value of $R$ (shown in solid colour) increases as $L$ increases, implying that the contribution from finite-size effects diminishes. The interaction values \( U \) were sampled with a resolution of \( 0.1 \) up to \( U = 1 \), and with increments of \( 0.2 \) thereafter. 

A pertinent question is whether the ferromagnetic state is fully or partially polarized. To determine, we compute the expectation value of the square of the total spin operator, \( \langle \vec{S}_{\text{tot}}^2 \rangle \), on the ground state. This quantity is defined as
\begin{equation}
    \langle \vec{S}_{\text{tot}}^2 \rangle = \sum_{i,j,\lambda,\lambda'} \langle \vec{S}_{i,\lambda} \cdot \vec{S}_{j,\lambda'} \rangle = S(S+1).
\end{equation}
Note that, while \( S(k) \) and the correlation ratio \( R \) were computed using spin correlations along a single leg of the ladder, here we consider contributions from both legs to capture the total spin of the system. For a fully polarized ferromagnetic state, the total spin quantum number is \( S = S_{\text{max}} = N/2 \), where \( N \) is the total number of electrons. In Fig.~\ref{fig5}(d), we plot this normalized ratio as a function of the interaction strength \( U \) for two large system sizes (\( L = 144 \) and \( L = 192 \)). For small interaction strengths (\( U \leq 0.2 \)), the normalized spin ratio remains small and tends toward zero with increasing system size, indicating the absence of full polarization. However, as \( U \) increases, the difference between the values obtained for the two system sizes diminishes, and the ratio converges toward unity, consistent with the emergence of a fully polarized ferromagnetic ground state. To minimize boundary effects and better reflect the bulk behavior, this calculation is performed in the central region of the ladder. Nevertheless, the convergence to unity, independently of the system size,  implies that it is a true thermodynamic phase.

In all, these results show that the ferromagnetic phase observed at strong coupling is a true thermodynamic phase characterized by full spin polarization.

We now examine the case with two Fermi points. In this regime as well, the spin–spin correlation exhibits antiferromagnetic behavior at smaller values of the interaction strength \( U \), and evolves into ferromagnetic behavior as \( U \) increases. This transition is illustrated in Fig.~\ref{fig5}(e), for parameters \( \gamma_0 = 0.37 \) and \( g = 0.3    \), which correspond to two Fermi points. In the antiferromagnetic phase, the spin structure factor displays a peak at \( k = 2\pi/12=2k_F \), as shown in the inset of the figure. This feature is consistent with predictions from weak-coupling bosonization. Note that the critical interaction strength \( U_s \) required for the onset of ferromagnetism can differ from the four Fermi point case because the bandwidth here is different. These observations collectively indicate that the emergence of ferromagnetism is not tied to the number of Fermi points.

Next, we inquire which aspect of the moir\'e potential is responsible for ferromagnetic correlations. We have already shown that when flux is absent, the periodic potential alone cannot drive the system to ferromangetic state, i.e., Case 1 of moir\'e does not present this novel behavior, only Case 2 exhibits this. One may wonder if this is intrinsically the physics related to the flux. To examine that effect, we switch off the periodicity in the interleg hopping but retain the flux, i.e., we put $g=0$ and $\Gamma=1$ in Eq.~\ref{intham}. We find the spin-spin correlation always to be antiferromagnetic as shown in Fig.~\ref{fig5}(f), for  $\gamma_{0}=0.49$ and $t=0.2$. This antiferromagnetic behavior persists across a wide range of parameters, even at large \( U \). This unequivocally confirms that ferromagnetism in our model arises from a combination of both flux and periodicity in the interleg hopping at the noninteracting level, with both features being equally vital.

Finally, to assess the generality of this behavior, we also explore other values of \( J \), such as \( J = 4 \) and \( J = 8 \), along with different combinations of \( \gamma_0 \) and \( g \) within the parameter regime where both flux and modulation are present. In all cases, we observe qualitatively similar behavior: spin–spin correlations transition from antiferromagnetic to ferromagnetic with increasing interaction strength \( U \). This demonstrates that the observed ferromagnetism is a generic feature of the model and is not tied to a specific value of \( J \), which sets the periodicity of the interleg modulation and the strength of the flux.

In summary, we find that ferromagnetism is a robust feature of the moir\'e ladder with flux in the strong-coupling regime. The transition from antiferromagnetic to ferromagnetic spin correlations, observed as the ratio $\langle U\rangle/\delta_{BW}$ increases, is independent of the number of Fermi points and persists across different parameter regimes at the single-particle level. Our analysis confirms that this behavior requires both a uniform flux and a spatially modulated interleg hopping—neither alone is sufficient. Diagnostics based on spin structure factors and spin polarization further demonstrate that the ferromagnetic phase exhibits true long-range order and survives in the thermodynamic limit.

%%%%%%%%%%%%%%%%%%%

\section{\label{otherfillings}Magnetic ground states at various fillings}

\begin{figure}
\centering
\includegraphics[width=1\columnwidth]{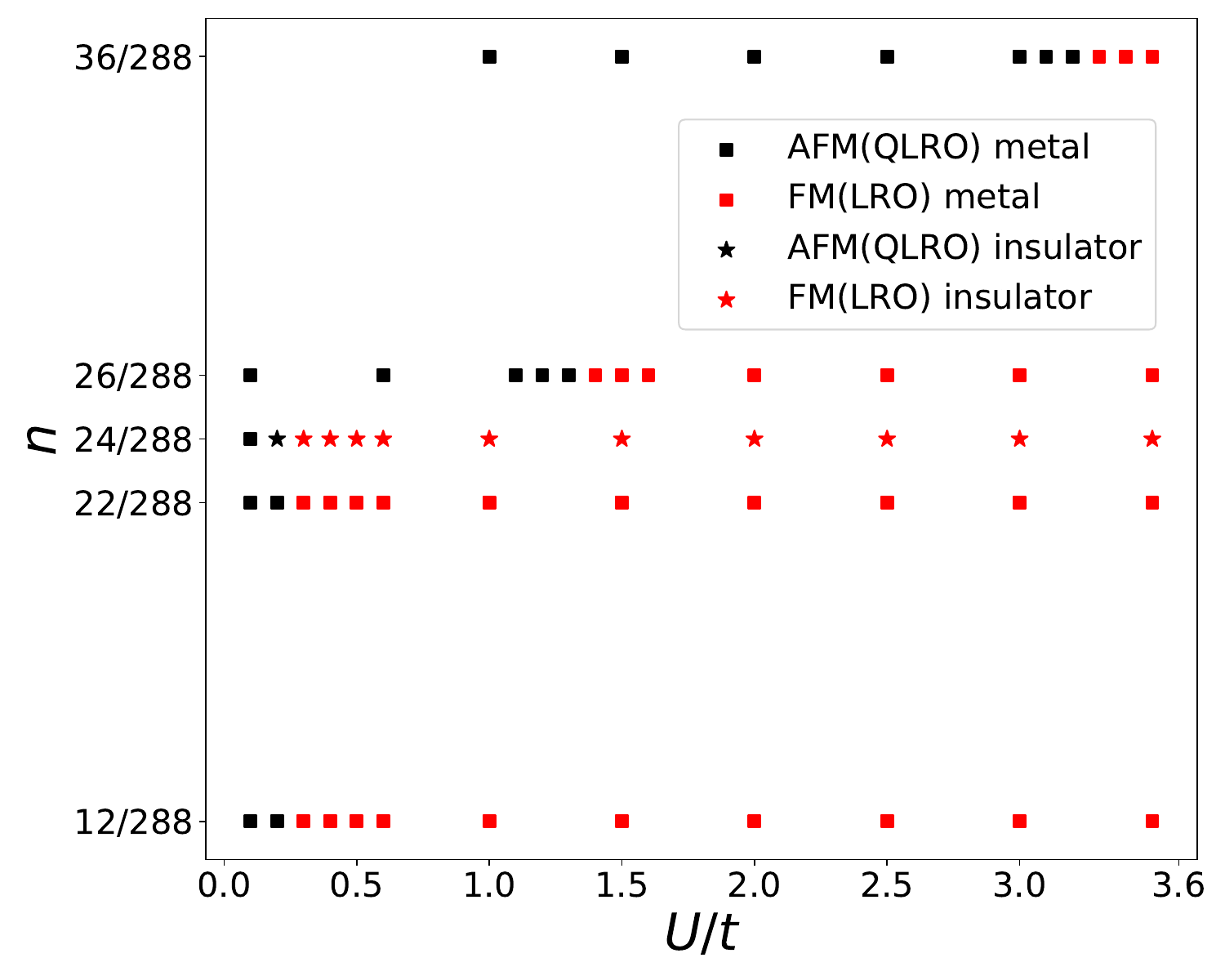}
\caption{The interacting phase diagram is presented as a function of electron density \( n \)  of the lowest band and interaction strength \( U/t \). For each electron filling---specifically, one-quarter ($n=1/4J=12/288$), slightly below half ($n=22/288$),half-filling ($n=1/2J=24/288$), slightly above half ($n=26/288$), three-quarters ($n=3/4J=36/288$)  of the lowest band(from bottom to top)---the critical interaction strength \( U \) required to induce a transition from antiferromagnetism (AFM) to ferromagnetism (FM) increases with filling. The spin-spin correlations for the data points (depicted as squares and stars) were computed numerically using DMRG. Squares (stars) indicate the metallic (insulating) phase. The black color indicates antiferromagnetic correlations, while the red color indicates ferromagnetic correlations. All calculations were performed for \( J = 6 \),  \( \gamma_{0} = 0.49 \),\( g = 0.1 \),\( t = 1 \) and a system size of \( L = 144 \).
} \label{fig6}
\end{figure}

\begin{table}
\centering
\renewcommand{\arraystretch}{1.5}
\setlength{\tabcolsep}{12pt}
\makebox[\linewidth]{%
\begin{tabular}{c c c c}
\toprule
\addlinespace[1pt] % small space between top lines
\toprule
Filling & $U_{s}$ & $\langle U_{s} \rangle / \delta_{BW}$ & $\langle U_{s} \rangle / \Delta_{BG}$ \\
\midrule
\midrule
12/288     & 0.3 & 0.51  & 0.08 \\
22/288 & 0.3 & 0.89  & 0.14 \\
24/288     & 0.3 & 0.96  & 0.15 \\
26/288 & 1.4 & 4.80  & 0.73 \\
36/288     & 3.2 & 14.93 & 2.28 \\
\bottomrule
\bottomrule

\end{tabular}%

}
\caption{This table shows the values of $\langle U_s \rangle / \delta_{BW}$ and $\langle U_s \rangle / \Delta_{BG}$ at various fillings, evaluated at their respective critical interaction strength $U_s$. Here $U_{s}$ represents the value of $U$ where the spin-spin correlation of the ground state becomes antiferromagnetic to ferromagnetic. The listed $U_s$ values are accurate within an error margin of $0.1$. All these calculations are performed for $L=144$. }\label{ustable}
\end{table}

After the extensive analysis at half-filling, a natural question arises: do the nontrivial features of the ground state—particularly ferromagnetism—persist at other electron fillings? Away from half-filling, the charge sector does not undergo a metal-insulator transition for the onsite Hubbard interaction considered here. As shown in the Appendix~\ref{gap}, the charge gap remains zero, indicating metallic behavior. The spin sector, however, remains open to further investigation. Specifically, we ask whether the ferromagnetic correlations observed at half-filling survive at other electron densities.

Recall that at half-filling, the emergence of ferromagnetism was shown to require both a uniform flux and spatially modulated interleg hopping, and to be independent of the number of Fermi points. Accordingly, we now study the full moir\'e model (Case 2: \( g \ne 0,\, \Gamma = 1 \)) using the same parameters as before—\( \gamma_0 = 0.49 \), \( g = 0.1 \), \( J = 6 \), and \( t = 1 \)—corresponding to four Fermi points. We consider electron densities corresponding to one-quarter filling (\(n=12/288\)), three-quarter filling (\(n=36/288\)), and two electrons above (\(n=26/288\)) and below (\(n=22/288\)) half-filling of the lowest band, using a fixed system size of \(L = 144\).

Our results are summarized in Fig.~\ref{fig6}. At each filling, the system undergoes a transition from an antiferromagnetic phase (shown in black) to a ferromagnetic phase (shown in red) as the interaction strength \( U \) increases. As in the half-filled case, we classify a phase as ferromagnetic when the spin structure factor shows a peak at \( k = 0 \), indicating net magnetization, and as antiferromagnetic when it exhibits a finite-momentum peak characteristic of spin-density-wave correlations and zero net magnetization (see Appendix~\ref{sfof} for details). The ferromagnetic phase exhibits long-range order (LRO), while the antiferromagnetic phase remains quasi-long-ranged (QLRO), consistent with expectations in one dimension.

Although ferromagnetism is observed across all fillings, a pronounced asymmetry emerges between electron densities below and above half-filling. For fillings below half, the critical interaction strength \( U_s \) required to induce ferromagnetism remains nearly constant. In contrast, \( U_s \) increases significantly as one moves above half-filling. Phase boundaries were determined by scanning \( U \) in steps of 0.1 near the transition point. To assess the relative strength of interactions at the transition, we compute the ratios \( \langle U \rangle / \delta_{\text{BW}} \) and \( \langle U \rangle / \Delta_{\text{BG}} \), evaluated at \( U = U_s \). These values, summarized in Table~\ref{ustable}, indicate that ferromagnetism below half-filling occurs in a regime with minimal band mixing, while the higher ratios above half-filling point to significant interband contributions.

As noted earlier, at half-filling the system also undergoes a metal-insulator transition at a lower critical interaction \( U_c \approx 0.1 < U_s \approx 0.3 \). At all other fillings studied, the charge sector remains metallic, and this information is included in the phase diagram for completeness.

Taken together, these results establish that ferromagnetism is a robust and persistent feature of the moir\'e ladder, extending well beyond half-filling and occurring independently of whether the charge sector is gapped or not.

%%%%%%%%%%%%%%%

\section{\label{discussion}Discussion}

The preceding sections offer a detailed, systematic analysis of how interactions shape the ground states of moir\'e minibands. Two markedly different many-body landscapes emerge, depending on whether a flux is present or not, in addition to the periodic interleg hopping. In Case 1, when flux is absent, the ground states mirror those of a conventional one-dimensional Hubbard model: no qualitatively new features are observed either in the charge or the spin sector.  BZ folding and bandwidth reduction alone do not yield novel correlated behaviour. By contrast, in Case 2, when flux is introduced, it fundamentally reorganizes the low-energy theory. Both charge and spin sectors acquire qualitatively new features: the metal-insulator transition changes character, and robust ferromagnetic spin–spin correlations arise at strong coupling across a variety of fillings, neither of which is expected in the standard 1D Hubbard model. Recall that the flux was added as a one-dimensional proxy for the momentum shift between the Dirac cones in TBG  [cf. Fig.~\ref{fig1}(d), Section II]. The momentum shift originates entirely due to the fact that Dirac cones in each layer of graphene arise at the $K$ point; had this not been the case, and the low energy sector of each layer resided at the $\Gamma$ point, this shift would be absent. Thus, we conclude that moir\'e physics is more than simple zone folding or band flattening; the microscopic route to miniband formation--specifically, whether a momentum shift precedes hybridization--determines which interaction channels become relevant and which phases emerge. In short, how the moir\'e band forms matters more than how flat it is.

Building on these insights, several generalizations of our model can be made that bring it closer to TBG, while remaining in one dimension. For example, introducing a second copy of the ladder carrying the opposite flux would form a time-reversal–invariant, effective two-valley system whose inter-ladder couplings emulate intervalley scattering in TBG. Additionally, incorporating sublattice degrees of freedom, such as by using Su-Schrieffer-Heeger (SSH) chains on each leg, would allow for further exploration of symmetry-protected and topology-influenced interaction physics in 1D moir\'e systems. Together, these extensions preserve the tractability of one dimension while importing the valley, sublattice, and topology ingredients that dominate the correlated physics of graphene-based moir\'e materials.

Beyond its implications for moir\'e systems, our findings raise fundamental questions about correlated physics in one dimension. While the behavior of the charge sector in our model is largely consistent with established expectations, the spin sector reveals surprising and unconventional features. As is well known, the standard one-dimensional Hubbard model does not exhibit ferromagnetism at any filling, thanks to Lieb-Mattis theorem.  There are three main approaches to circumventing this and realizing ferromagnetism in such systems. First, Nagaoka ferromagnetism arises when the system has one less electron than half-filling of the full lattice ($n \to 1$) and the onsite interaction is infinite~\cite{Nagaoka:1966zz, PhysRevB.40.9192, 10.1143/PTP.99.489}. Since our model involves finite interaction strength and a low filling of the full lattice, it does not fall into the Nagaoka regime. Second, Muller-Hartmann extended Nagaoka's ferromagnetism to finite $U$  and finite hole densities, showing that ferromagnetism can emerge  at low electron densities ($n \to 0$)~\cite{ Müller-Hartmann1995}. In this class, ferromagnetism emerges from inter-valley scatterings present in the two band minima, hence it depends on the number of Fermi points at a specific filling. However, as we show, ferromagnetism in our model does not depend on a number of Fermi points, and therefore it does not belong to this class either. Third, Mielke and Tasaki have shown that lattice models with a flat band and certain connectivity conditions can exhibit fully polarized ferromagnetism at specific fillings, even for infinitesimal on-site interaction $U$~\cite{MIELKE1993443, Mielke1993, Mielke_1991, Mielke_1999, PhysRevLett.69.1608}. Tasaki illustrated this in a one-dimensional multiband model made of periodically repeated triangular units, where suppressing hopping between even sublattices yields an exactly flat lowest band and ferromagnetism. Even when the band is dispersive, saturated insulating ferromagnetism can still emerge at half-filling over a broad parameter range for sufficiently large $U$~\cite{ PhysRevLett.75.4678, Tasaki1996}. 
It remains unclear whether our model fits within the Mielke–Tasaki class. While the Mielke–Tasaki framework permits an exact rewriting of the interaction term in a localized basis—enabling analytical treatment—such a simplification is difficult in our model because of the inherent large periodicity. Determining this connection remains an open question for future work.

%%%%%%%%%%%%%%%%%%%%%

\section{\label{conclusion}Concluding remarks}

In this work, we have carried out a comprehensive study of interaction-driven phenomena in a one-dimensional model—the moir\'e ladder—that serves as a minimal analog of moir\'e systems such as twisted bilayer graphene. The model incorporates two key ingredients: a spatially modulated interleg hopping and a uniform magnetic flux. While previous work showed that this system supports a quasiflat band and unconventional many-body phases at half-filling, the microscopic origin of these phases and the individual roles of the two ingredients remained unclear. 

By combining bosonization and DMRG techniques, we have revisited the ground state at half filling and performed a comprehensive analysis of the interacting phase diagram. Our results show that the metal-insulator transition at finite interaction strength can be understood within a perturbative framework, governed by the number of Fermi points in the noninteracting band structure. In contrast, the ferromagnetic correlations that arise at strong coupling are genuinely nonperturbative in nature. Crucially, our analysis demonstrates that both the uniform flux and the spatially modulated interleg hopping are necessary to stabilize the ferromagnetic phase—neither ingredient alone is sufficient. Overall, in the absence of flux, the periodic modulation of the interleg hopping, while flattening the band, does not give rise to any qualitatively new correlated behavior.

We have also extended our study beyond half filling, examining the ground state at quarter, three-quarter, and slightly doped half fillings. We find that at intermediate interaction strengths, the system exhibits ferromagnetic correlations below half-filling and antiferromagnetic correlations above. At stronger interactions, where interband mixing becomes significant, ferromagnetic correlations dominate across all studied fillings. Furthermore, in all cases where ferromagnetism appears, reducing the interaction strength relative to the minibandwidth leads to a transition to antiferromagnetic correlations.

Our results underscore that flatness of the band is not the only determinant of correlated behavior; the detailed structure of the band—particularly those encoded by the flux—play a decisive role. The moir\'e ladder thus provides a valuable framework for isolating and understanding the mechanisms underlying correlation-driven phases in more complex moir\'e systems. Beyond its implications for moir\'e physics, our findings also contribute to the broader understanding of unconventional ferromagnetism in one dimension, highlighting the potential of band-structure engineering to circumvent standard no-go theorems and realize exotic many-body ground states.

%%%%%%%%%%%%%%%%%

\begin{acknowledgments}
   We thank S. Bera and S. Pujari for useful discussions. H.K.P. would like to thank DST SERB, India for financial support via Grant No. CRG/2021/005453. 
   R.R.K. would like to acknowledge the financial support from Indian Institute
of Technology Bombay through Institute Post Doctoral
Fellowship. P.K.P. and H.K.P. thank the National Supercomputing Mission (NSM) for providing computing resources
of “PARAM Rudra” at IITB, implemented by C-DAC and
supported by the Ministry of Electronics and Information Technology (MeitY) and the Department of Science and Technology (DST), India. Y.H. was supported by the U.S. DOE NNSA under Contract No. 89233218CNA000001 and by the Center for Integrated Nanotechnologies, a DOE BES user facility, in partnership with the LANL Institutional Computing Program for computational resources.
\end{acknowledgments}

\vspace{5mm}

\noindent $^\flat$ These authors contributed equally to this work.

%%%%%%%%%%%%%%%%%%%%%%%%%%%%%%%%%%%%%%%%%%%%%%%%%%%%%%%%%%%

\newpage

\bibliography{ref.bib}

\newpage

\appendix

\section{Bosonization details}\label{boso}
Here we discuss the bosonization details of two Fermi points, in both Case 1 and Case 2, and four Fermi points in Case 2. We consider the lower band, $\epsilon_{\alpha}$ in Eq.~\ref{EDC2}
{\small\begin{align}
\epsilon_{\alpha}(k) 
&= -2t \left[ \cos k \cos\left(\frac{\phi}{2}\right) + 
\sqrt{ \sin^2 k \sin^2\left(\frac{\phi}{2}\right) + \left(\frac{\gamma_0}{2t}\right)^2 } \right]
\end{align}}
Now, introducing periodic modulation ($g\neq0$) will induce a gap at $\pm\pi/aJ$, which results in the lowest miniband derived from $\epsilon_{\alpha}$. The half-filling of this band, $\epsilon_{\alpha}$ in the reduced BZ, has both two and four Fermi points depending on the parameter $\gamma_0$, as shown in Fig.~\ref{fig4}(d)-(I),(II),(III). Two Fermi points appear when $\gamma_0<\gamma_0^{c1}$ and $\gamma_0>\gamma_0^{c2}$, whereas four Fermi points occur when $\gamma_0^{c1}<\gamma_0<\gamma_0^{c2}$. Two Fermi points for Case 1, i.e., $\Gamma=0$ appear for all $\gamma_0$, as shown in Fig.~\ref{fig4}(a).

\subsection{Two Fermi points}\label{boso2fp}
The half-filling of band $\epsilon_{\alpha}$ in the reduced BZ, has two Fermi points at $k_F=\pm\pi/2aJ$. Upon linearizing the spectrum near these Fermi points, for $\gamma_0>\gamma_0^{c2}$, we write the annihilation operator in terms of the left and right movers for the lower band
\begin{equation}
c_{\lambda,\sigma}= \zeta_{\lambda,k_F} e^{ik_F x} \alpha_{R\sigma}(x)+ \zeta_{\lambda,-k_F} e^{-ik_F x} \alpha_{L\sigma}(x)
\label{RLM1}
\end{equation}
where $\zeta_{1,\pm k_F}=u_{\pm k_F}$ and $\zeta_{2,\pm k_F}=v_{\pm k_F}$. The non-interacting part can be written as
\begin{equation}
    H_{non-int}=-iv_F^{\alpha}\sum_{\sigma}\int dx \left( \alpha_{R\sigma}^{\dagger}\partial_x\alpha_{R\sigma}-\alpha_{L\sigma}^{\dagger}\partial_x\alpha_{L\sigma}\right)
\end{equation}
The interacting part, i.e., $:H_\mathrm{int}: = U \sum_{j,\lambda} n_{j\uparrow\lambda} n_{j\downarrow\lambda}$ (where $::$ denotes normal ordering), has momentum conserving ($H_{d}, H_{od}$ and $H_{b}$) and Umklapp ($H_{u}$) scattering terms,
\begin{align}
    H_d&= U(u_{k_F}^2+v_{k_F}^2) \nonumber\\
    &\int dx \left( \alpha_{R\uparrow}^{\dagger}\alpha_{R\uparrow}\alpha_{R\downarrow}^{\dagger}\alpha_{R\downarrow}+\alpha_{L\uparrow}^{\dagger}\alpha_{L\uparrow}\alpha_{L\downarrow}^{\dagger}\alpha_{L\downarrow}\right)\\
    H_{od}&= 2U(u_{k_F}v_{k_F})\nonumber\\
    &\int dx\left( \alpha_{R\uparrow}^{\dagger}\alpha_{R\uparrow}\alpha_{L\downarrow}^{\dagger}\alpha_{L\downarrow}+\alpha_{L\uparrow}^{\dagger}\alpha_{L\uparrow}\alpha_{R\downarrow}^{\dagger}\alpha_{R\downarrow}\right)\\
    H_{b}&= 2U(u_{k_F}v_{k_F})^2\nonumber\\
    &\int dx\left( \alpha_{R\uparrow}^{\dagger}\alpha_{L\uparrow}\alpha_{L\downarrow}^{\dagger}\alpha_{R\downarrow}+H.c\right)\\
    H_{u}&=2U(u_{k_F}v_{k_F})^2\nonumber\\
    &\int dx\left( e^{-4ik_FJx}\alpha_{R\uparrow}^{\dagger}\alpha_{L\uparrow}\alpha_{R\downarrow}^{\dagger}\alpha_{L\downarrow}+H.c\right) 
\end{align}
where $u_{k_F}$ and $v_{k_F}$ are given in Eq.~\ref{ukvk}. Now, we use the usual bosonization rules, 
\begin{equation}
    \alpha_{R\sigma}(x)=F_{R\sigma} \frac{e^{-i2\sqrt{\pi}\phi_{R\sigma}(x)}}{\sqrt{2\pi\eta}}, \;\; \alpha_{L\sigma}(x)=F_{L\sigma} \frac{e^{i2\sqrt{\pi}\phi_{L\sigma}(x)}}{\sqrt{2\pi\eta}}
\end{equation}
where, $F_{R\sigma,L\sigma}$ are Klein factors that ensure anticommutation of fermion operators and $\eta$ is the short distance cut-off. 
The bosonic fields can be re-written using $\phi_{R\sigma(L\sigma)}=(\phi_{\sigma}\mp\theta_{\sigma})/2$ and $\phi_{\uparrow(\downarrow)}=(\phi_c\pm\phi_s)/\sqrt{2}$ (similarly for $\theta$ field), where $\phi_{\sigma}$ and $\theta_{\sigma}$ are the nonchial boson fields and $\phi_{c/s}$ and $\theta_{c/s}$ charge ($c$)-spin ($s$) separated bosonic fields, with $R\sigma(L\sigma)$ corresponding to $-(+)$ and $\uparrow(\downarrow)$ corresponding to $+(-)$. Now, the non-interacting Hamiltonian can be written as
\begin{equation}
    H_{non-int}= \frac{v_F^{\alpha}}{2}\sum_{\kappa=c,s}\int dx \left[(\partial_x\phi_{\kappa})^2+ (\partial_x\theta_{\kappa})^2\right]
\end{equation}
The $H_{d}$ and $H_{od}$ terms can be written as
\begin{align}
    H_{d}&= g_d \int dx \left[(\partial_x\phi_{c})^2+ (\partial_x\theta_{c})^2-(\partial_x\phi_{s})^2- (\partial_x\theta_{s})^2\right]\\
    H_{od}&=g_{od} \int dx \left[(\partial_x\phi_{c})^2- (\partial_x\theta_{c})^2-(\partial_x\phi_{s})^2+ (\partial_x\theta_{s})^2\right]
\end{align}
where $g_{d}=U(u_{k_F}^2+v_{k_F}^2)/4\pi$ and $g_{od}=U(u_{k_F}v_{k_F})/2\pi$. Therefore, the quadratic part ($H_q=H_{non-int}+H_{d}+H_{od}$) of the bosonized Hamiltonian  gives charge and spin Luttinger liquids
\begin{equation}
    H_{q}= \sum_{\kappa=c,s}\frac{v_{\kappa}}{2}\int dx \left[ K_{\kappa}(\partial_x\phi_{\kappa})^2+ \frac{1}{K_{\kappa}}(\partial_x\theta_{\kappa})^2\right]
    \label{BQ2FP}
\end{equation}
where $v_{\kappa}=\sqrt{(v_F^{\alpha}\pm2g_d)^2-(2g_{od})^2}$ and $K_{\kappa}=\sqrt{\frac{v_F^{\alpha}\pm2g_d\pm2g_{od}}{v_F^{\alpha}\pm2g_d\mp2g_{od}}}$.

The backward and Umklapp scattering terms can be written as
\begin{align}
H_{b}&=g_b \int dx \cos[2\sqrt{2\pi}\phi_{s}] \label{gb}\\
H_{u}&=g_u \int dx \cos[2\sqrt{2\pi}\phi_{c}]
\label{gu}
\end{align} 
where, $g_b=g_u=U(u_{k_F}v_{k_F})^2/\pi^2\eta^2$. The Umklapp term $H_{u}$ appears only at half-filling, where $e^{\pm4ik_FJx}$ becomes non-oscillatory. For any $U>0$ we have $K_c>1$, therefore, the scaling dimension of the Umklapp term $\Delta_{g_u}=2/K_c$ becomes relevant. Therefore, the bosonic field $\phi_c$ is pinned at the minima, and the charge sector opens up a gap for infinitesimal $U$. The spin sector is gapless since the coupling $g_b$ is marginally irrelevant due to $SU(2)$ symmetry ($K_s=1$), which leads to the scaling dimension of $H_b$, $\Delta_{g_b}=2/K_s=2$ for all the fillings.
For $\gamma_0<\gamma_0^{c1}$, the signs of $k_F$ in Eq.~\ref{RLM1} exchange between right and left movers. However, the bosonized Hamiltonians remain the same as in Eq.~\ref{BQ2FP}, Eq.~\ref{gb}, and Eq,\ref{gu}. 

The bosonization for Case 1: $g\neq0$, $\Gamma=0$ (no flux), can be obtained by considering $\Gamma=0$ in the above case (with flux). In this case, the dispersion is 
\begin{equation}
\epsilon_{\alpha}(k)=-2t \cos k - \gamma_{0}
\end{equation}
Introducing periodic modulation ($g\neq 0$) will induce a gap at $\pm\pi/aJ$, which results in the lowest miniband derived from $\epsilon_{\alpha}$. The half-filling of this band in the reduced BZ always has two Fermi points at $k_F=\pm\pi/2aJ$. 
The zero flux limit i.e., $\Gamma=0$, implies that $u_{k_F}=u_{k_F}=1/\sqrt{2}$. Threfore, it modifies the couplings  and Luttinger parameters as $g_{d}=g_{od}=U/4\pi$, $g_{b}=g_{u}=U/(2\pi\eta)^2$ and $v_{\kappa}=\sqrt{\left(v_{F}^{\alpha}\pm\frac{U}{2\pi}\right)^2-\left(\frac{U}{2\pi}\right)^2}$, $K_{\kappa}=\sqrt{1\pm\frac{U}{v_F\pi}}$. However, the bosonized Hamiltonians are obtained to be the same as in Eq.~\ref{BQ2FP}, Eq.~\ref{gb}, and Eq,\ref{gu}. Therefore, the results qualitatively remain the same as discussed previously.

\subsection{Four Fermi points}\label{boso4fp}

The half-filling of $\epsilon_{\alpha}$ in the reduced BZ has four Fermi points when $\gamma_0^{c1}<\gamma_0<\gamma_0^{c2}$, appearing at $k_1,k_2,k_3$ and $k_4$. We have
\begin{align}
k_1-k_2&=\frac{\pi n}{aJ}\\
\sin \left( \frac{k_1+k_2}{2}\right)&=\sqrt{\frac{\left(\frac{\gamma_0}{2t}\right)^2 \cos^2\left(\frac{\phi}{2}\right)}{\sin^2\frac{n\pi}{2aJ}-\sin^2\left(\frac{\phi}{2}\right)}+\sin^2\left(\frac{\phi}{2}\right)}  
\end{align}
where $n$ is the electronic density. The Fermi points can be obtained as
\begin{align}
k_1&=\frac{n\pi}{2aJ}+\sin^{-1}\left(\sqrt{\sin^2\left(\frac{\phi}{2}\right)-\frac{\left( \frac{\gamma_0}{2t}\right) ^2\cos^2(\frac{\phi}{2})}{\sin^2(\frac{\phi}{2})-\sin^2(\frac{n\pi}{2aJ})}} \right) \\
k_2&=-\frac{n\pi}{2aJ}+\sin^{-1}\left(\sqrt{\sin^2\left(\frac{\phi}{2}\right)-\frac{\left( \frac{\gamma_0}{2t}\right) ^2\cos^2(\frac{\phi}{2})}{\sin^2(\frac{\phi}{2})-\sin^2(\frac{n\pi}{2aJ})}} \right)
\end{align}
Note that the symmetric spectrum, in Fig.~\ref{fig4}(d)-(II), along $k=0$ implies $k_1=-k_4$ and $k_2=-k_3$. Now we linearize the spectrum in the vicinity of these four Fermi points and take the low-energy continuum limit where the slowly varying right and left moving fermionic fields are identified with positive and negative Fermi velocities, respectively. We write the annihilation operator in terms of the left and right movers for the lower band as
\begin{align}
    c_{\lambda,\sigma}&= \zeta_{\lambda,k_1} e^{ik_1 x} \alpha_{1\sigma}(x)
    +\zeta_{\lambda,k_2} e^{ik_2 x} \alpha_{2\sigma}(x)\nonumber\\
    &+\zeta_{\lambda,k_3} e^{ik_3 x} \alpha_{3\sigma}(x)
    +\zeta_{\lambda,k_4} e^{ik_4 x} \alpha_{4\sigma}(x)
\end{align}
where $\zeta_{1,k_{i}}=u_{k_{i}}$, $\zeta_{2,k_{i}}=v_{k_{i}}$, $\alpha_{i\sigma}$ are the slowly varying fields in the vicinity of the Fermi points at $k_{i}$ respectively, with $i=1,2,3,4$. Therefore, the non-interacting part of the Hamiltonian becomes
\begin{align}
H_{non-int}=-i \sum_{\sigma}\int &dx (v_{F_1}^{\alpha}\alpha_{1\sigma}^{\dagger}\partial_x \alpha_{1\sigma} - v_{F_2}^{\alpha}\alpha_{2\sigma}^{\dagger}\partial_x \alpha_{2\sigma} \nonumber\\
&+ v_{F_3}^{\alpha}\alpha_{3\sigma}^{\dagger}\partial_x \alpha_{3\sigma} - v_{F_4}^{\alpha}\alpha_{4\sigma}^{\dagger}\partial_x \alpha_{4\sigma} ) 
\end{align}
where $v_{F_i}^{\alpha}$ are the Fermi velocities at $k_i$ respectively. Now we bosonize the fermionic fields into the bosonic fields $\phi_{i\sigma}(x)$, according to 
\begin{align}
\alpha_{1\sigma}&=F_{1\sigma} \frac{1}{\sqrt{2\pi \eta}}e^{-2i\sqrt{\pi}\phi_{1\sigma}}, \hspace{0.2cm} \alpha_{2\sigma}=F_{2\sigma} \frac{1}{\sqrt{2\pi \eta}}e^{2i\sqrt{\pi}\phi_{2\sigma}} \nonumber \\
\alpha_{3\sigma}&=F_{3\sigma} \frac{1}{\sqrt{2\pi \eta}}e^{-2i\sqrt{\pi}\phi_{3\sigma}}, \hspace{0.2cm} \alpha_{4\sigma}=F_{4\sigma} \frac{1}{\sqrt{2\pi \eta}}e^{2i\sqrt{\pi}\phi_{4\sigma}}
\end{align}
where $\eta$ is the short distance cut-off and $F_{i\sigma}$ are the Klein factors which ensure the anticommutation
relations between the fermionic species. The bosonized form of the non-interacting part is
\begin{align}
H_{non-int}= \sum_{\sigma}\int &dx [v_{F_1}^{\alpha}(\partial_x \phi_{1\sigma})^2 + v_{F_2}^{\alpha}(\partial_x \phi_{2\sigma})^2 \nonumber\\
&+ v_{F_3}^{\alpha}(\partial_x \phi_{3\sigma})^2 + v_{F_4}^{\alpha}(\partial_x \phi_{4\sigma})^2 ]
\end{align}

Now we bosonize the interacting Hamiltonian i.e., $:H_\mathrm{int}: = U \sum_{j,\lambda} n_{j\uparrow\lambda} n_{j\downarrow\lambda}$. As there are four Fermi points, the interaction term yields many possible scattering terms. We will categorize relevant scatterings and ignore all the oscillating terms as they will average to zero on integration. At first, we consider the momentum conserving terms %(i.e. $\delta k=0$) 
which are further categorized into four different components. Two among them are diagonal ($H_{d}$) and off-diagonal ($H_{od}$) terms,
\begin{align}
H_{d}&=U\int dx[(u_{k_1}^4+v_{k_1}^4) \alpha_{1\uparrow}^{\dagger}\alpha_{1\uparrow}\alpha_{1\downarrow}^{\dagger}\alpha_{1\downarrow}\nonumber\\
&+(u_{k_2}^4+v_{k_2}^4) \alpha_{2\uparrow}^{\dagger}\alpha_{2\uparrow}\alpha_{2\downarrow}^{\dagger}\alpha_{2\downarrow}\nonumber\\
&+(u_{k_3}^4+v_{k_3}^4) \alpha_{3\uparrow}^{\dagger}\alpha_{3\uparrow}\alpha_{3\downarrow}^{\dagger}\alpha_{3\downarrow}\nonumber\\
&+(u_{k_4}^4+v_{k_4}^4) \alpha_{4\uparrow}^{\dagger}\alpha_{4\uparrow}\alpha_{4\downarrow}^{\dagger}\alpha_{4\downarrow}] 
 \end{align}
\begin{align}
H_{od}&=U\int dx[(u_{k_1}^2 u_{k_2}^2+v_{k_1}^2 v_{k_2}^2)  \nonumber\\&(\alpha_{1\uparrow}^{\dagger}\alpha_{1\uparrow}\alpha_{2\downarrow}^{\dagger}\alpha_{2\downarrow}+\alpha_{2\uparrow}^{\dagger}\alpha_{2\uparrow}\alpha_{1\downarrow}^{\dagger}\alpha_{1\downarrow})\nonumber\\
&+(u_{k_1}^2 u_{k_3}^2+v_{k_1}^2 v_{k_3}^2)  (\alpha_{1\uparrow}^{\dagger}\alpha_{1\uparrow}\alpha_{3\downarrow}^{\dagger}\alpha_{3\downarrow}+\alpha_{3\uparrow}^{\dagger}\alpha_{3\uparrow}\alpha_{1\downarrow}^{\dagger}\alpha_{1\downarrow})\nonumber\\
&+(u_{k_1}^2 u_{k_4}^2+v_{k_1}^2 v_{k_4}^2)  (\alpha_{1\uparrow}^{\dagger}\alpha_{1\uparrow}\alpha_{4\downarrow}^{\dagger}\alpha_{4\downarrow}+\alpha_{4\uparrow}^{\dagger}\alpha_{4\uparrow}\alpha_{1\downarrow}^{\dagger}\alpha_{1\downarrow})\nonumber\\
&+(u_{k_2}^2 u_{k_3}^2+v_{k_2}^2 v_{k_3}^2)  (\alpha_{2\uparrow}^{\dagger}\alpha_{2\uparrow}\alpha_{3\downarrow}^{\dagger}\alpha_{3\downarrow}+\alpha_{3\uparrow}^{\dagger}\alpha_{3\uparrow}\alpha_{2\downarrow}^{\dagger}\alpha_{2\downarrow})\nonumber\\
&+(u_{k_2}^2 u_{k_4}^2+v_{k_2}^2 v_{k_4}^2)  (\alpha_{2\uparrow}^{\dagger}\alpha_{2\uparrow}\alpha_{4\downarrow}^{\dagger}\alpha_{4\downarrow}+\alpha_{4\uparrow}^{\dagger}\alpha_{4\uparrow}\alpha_{2\downarrow}^{\dagger}\alpha_{2\downarrow})\nonumber\\
&+(u_{k_3}^2 u_{k_4}^2+v_{k_3}^2 v_{k_4}^2)  (\alpha_{3\uparrow}^{\dagger}\alpha_{3\uparrow}\alpha_{4\downarrow}^{\dagger}\alpha_{4\downarrow}+\alpha_{4\uparrow}^{\dagger}\alpha_{4\uparrow}\alpha_{3\downarrow}^{\dagger}\alpha_{3\downarrow})]
\end{align}
The bosonized forms are 
\begin{align}
H_{d}&=\int dx[g_{11}\partial_x\phi_{1\uparrow}\partial_x\phi_{1\downarrow} 
+g_{22} \partial_x\phi_{2\uparrow}\partial_x\phi_{2\downarrow} \nonumber\\
&+g_{33} \partial_x\phi_{3\uparrow}\partial_x\phi_{3\downarrow} 
+g_{44} \partial_x\phi_{4\uparrow}\partial_x\phi_{4\downarrow}]
\end{align}
where the couplings $g_{ii}=U(u_{k_i}^4+v_{k_i}^4)/\pi$, and
\begin{align}
H_{od}&=\int dx[g_{12} (\partial_x\phi_{1\uparrow}\partial_x\phi_{2\downarrow}+\partial_x\phi_{2\uparrow}\partial_x\phi_{1\downarrow})\nonumber\\
&+g_{13} (\partial_x\phi_{1\uparrow}\partial_x\phi_{3\downarrow}+\partial_x\phi_{3\uparrow}\partial_x\phi_{1\downarrow})\nonumber\\
&+g_{14} (\partial_x\phi_{1\uparrow}\partial_x\phi_{4\downarrow}+\partial_x\phi_{4\uparrow}\partial_x\phi_{1\downarrow})\nonumber\\
&+g_{23} (\partial_x\phi_{2\uparrow}\partial_x\phi_{3\downarrow}+\partial_x\phi_{3\uparrow}\partial_x\phi_{2\downarrow})\nonumber\\
&+g_{24} (\partial_x\phi_{2\uparrow}\partial_x\phi_{4\downarrow}+\partial_x\phi_{4\uparrow}\partial_x\phi_{2\downarrow})\nonumber\\
&+g_{34} (\partial_x\phi_{3\uparrow}\partial_x\phi_{4\downarrow}+\partial_x\phi_{4\uparrow}\partial_x\phi_{3\downarrow})]
\end{align}
where the couplings $g_{ii^{\prime}}=U(u_{k_i}^2 u_{k_{i^{\prime}}}^2+v_{k_i}^2 v_{k_{i^{\prime}}}^2)/\pi$,
with $i,i^{\prime}=1,2,3,4$.
We introduce the charge ($c$) and spin ($s$) modes through the fields, $\phi_{ic}=(\phi_{i\uparrow}+\phi_{i\downarrow})/\sqrt{2}$ and $\phi_{is}=(\phi_{i\uparrow}-\phi_{i\downarrow})/\sqrt{2}$ %where $i=1,2,3,4$ 
and also introduce linear combinations of these fields, $\phi_{\kappa}^{1\pm}=\phi_{4\kappa}\pm \phi_{1\kappa}$ and $\phi_{\kappa}^{2\pm}=\phi_{2\kappa}\pm \phi_{3\kappa}$ where $\kappa=c,s$. The symmetric spectrum along $k=0$ (i.e., $k_1=-k_4$ and $k_2=-k_3$) implies $|v_{F_1}^{\alpha}|=|v_{F_4}^{\alpha}|$, $|v_{F_2}^{\alpha}|=|v_{F_3}^{\alpha}|$, $u_{k_{1(4)}}^2=v_{k_{4(1)}}^2$, $u_{k_{2(3)}}^2=v_{k_{3(2)}}^2$, $g_{12}=g_{34}$, $g_{13}=g_{24}$, $g_{11}=g_{44}$ and  $g_{22}=g_{33}$.

Finally, the quadratic part of the bosonic Hamiltonian can be diagonalized and the Luttinger parameters can be obtained using the Bogoliubov (BdG) transformation of the bosonic fields such that $\phi_{\kappa}^{1(2)+}=[(1+K_{\kappa}^+)\tilde{\phi}_{\kappa}^{1(2)+}-(1-K_{\kappa}^+)\tilde{\phi}_{\kappa}^{2(1)+}]/2\sqrt{K_{\kappa}^+}$ and $\phi_{\kappa}^{1(2)-}=[(1+K_{\kappa}^-)\tilde{\phi}_{\kappa}^{1(2)-}-(1-K_{\kappa}^-)\tilde{\phi}_{\kappa}^{2(1)-}]/2\sqrt{K_{\kappa}^-}$ where $K_{\kappa}^{\pm}$ are the Luttinger parameters. Therefore, the quadratic part of the Hamiltonian ($H_{q}=H_{non-int}+H_{d}+H_{od}$) can be written as
\begin{align}
H_{q}&= \sum_{\kappa=c,s}v_{F_{\kappa}}^{+} \int dx \left[ K_{\kappa}^{+} (\partial_x\phi_{\kappa+})^2 + \frac{1}{K_{\kappa}^{+}} (\partial_x\theta_{\kappa+})^2\right] \nonumber\\
&+\sum_{\kappa=c,s}v_{F_{\kappa}}^{-} \int dx\left[ K_{\kappa}^{-} (\partial_x\phi_{\kappa-})^2 + \frac{1}{K_{\kappa}^{-}} (\partial_x\theta_{\kappa-})^2\right] 
\end{align}
where, $\phi_{\kappa\pm}=\tilde{\phi}_{\kappa}^{1\pm}+\tilde{\phi}_{\kappa}^{2\pm}$ and $\theta_{\kappa\pm}=\tilde{\phi}_{\kappa}^{1\pm}-\tilde{\phi}_{\kappa}^{2\pm}$. The Luttinger parameters are obtained as
\begin{equation}
K_{\kappa}^{\pm}= \sqrt{\frac{v_{1\kappa}^{\pm}+v_{2\kappa}^{\pm}-2\Lambda^{\pm}_{\kappa}}{v_{1\kappa}^{\pm}+v_{2\kappa}^{\pm}+2\Lambda^{\pm}_{\kappa}}}
\label{LLP}
\end{equation}
\begin{equation}
v_{F_{\kappa}}^{\pm}= \frac{v_{1\kappa}^{\pm}+v_{2\kappa}^{\pm}}{2} \sqrt{1-\left(\frac{2\Lambda^{\pm}_{\kappa}}{v_{1\kappa}^{\pm}+v_{2\kappa}^{\pm}} \right)^2 }
\end{equation}
where, $v_{1c}^{\pm}=(v_{F_{1}}^{\alpha}+v_{F_{4}}^{\alpha}+g_{11}+g_{44}\pm 2g_{14})/4$, $v_{2c}^{\pm}=(v_{F_{2}}^{\alpha}+v_{F_{3}}^{\alpha}+g_{22}+g_{33}\pm 2g_{23})/4$, $\Lambda^{\pm}_{c}=(\pm g_{12}+g_{13}+g_{24}\pm g_{34})/4$ for charge sector and $v_{1s}^{\pm}=(v_{F_{1}}^{\alpha}+v_{F_{4}}^{\alpha}-g_{11}-g_{44}\mp 2g_{14})/4$, $v_{2s}^{\pm}=(v_{F_{2}}^{\alpha}+v_{F_{3}}^{\alpha}-g_{22}-g_{33}\mp 2g_{23})/4$, $\Lambda^{\pm}_{s}=(\mp g_{12}-g_{13}-g_{24}\mp g_{34})/4$ for spin sector.

The other two momentum-conserving components are four Fermi points ($H_{ff}$) and backward ($H_{b}$) scattering terms 
\begin{align}
H_{ff}&=U\int dx(u_{k_1} u_{k_2} u_{k_3} u_{k_4}+v_{k_1} v_{k_2} v_{k_3} v_{k_4}) \nonumber\\&[\alpha_{1\uparrow}^{\dagger}\alpha_{2\uparrow}\alpha_{4\downarrow}^{\dagger}\alpha_{3\downarrow}+\alpha_{1\uparrow}^{\dagger}\alpha_{3\uparrow}\alpha_{4\downarrow}^{\dagger}\alpha_{2\downarrow} 
+\alpha_{2\uparrow}^{\dagger}\alpha_{4\uparrow}\alpha_{3\downarrow}^{\dagger}\alpha_{1\downarrow}\nonumber\\
&+\alpha_{3\uparrow}^{\dagger}\alpha_{4\uparrow}\alpha_{2\downarrow}^{\dagger}\alpha_{1\downarrow}+H.c]
\end{align}
\begin{align}
H_{b}&=U\int dx[(u_{k_1}^2 u_{k_2}^2 +v_{k_1}^2 v_{k_2}^2)\alpha_{1\uparrow}^{\dagger}\alpha_{2\uparrow}\alpha_{2\downarrow}^{\dagger}\alpha_{1\downarrow}\nonumber\\
&+(u_{k_1}^2 u_{k_3}^2 +v_{k_1}^2 v_{k_3}^2)\alpha_{1\uparrow}^{\dagger}\alpha_{3\uparrow}\alpha_{3\downarrow}^{\dagger}\alpha_{1\downarrow}\nonumber\\
&+(u_{k_1}^2 u_{k_4}^2 +v_{k_1}^2 v_{k_4}^2)\alpha_{1\uparrow}^{\dagger}\alpha_{4\uparrow}\alpha_{4\downarrow}^{\dagger}\alpha_{1\downarrow}\nonumber\\
&+(u_{k_2}^2 u_{k_3}^2 +v_{k_2}^2 v_{k_3}^2)\alpha_{2\uparrow}^{\dagger}\alpha_{3\uparrow}\alpha_{3\downarrow}^{\dagger}\alpha_{2\downarrow}\nonumber\\
&+(u_{k_2}^2 u_{k_4}^2 +v_{k_2}^2 v_{k_4}^2)\alpha_{2\uparrow}^{\dagger}\alpha_{4\uparrow}\alpha_{4\downarrow}^{\dagger}\alpha_{2\downarrow}\nonumber\\
&+(u_{k_3}^2 u_{k_4}^2 +v_{k_3}^2 v_{k_4}^2)\alpha_{3\uparrow}^{\dagger}\alpha_{4\uparrow}\alpha_{4\downarrow}^{\dagger}\alpha_{3\downarrow} 
+ H.c]
\end{align}
The bosonized forms are 
\begin{align}
    H_{ff}&=g_{ff} \int dx[\cos[2\sqrt{\pi}(\phi_{1\uparrow}+\phi_{2\downarrow}-\phi_{3\uparrow}-\phi_{4\downarrow})]\nonumber\\
-&\cos[2\sqrt{\pi}(\phi_{1\uparrow}+\phi_{2\uparrow}-\phi_{3\downarrow}-\phi_{4\downarrow})]\nonumber\\
+&\cos[2\sqrt{\pi}(\phi_{1\downarrow}+\phi_{2\uparrow}-\phi_{3\downarrow}-\phi_{4\uparrow})]\nonumber\\
-&\cos[2\sqrt{\pi}(\phi_{1\downarrow}+\phi_{2\downarrow}-\phi_{3\uparrow}-\phi_{4\uparrow})]]
\end{align}
where $g_{ff}=U(u_{k_1} u_{k_2} u_{k_3} u_{k_4}+v_{k_1} v_{k_2} v_{k_3} v_{k_4})/2\pi^2\eta^2$, and
\begin{align}
H_{b}&= \int dx[g_{b12} \cos[2\sqrt{\pi}(\phi_{1\uparrow}-\phi_{1\downarrow}+\phi_{2\uparrow}-\phi_{2\downarrow})]\nonumber\\
&+g_{b13} \cos[2\sqrt{\pi}(\phi_{1\uparrow}-\phi_{1\downarrow}+\phi_{3\downarrow}-\phi_{3\uparrow})]\nonumber\\
&+g_{b14} \cos[2\sqrt{\pi}(\phi_{1\uparrow}-\phi_{1\downarrow}+\phi_{4\uparrow}-\phi_{4\downarrow})]\nonumber\\
&+g_{b23} \cos[2\sqrt{\pi}(\phi_{2\uparrow}-\phi_{2\downarrow}+\phi_{3\uparrow}-\phi_{3\downarrow})]\nonumber\\
&+g_{b24} \cos[2\sqrt{\pi}(\phi_{2\uparrow}-\phi_{2\downarrow}+\phi_{4\uparrow}-\phi_{4\downarrow})]\nonumber\\
&+g_{b34} \cos[2\sqrt{\pi}(\phi_{3\uparrow}-\phi_{3\downarrow}+\phi_{4\uparrow}-\phi_{4\downarrow})]]
\end{align}
where $g_{bii^{\prime}}=-U(u_{k_i}^2 u_{k_{i^{\prime}}}^2 +v_{k_1}^2 v_{k_{i^{\prime}}}^2)/2\pi^2\eta^2$.
At half-filling, apart from the above scattering terms, new Umklapp scattering terms where the change in the momentum is $2\pi/aJ$ can be obtained. However, among such scatterings, only higher-order terms are allowed since $4|k_{1(3)}-k_{2(4)}|=2\pi/aJ$. Such Umklapp terms appear due to scatterings that involve two and four Fermi points. Therefore, the relevant Umklapp ($H_{u}$) terms are,
\begin{align}
H_{u}&=U^2 \int dxe^{-i4(k_1-k_2)Jx} [(u_{k_1}^2 u_{k_2}^2 +v_{k_1}^2 v_{k_2}^2)^2 \nonumber\\
&(\alpha_{1\uparrow}^{\dagger}\alpha_{2\uparrow}\alpha_{1\downarrow}^{\dagger}\alpha_{2\downarrow})^2+(u_{k_3}^2 u_{k_4}^2 +v_{k_3}^2 v_{k_4}^2)^2 (\alpha_{3\uparrow}^{\dagger}\alpha_{4\uparrow}\alpha_{3\downarrow}^{\dagger}\alpha_{4\downarrow})^2\nonumber\\
&+(u_{k_1} u_{k_2} u_{k_3} u_{k_4}+v_{k_1} v_{k_2} v_{k_3} v_{k_4})^2 
(\alpha_{1\uparrow}^{\dagger}\alpha_{2\uparrow}\alpha_{3\downarrow}^{\dagger}\alpha_{4\downarrow})^2\nonumber\\
&+(u_{k_1} u_{k_2} u_{k_3} u_{k_4}+v_{k_1} v_{k_2} v_{k_3} v_{k_4})^2 
(\alpha_{1\uparrow}^{\dagger}\alpha_{4\uparrow}\alpha_{3\downarrow}^{\dagger}\alpha_{2\downarrow})^2\nonumber\\
&+(u_{k_1} u_{k_2} u_{k_3} u_{k_4}+v_{k_1} v_{k_2} v_{k_3} v_{k_4})^2 
(\alpha_{3\uparrow}^{\dagger}\alpha_{2\uparrow}\alpha_{1\downarrow}^{\dagger}\alpha_{4\downarrow})^2\nonumber\\
&+(u_{k_1} u_{k_2} u_{k_3} u_{k_4}+v_{k_1} v_{k_2} v_{k_3} v_{k_4})^2 
(\alpha_{3\uparrow}^{\dagger}\alpha_{4\uparrow}\alpha_{1\downarrow}^{\dagger}\alpha_{3\downarrow})^2] \nonumber\\
&+ H.c
\end{align}
The bosonic form is
\begin{align}
H_{u}&= \int dx[g_{u12} \cos[4\sqrt{\pi}(\phi_{1\uparrow}+\phi_{1\downarrow}+\phi_{2\uparrow}+\phi_{2\downarrow})]\nonumber\\
&+g_{u34} \cos[4\sqrt{\pi}(\phi_{3\uparrow}+\phi_{3\downarrow}+\phi_{4\uparrow}+\phi_{4\downarrow})]\nonumber\\
&+g_{uff} [\cos[4\sqrt{\pi}(\phi_{1\uparrow}+\phi_{2\uparrow}+\phi_{3\downarrow}+\phi_{4\downarrow})]\nonumber\\
&+\cos[4\sqrt{\pi}(\phi_{1\uparrow}+\phi_{2\downarrow}+\phi_{3\downarrow}+\phi_{4\uparrow})]\nonumber\\
&+\cos[4\sqrt{\pi}(\phi_{1\downarrow}+\phi_{2\uparrow}+\phi_{3\uparrow}+\phi_{4\downarrow})]\nonumber\\
&+\cos[4\sqrt{\pi}(\phi_{1\downarrow}+\phi_{2\downarrow}+\phi_{3\uparrow}+\phi_{4\uparrow})]]]
\label{um}
\end{align}
where $g_{u12}=2U^2(u_{k_1}^2 u_{k_2}^2 +v_{k_1}^2 v_{k_2}^2)^2/(2\pi\eta)^4$, $g_{u34}=2U^2(u_{k_3}^2 u_{k_4}^2 +v_{k_3}^2 v_{k_4}^2)^2/(2\pi\eta)^4$ and $g_{uff}=2U^2(u_{k_1}^2 u_{k_2}^2 u_{k_3}^2 u_{k_4}^2+v_{k_1}^2 v_{k_2}^2 v_{k_3}^2 v_{k_4}^2)^2/(2\pi\eta)^4$.
Therefore, the bosonic version of the total Hamiltonian ($H_{t}=H_{q}+H_{ff}+H_{b}+H_{u}$) reads,
\begin{align}
H_{t}&=H_{q}+\int dx[2g_{ff} \cos[2\sqrt{\pi}\theta_{c-}]\nonumber\\
&(\cos[2\sqrt{\pi}\theta_{s+}]-\cos[2\sqrt{\pi}\phi_{s+}])\nonumber\\
&+ 2g_{b12} \cos[2\sqrt{\pi}\phi_{s+}]\cos[2\sqrt{\pi}\theta_{s-}] \nonumber\\
&+ 2g_{b13} \cos[2\sqrt{\pi}\theta_{s+}]\cos[2\sqrt{\pi}\theta_{s-}] \nonumber\\
&+ g_{b14} \cos[2\sqrt{\pi}(\theta_{s+}+\phi_{s+})]\nonumber\\
&+ g_{b23} \cos[2\sqrt{\pi}(\theta_{s+}-\phi_{s+})] \nonumber\\
&+ 2g_{u12} \cos[2\sqrt{2\pi}\phi_{c+}]\cos[2\sqrt{2\pi}\theta_{c-}] \nonumber\\
&+ 2g_{uff} \cos[2\sqrt{2\pi}\phi_{c+}]\nonumber\\
&(\cos[2\sqrt{2\pi}\theta_{s-}]-\cos[2\sqrt{2\pi}\theta_{s+}])]
\label{boso-total}
\end{align}
Note that the terms $g_{b12}=g_{b34}$, $g_{b13}=g_{b24}$ and $g_{u12}=g_{u34}$. As it can be seen from the Eq. \ref{boso-total}, the coupling of various bosonic fields in the cosine terms makes it complicated to identify the bosonic fields that are pinned at the minima of the potentials. However, to deduce the interaction effects, the perturbative RG transformation by iterative coarse-graining and rescaling of the fields can be employed \cite{giamarchi2004quantum}. This gives a set of RG equations for the various coupling constants, which generally have the form $dg_i/dl=(2-\Delta_{g_i})g_i-Cg_jg_k$, using operator product expansion up to the second order \cite{cardy1996scaling}, where $\Delta_{g_i}$ is the scaling dimension of the interacting terms and $C$ depends on Luttinger liquids. Up to the first order in RG, when the scaling dimension $\Delta_{g_i}<2$, the corresponding coupling $g_i$ becomes relevant and grows to the strong coupling limit. However, when $\Delta_{g_i}=2$ and $\Delta_{g_i}>2$, the couplings remain marginal and irrelevant respectively. In the case of marginal couplings, the second-order contribution in RG equations dictates the flow to be relevant or irrelevant.

Therefore, we calculate the scaling dimensions of the cosine operators in Eq.~\ref{boso-total} except Umklapp terms and obtain the RG equations up to the second order as follows,
{\small\begin{align}
\frac{dg_{ff}^{(1)}}{dl}&=\left[2-\left( \frac{1}{K_{c}^{-}}+\frac{1}{K_{s}^{+}}\right)\right] g_{ff}^{(1)} - \left(\frac{K_{s}^{+}}{2}\right)g_{ff}^{(2)}(g_{b14}+g_{b23})\label{AC1}\\
\frac{dg_{ff}^{(2)}}{dl}&=\left[2-\left( \frac{1}{K_{c}^{-}}+K_{s}^{+}\right) \right]g_{ff}^{(2)}+\left(\frac{1}{2K_{s}^{+}}\right)g_{ff}^{(1)}(g_{b14}+g_{b23})\\
\frac{dg_{b12}}{dl}&=\left[2-\left( K_{s}^{+}+\frac{1}{K_{s}^{-}}\right)\right] g_{b12} - \left(\frac{2}{K_{s}^{+}}\right)g_{b13}(g_{b14}+g_{b23})\\
\frac{dg_{b13}}{dl}&=\left[2-\left( \frac{1}{K_{s}^{+}}+\frac{1}{K_{s}^{-}}\right) \right]g_{b13} - \left(2K_{s}^{+}\right)g_{b12}(g_{b14}+g_{b23})\\
\frac{dg_{b14}}{dl}&=\left[2-\left( K_{s}^{+}+\frac{1}{K_{s}^{+}}\right)\right] g_{b14}-\frac{1}{4}\left( \frac{g_{b12}g_{b13}}{K_{s}^{-}}-\frac{g_{ff}^{(1)}g_{ff}^{(2)}}{2K_{c}^{-}}\right)\\
\frac{dg_{b23}}{dl}&=\left[2-\left( K_{s}^{+}+\frac{1}{K_{s}^{+}}\right)\right] g_{b23}-\frac{1}{4}\left( \frac{g_{b12}g_{b13}}{K_{s}^{-}}-\frac{g_{ff}^{(1)}g_{ff}^{(2)}}{2K_{c}^{-}}\right).\\
\frac{dK_{c}^{-}}{dl}&=-\frac{(g_{ff}^{(1)})^2}{2}\left(1+\frac{K_{c}^{-}}{K_{s}^{+}}\right)-\frac{(g_{ff}^{(2)})^2}{2}\left(1+K_{c}^{-}K_{s}^{+}\right)
\label{AC6}
\end{align}}
We numerically integrate these RG equations for initial values of various coupling constants to identify their contribution to opening a gap in the charge and spin sectors.
The $SU(2)$ symmetry of our model constrain the spin Luttinger liquids $K_{s}^{\pm}=1$. Therefore, the scaling dimensions of the couplings $\Delta_{g_{b12}}=\Delta_{g_{b13}}=\Delta_{g_{b14}}=\Delta_{g_{b23}}=2$ for $U>0$. However, the RG flow of second-order terms turns out to be irrelevant. The couplings $g_{ff}^{(1)}$ and $g_{ff}^{(2)}$ are weakly relevant in the first order RG as their scaling dimensions $\Delta_{g_{ff}^{(1)}}=\Delta_{g_{ff}^{(2)}}<2$ for $U>0$. They pin the dual bosonic fields $\phi_{s+}$ and $\theta_{s+}$ simultaneously. Therefore, the second-order terms in the RG equations dictate the dominant bosonic field among them. We identify the RG equation for $g_{ff}^{(2)}$ flowing to the strong coupling limit faster than $g_{ff}^{(1)}$. Therefore, these terms pin the fields $\theta_{c-}$ and $\phi_{s+}$ and open up gaps in one charge and one spin sectors.

Moreover, the RG equations (up to first order) for the Umklapp terms are,
\begin{align}
\frac{dg_{u12}}{dl}&=\left[2-2\left( K_{c}^{+}+\frac{1}{K_{c}^{-}}\right)\right] g_{u12} \label{gu12}\\
\frac{dg_{uff}^{(1)}}{dl}&=\left[2-2\left( K_{c}^{+}+\frac{1}{K_{s}^{-}}\right) \right]g_{uff}^{(1)}\\
\frac{dg_{uff}^{(2)}}{dl}&=\left[2-2\left( K_{c}^{+}+\frac{1}{K_{s}^{+}}\right) \right]g_{uff}^{(2)}.
\end{align}
The couplings $g_{uff}^{(1)}$ and $g_{uff}^{(2)}$ contribute to both charge and spin sectors. However, the scaling dimensions of these couplings are always above two, therefore, they remain irrelevant as RG flow does not grow with the length scale. 
The charge contributing term $g_{u12}$ turns relevant only above $U=U_c$ and RG flows into the strong coupling regime. This leads to a gap opening at the charge sector that dictates the metal-insulator transition.

\subsection{Density operator}\label{denop}
We write the density operator up to $2k_F$ terms,
\begin{align}
\rho_{\lambda,\sigma}(x)&=\zeta_{\lambda,k}\sum_{\sigma}\alpha^{\dagger}_{\sigma}(x)\alpha_{\sigma}(x) \nonumber\\
&=\sum_{\sigma}\zeta^2_{\lambda,k_1}\alpha_{1\sigma}^{\dagger}\alpha_{1\sigma}+\zeta^2_{\lambda,k_2}\alpha_{2\sigma}^{\dagger}\alpha_{2\sigma}+\zeta^2_{\lambda,k_3}\alpha_{3\sigma}^{\dagger}\alpha_{3\sigma} \nonumber\\
&+\zeta^2_{\lambda,k_4}\alpha_{4\sigma}^{\dagger}\alpha_{4\sigma}
+ (\zeta_{\lambda,k_1}\zeta_{\lambda,k_2} e^{-i(k_1-k_2)Jx}\alpha_{1\sigma}^{\dagger}\alpha_{2\sigma} \nonumber\\
&+\zeta_{\lambda,k_1}\zeta_{\lambda,k_3} e^{-i(k_1+k_2)Jx}\alpha_{1\sigma}^{\dagger}\alpha_{3\sigma}\nonumber\\
&+\zeta_{\lambda,k_1}\zeta_{\lambda,k_4} e^{-i2k_1Jx}\alpha_{1\sigma}^{\dagger}\alpha_{4\sigma}\nonumber\\
&+\zeta_{\lambda,k_2}\zeta_{\lambda,k_3} e^{-i2k_2Jx}\alpha_{2\sigma}^{\dagger}\alpha_{3\sigma}\nonumber\\
&+\zeta_{\lambda,k_2}\zeta_{\lambda,k_4} e^{-i(k_1+k_2)Jx}\alpha_{2\sigma}^{\dagger}\alpha_{4\sigma}\nonumber\\
&+\zeta_{\lambda,k_3}\zeta_{\lambda,k_4} e^{-i(k_1-k_2)Jx}\alpha_{3\sigma}^{\dagger}\alpha_{4\sigma}+H.c) 
\end{align}
where $\zeta_{1,k_i}=u_{k_i}$ and $\zeta_{2,k_i}=v_{k_i}$ ($u_{k_i}$ and $v_{k_i}$ are given in Eq.~\ref{ukvk}). In the four Fermi point regime, the $2k_F$ terms $\alpha_{1\sigma}^{\dagger}\alpha_{2\sigma}$ and $\alpha_{3\sigma}^{\dagger}\alpha_{4\sigma}$ dominates the power-law decay of the charge and spin correlations with an oscillatory coefficient $e^{\pm i(k_1-k_2)Jx}$ leading to the antiferromagnetic correlations with the exponent $\frac{1}{8}\left(K_{c}^{+}+\frac{1}{K_{s}^{-}}\right)$ below $U_c$ and $\left(\frac{1}{8K_{s}^{-}}\right)$ above $U_c$.

\section{Numerical results convergence}
\label{converge}
We verified the convergence of our numerical results by increasing the number of states. As shown in Fig.~\ref{fig7}, the ground state energy extrapolated to infinite bond dimension closely matches the energy obtained at the largest bond dimension, indicating that the energy is well converged. The figure shown is for three-quarters filling. Similar convergence is achieved for other fillings as well.

\begin{figure}[h]
\centering
\includegraphics[width=1\linewidth]{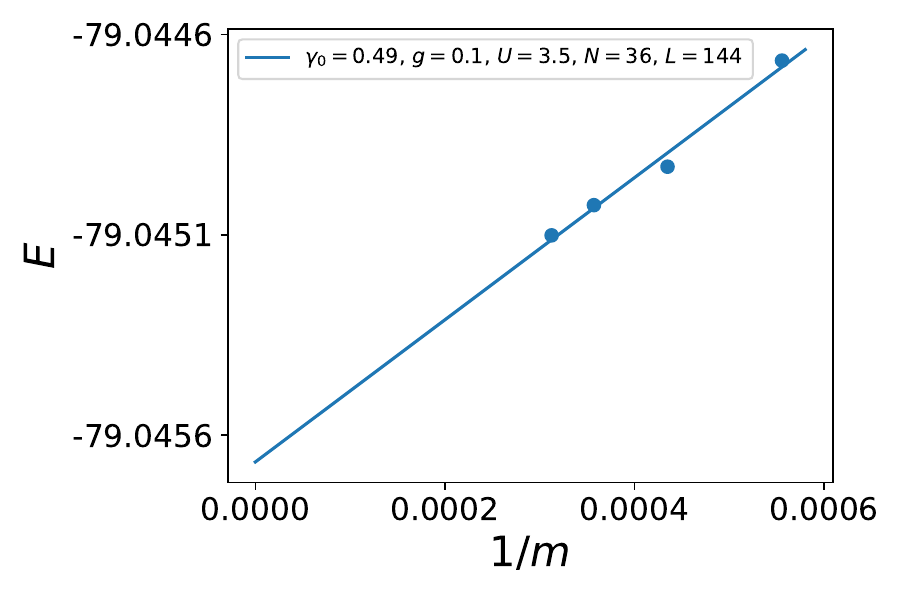}
\caption{\label{fig7}This figure shows the finite-size scaling of the ground state energy with respect to the bond dimension, where $E$ denotes the energy of the ground state and $m$ is the number of states retained. The maximum number of states used is 3200. Extrapolation to infinite bond dimension yields an energy difference of $10^{-4}$. This plot corresponds to the parameters: $\gamma_0 = 0.49$, $g = 0.1$, $U = 3.5$, $N = 36$, and $L = 144$ for three-quarter filling of the lowest band. Similar convergence is achieved for other fillings across different parameters as well.
} 
\end{figure}

\section{Charge density and charge gap}
\label{gap}
The charge density at various fillings considered in the main text is shown in Fig~\ref{fig8}. The charge gap, defined as  $\Delta_c=\lim_{L\to \infty}[E_{0}(N_e=N+2)+E_0(N_e=N-2)-2E_0(N_e=N)]$, where $E_0(N_e)$ is the ground state energy for $N_e$ electrons, is also calculated for the cases of one-quarter and three-quarters fillings (calculation of the charge gap at fillings which are slightly away from half filling is prohibitively expensive computationally and not shown). For onsite interaction, no gap is expected. Fig.~\ref{fig9}, showing the finite size scaling of the charge gap, corroborates this. We have confirmed that the results do not change even if $U$ is increased substantially.

\begin{figure}
\centering
\subfigure[]{\includegraphics[width=0.48\columnwidth]{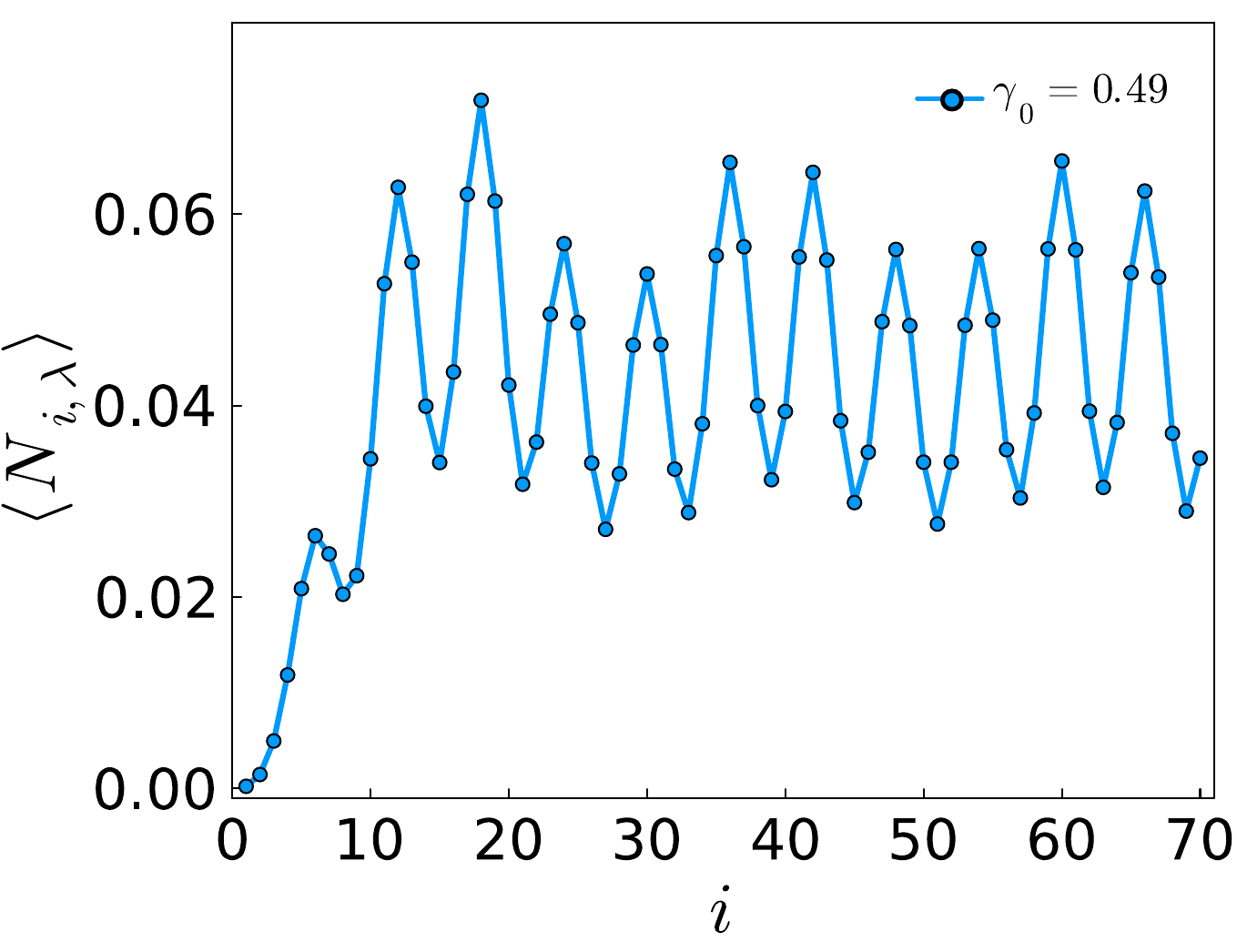}
\label{fig8a}}
\subfigure[]{\includegraphics[width=0.48\columnwidth]{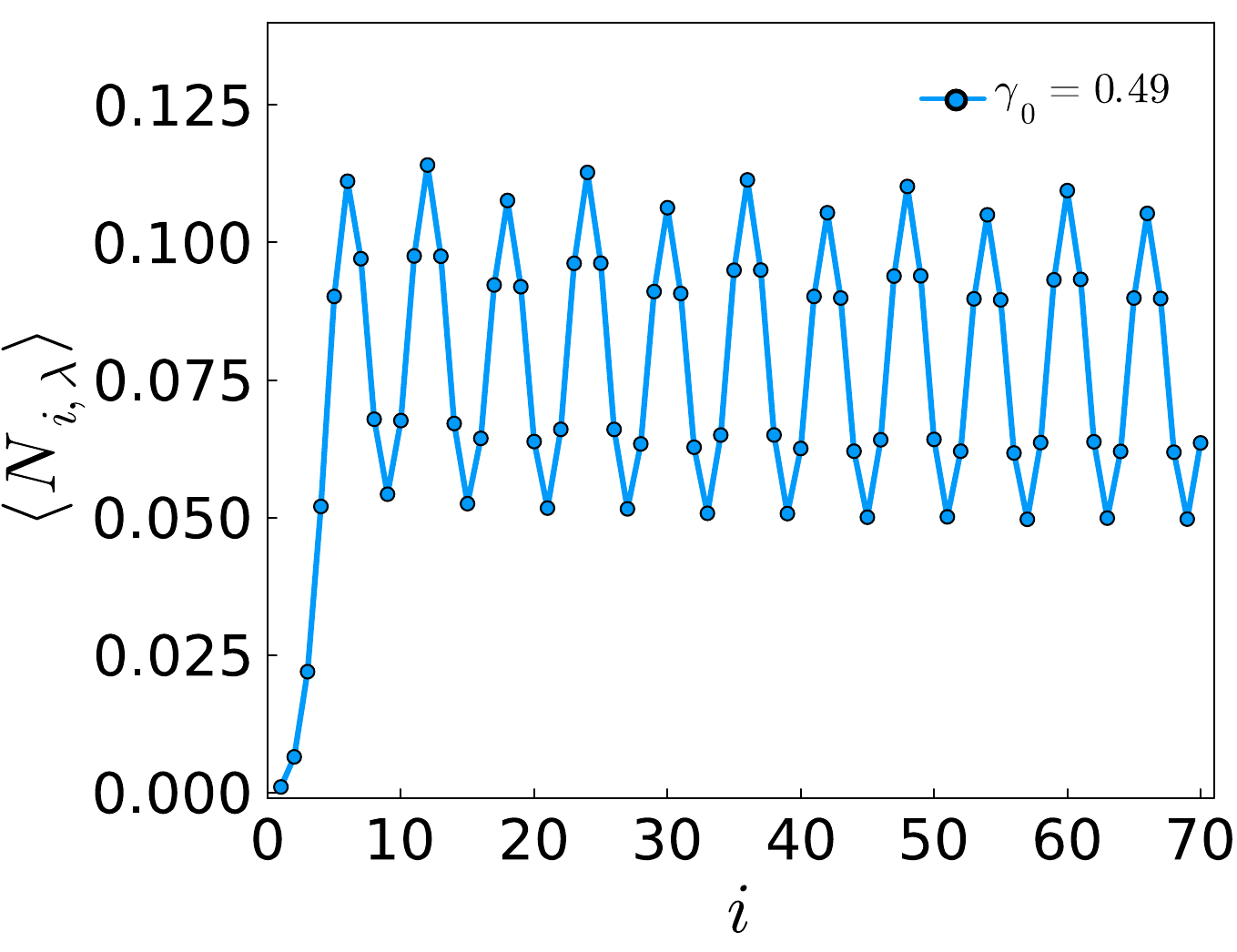}
\label{fig8b}}
\subfigure[]{\includegraphics[width=0.48\columnwidth]{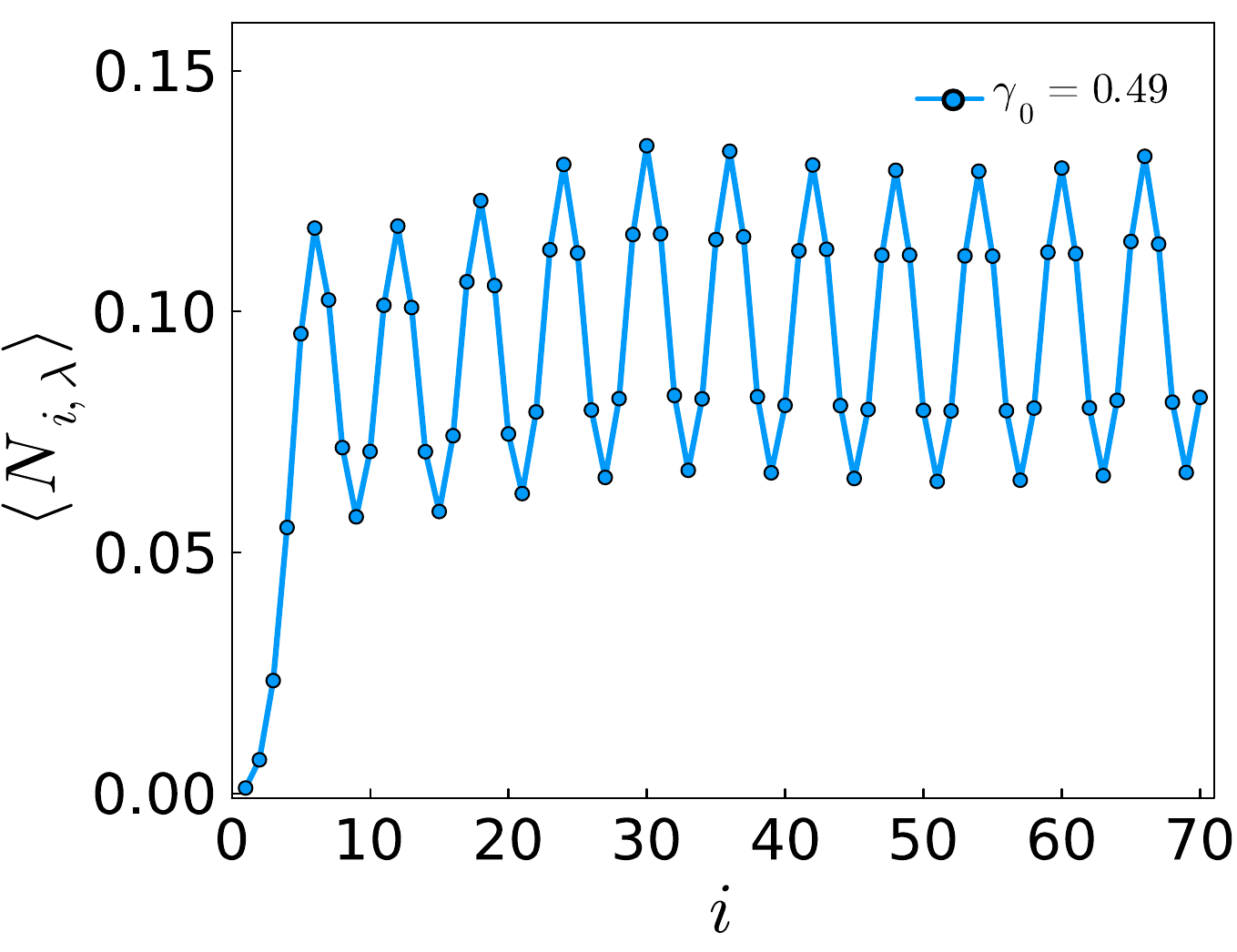}
\label{fig8c}}
\subfigure[]{\includegraphics[width=0.48\columnwidth]{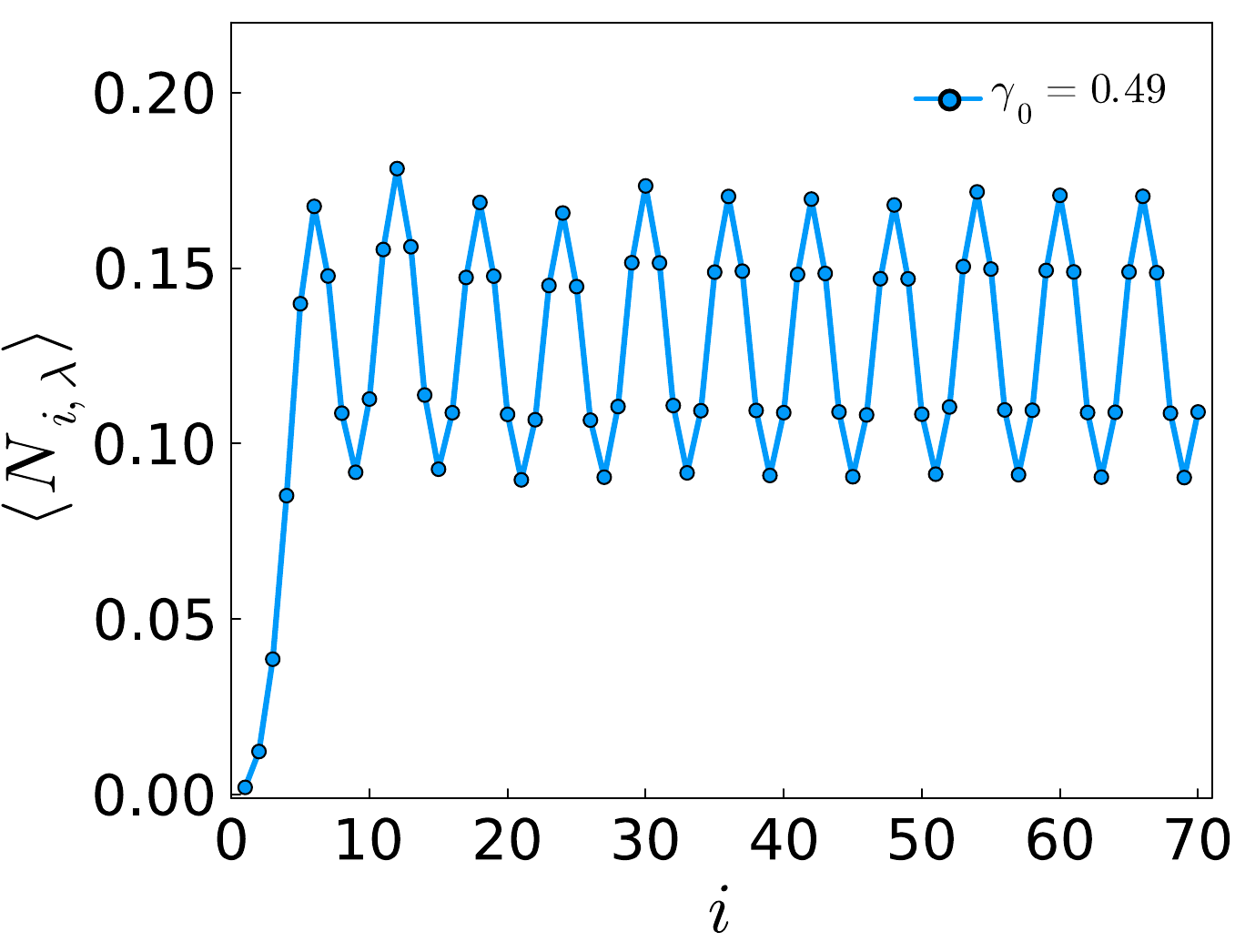}
\label{fig8d}}
\caption{The charge density at various fillings of the lowest band at $\gamma_{0}=0.49$, $g=0.1$. (a) One-quarter filling.(b) Half-filling minus two electrons. (c) Half filling plus two electrons. (d) Three-quarters filling.  The parameters used are the same as the ones in the corresponding figures.}\label{fig8} 
\end{figure}

\begin{figure}
\centering
\subfigure[]{\includegraphics[width=0.48\columnwidth]{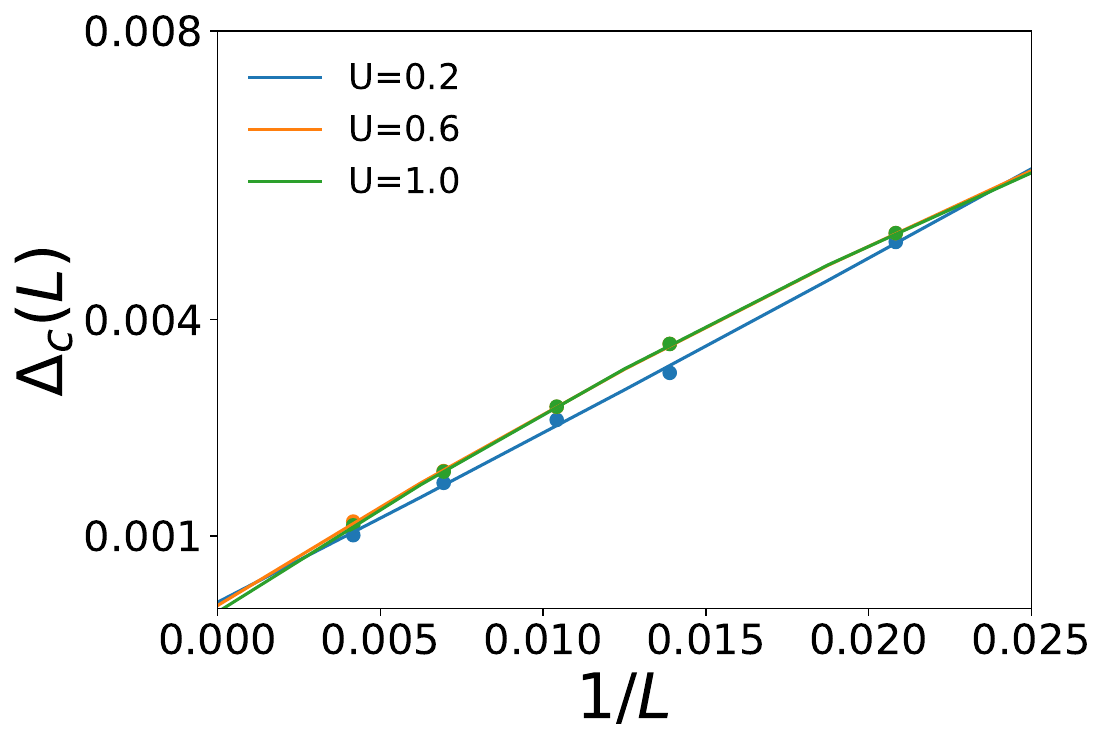}
\label{fig9a}}
\subfigure[]{\includegraphics[width=0.48\columnwidth]{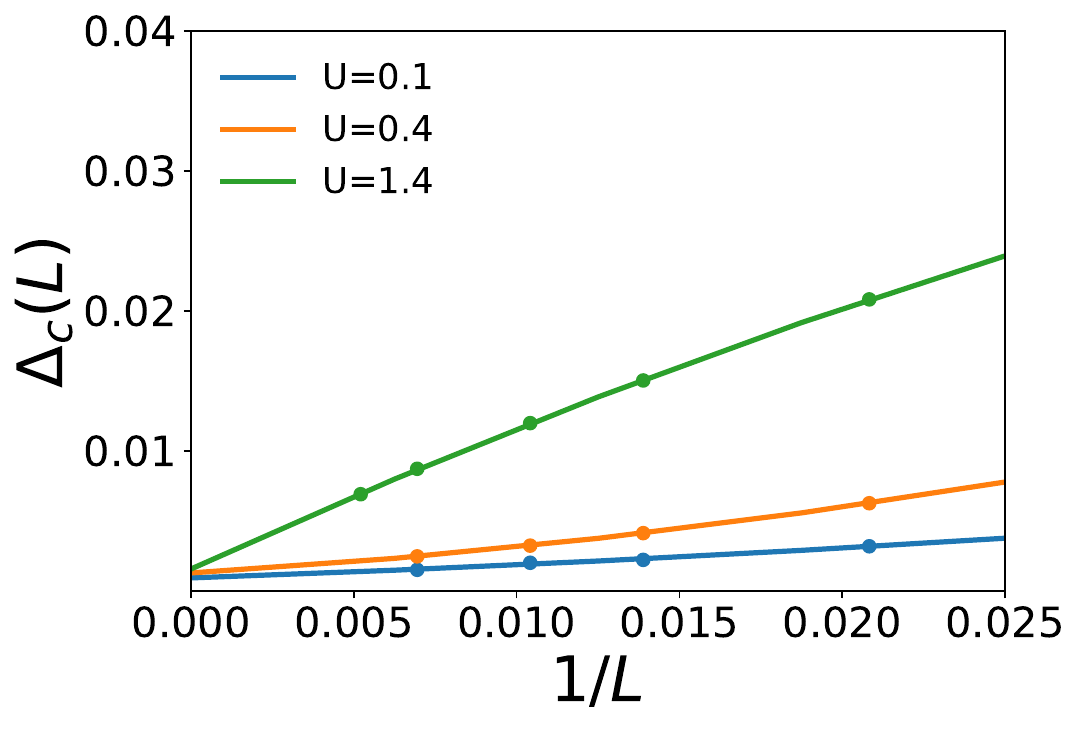}
\label{fig9b}}
\caption{The finite size scaling of charge gap for various onsite interaction strengths in the quasi-flat band regime for (a) 1/4 filling and (b) 3/4 filling. We
have used a least-square fit to the second order of polynomials in 1/L} \label{fig9}
\end{figure}

\section{\label{sfof} Spin gap and Spin-spin correlation for various fillings}

We demonstrate that the ferromagnetic spin-spin correlation at half-filling, for both the full-flux and periodic cases, is independent of the choice of reference site, as shown in Fig.~\ref{fig10}. The spin gap is defined as $
\Delta_s = E_0(S_z^{\text{tot}} = 1) - E_0(S_z^{\text{tot}} = 0).$
Numerical results show that $\Delta_s \approx 0$ (below $10^{-6}$), in agreement with bosonization predictions at half-filling. We also obtain $\Delta_s \approx 0$ for the other fillings discussed in the text.

\begin{figure*}
\centering
\subfigure[]{\includegraphics[width=0.6\columnwidth]{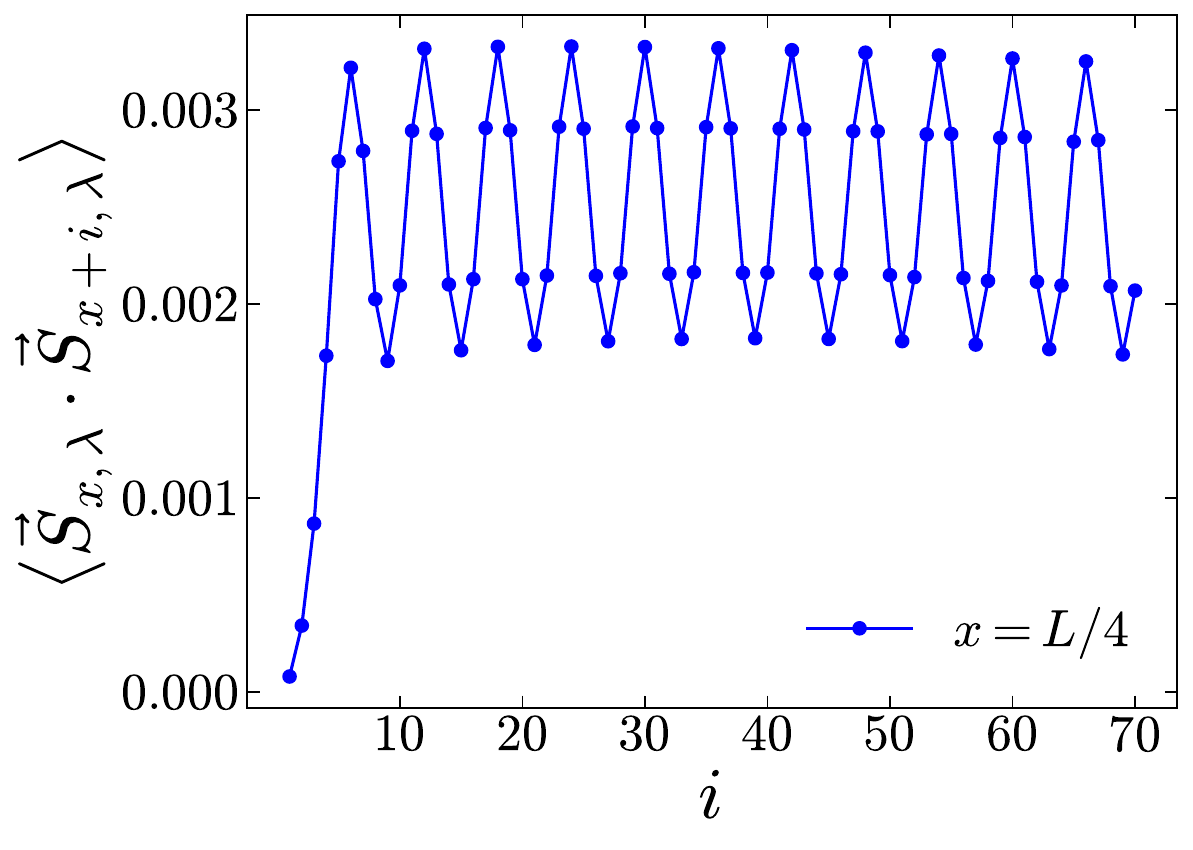}
\label{fig10a}}
\subfigure[]{\includegraphics[width=0.6\columnwidth]{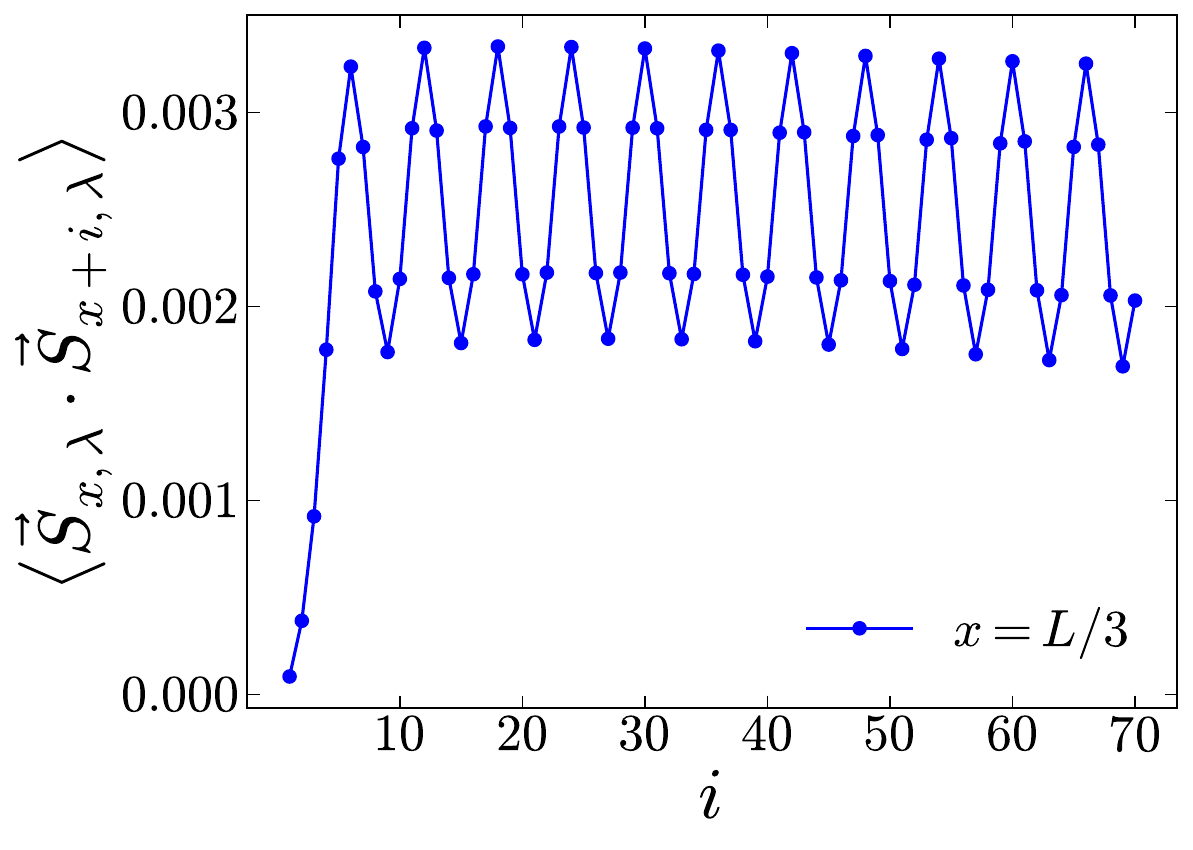}
\label{fig10b}}
\subfigure[]{\includegraphics[width=0.6\columnwidth]{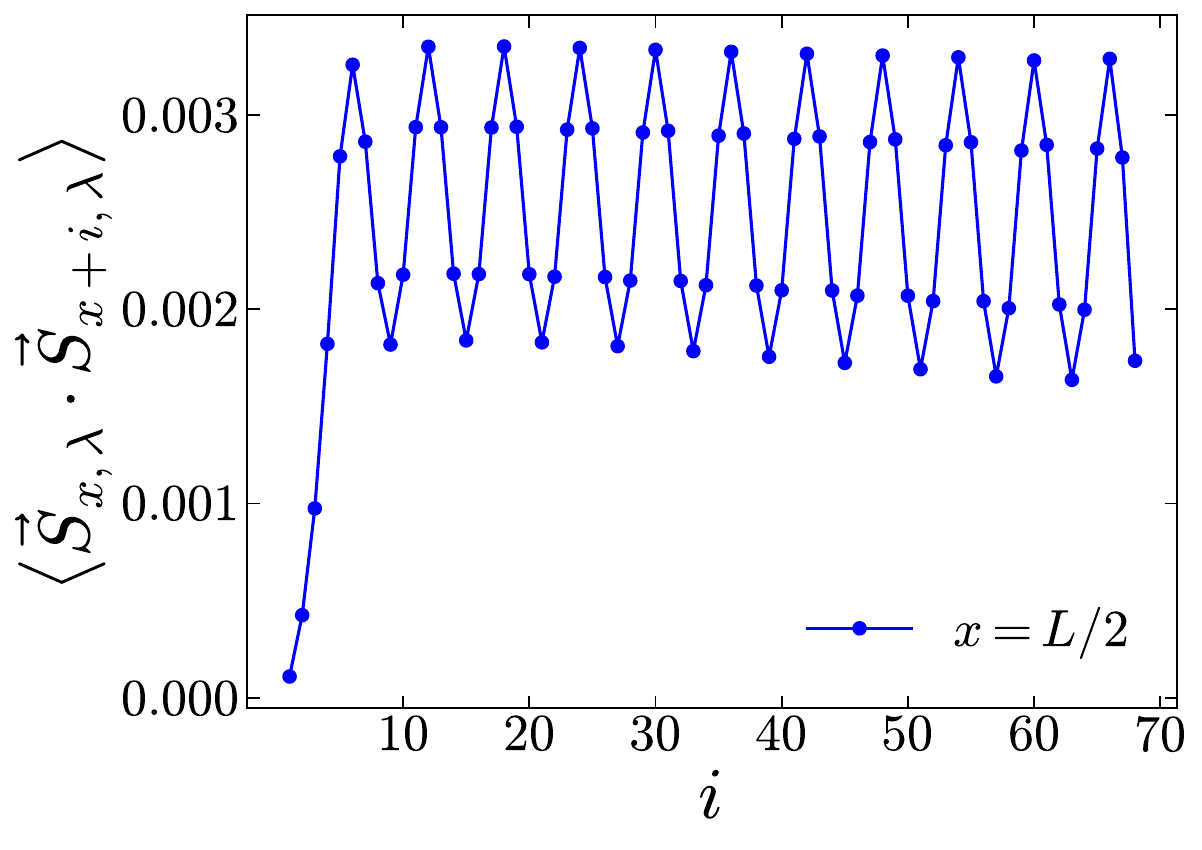}
\label{fig10c}}
\subfigure[]{\includegraphics[width=0.6\columnwidth]{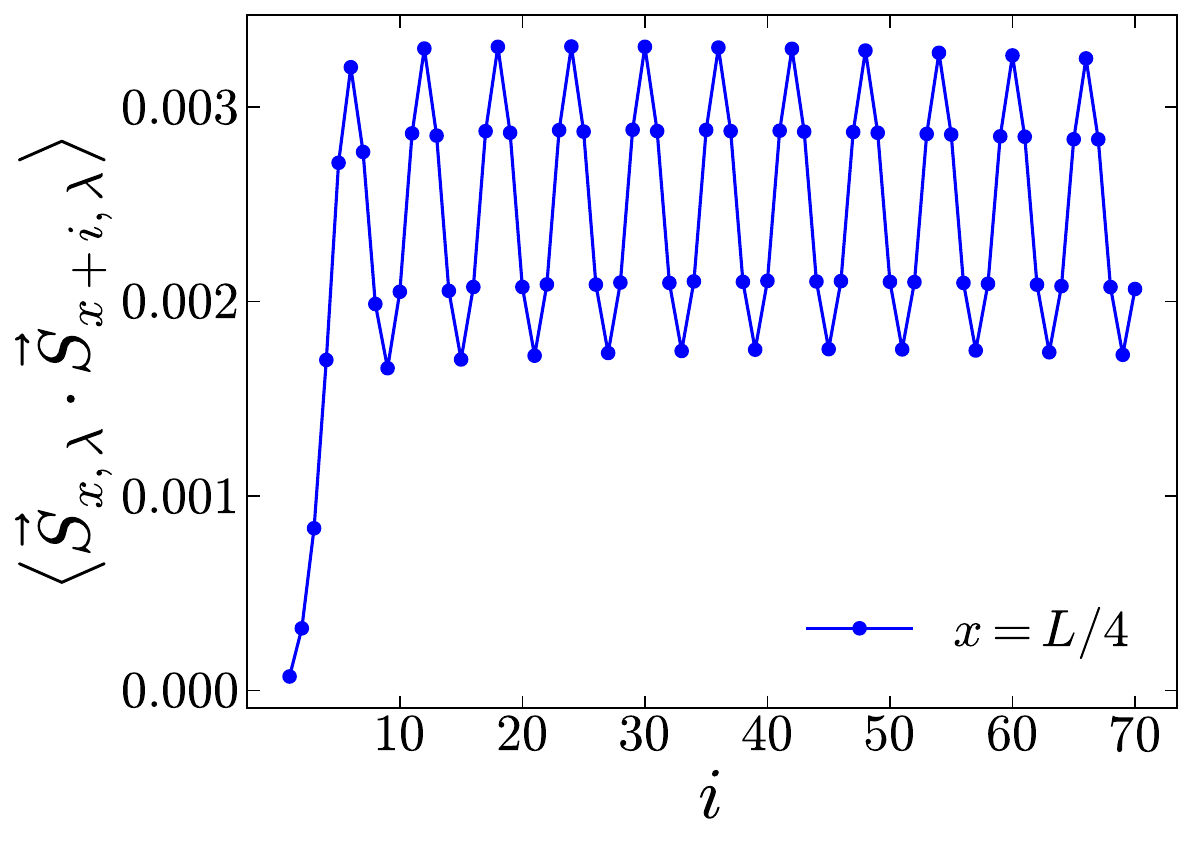}
\label{fig10d}}
\subfigure[]{\includegraphics[width=0.6\columnwidth]{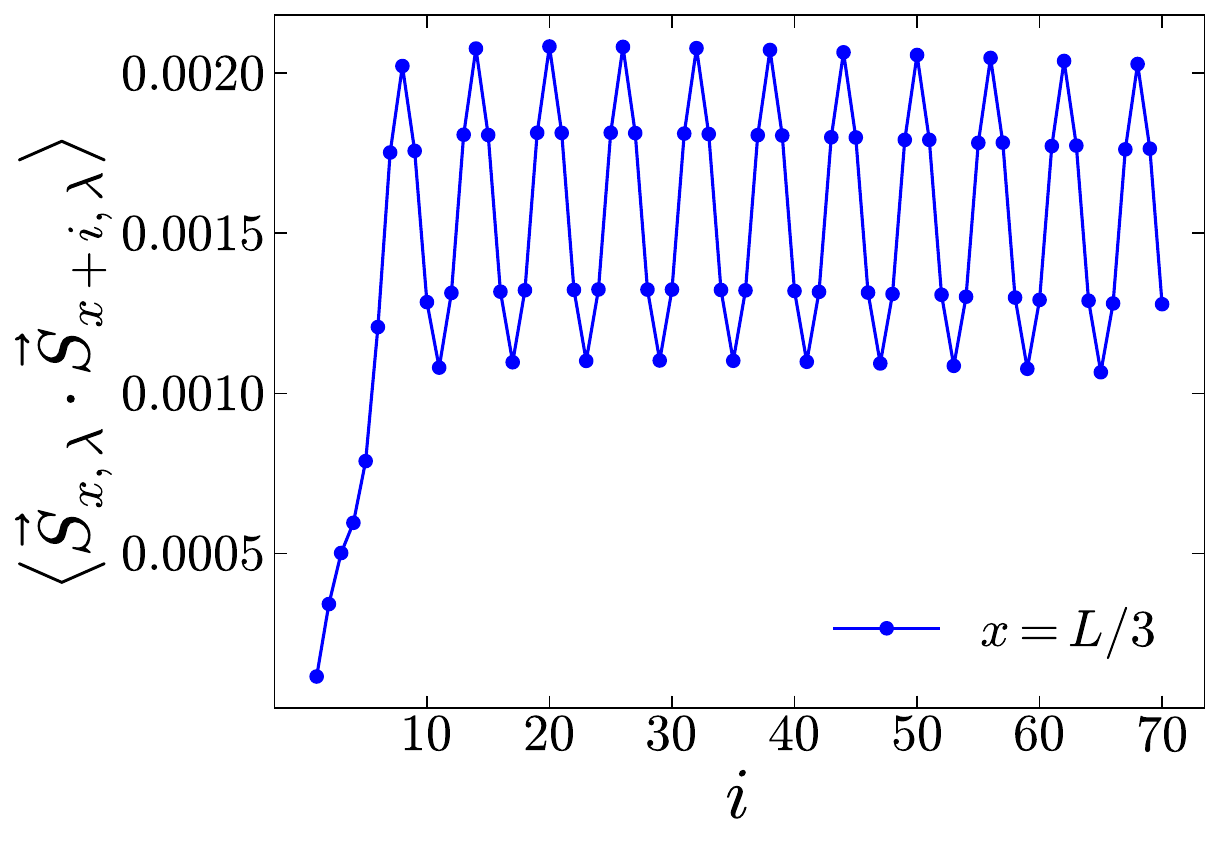}
\label{fig10e}}
\subfigure[]{\includegraphics[width=0.6\columnwidth]{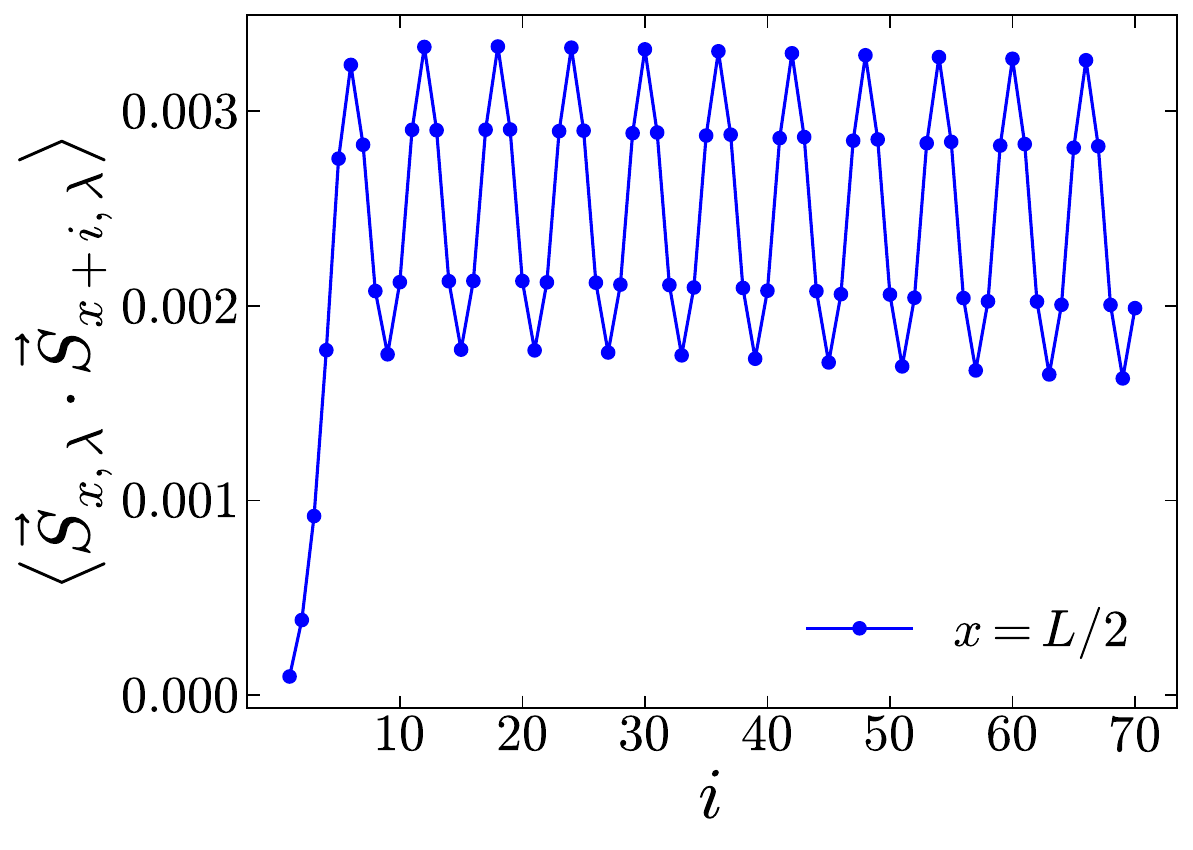}
\label{fig10f}}
\caption{Interleg spin–spin correlations are shown for different choices of reference points for L=144 and L=192.  (a), (b), and (c) correspond to a system size \( L = 144 \), with the reference site \( x \) taken at \( L/4 \), \( L/3 \), and \( L/2 \), respectively, for interaction strength \( U = 1.4 \). Similarly, (d), (e), and (f) correspond to \( L = 192 \) with the same set of reference points and the same value of \( U = 1.4 \). In all cases, the correlations consistently indicate ferromagnetic behavior. Notably, this behavior does not change qualitatively with the choice of reference point, demonstrating that the observed ferromagnetic correlations are robust and not influenced by finite-size effects.
 } \label{fig10}
\end{figure*}

We also present spin-spin correlation functions for electron densities corresponding to one-quarter filling (\(n=12/288\)), three-quarter filling (\(n=36/288\)), and two electrons above (\(n=26/288\)) and below (\(n=22/288\)) half-filling, all for a fixed system size of \(L = 144\). Similar to the half-filled case, these results reveal a transition from antiferromagnetic to ferromagnetic correlations as \(U\) increases. This behavior is illustrated in Fig.~\ref{fig11}.

\begin{figure}
\centering
\subfigure[]{\includegraphics[width=0.48\columnwidth]{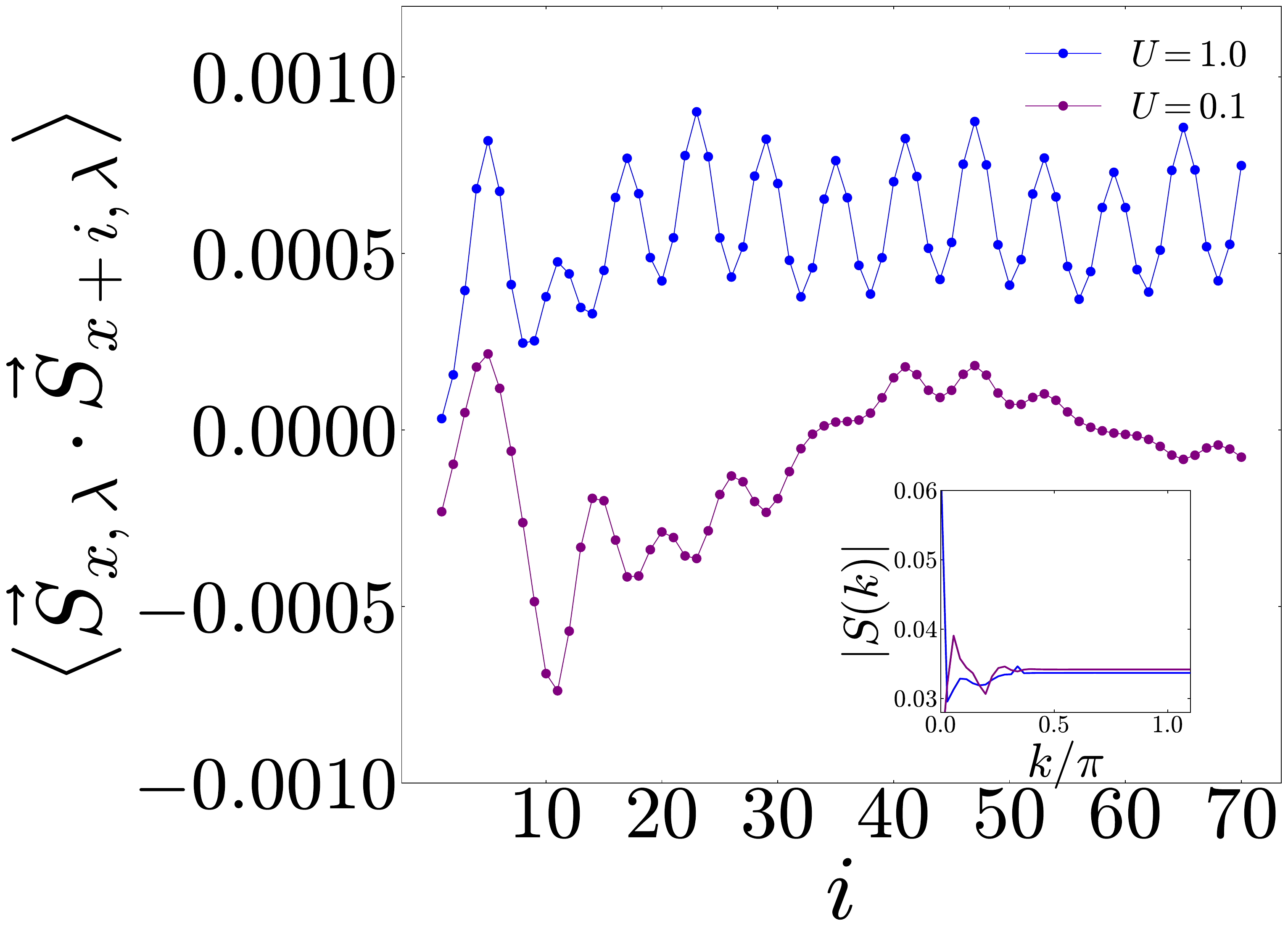}
\label{fig11a}}
\subfigure[]{\includegraphics[width=0.48\columnwidth]{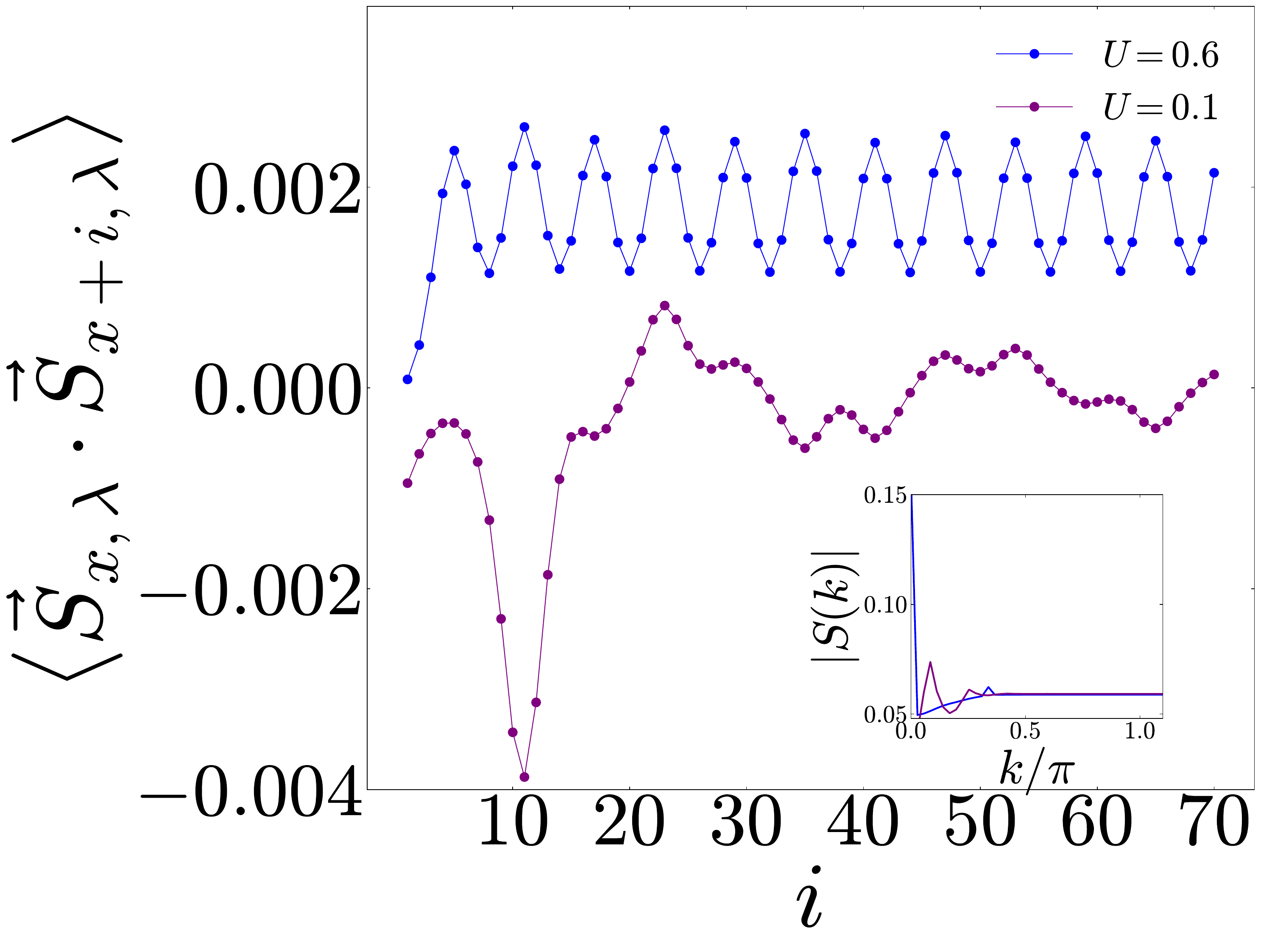}
\label{fig11b}}
\subfigure[]{\includegraphics[width=0.48\columnwidth]{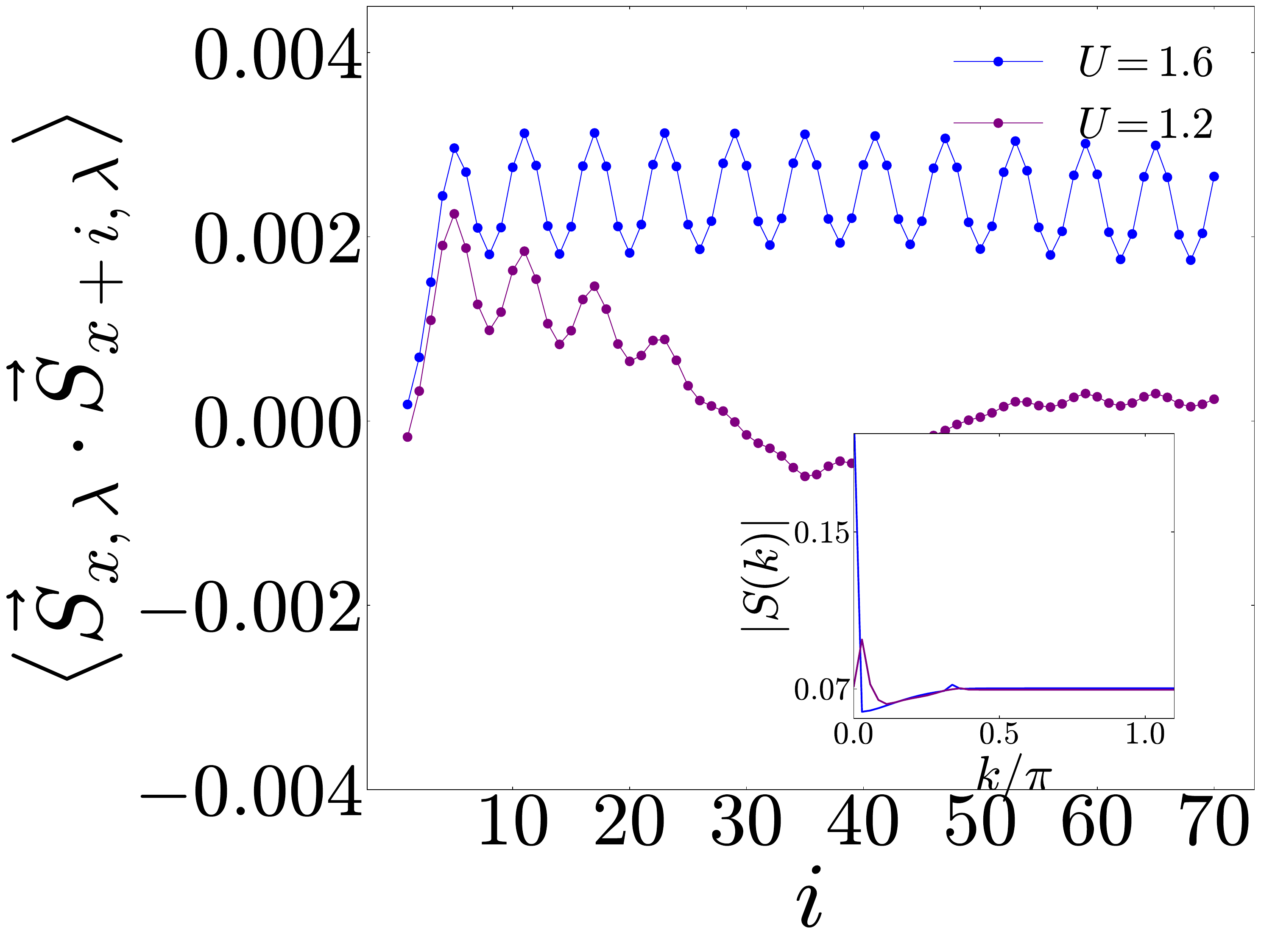}
\label{fig11c}}
\subfigure[]{\includegraphics[width=0.48\columnwidth]{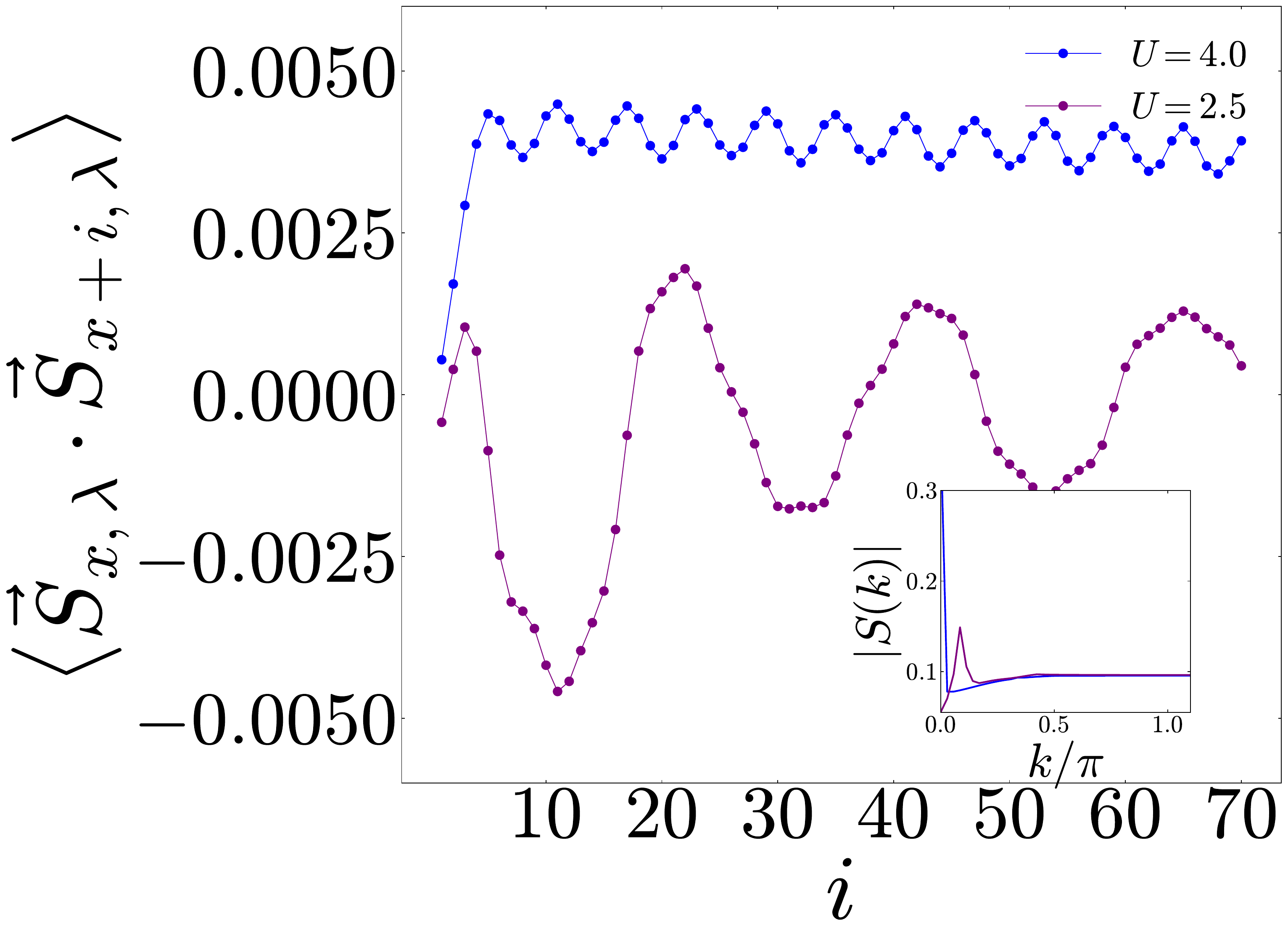}
\label{fig11d}}
\caption{Intraleg spin-spin correlation for different fillings of the lowest band with $\gamma_{0}=0.49$ and $g=0.1$ for (a)  one-quarter filling($n=1/4J=12/288$), (b) slighly below half-filling($n=22/288$), (c) slightly above half-filling($n=26/288$), and (d) three quarter filling($n=3/4J=36/288$). In the inset of each plot corresponding structure factor is plotted. These are all calculated for fixed length $L=144$ and $x=L/4$.} \label{fig11}
\end{figure}

\end{document}